\documentclass[pra,aps,twocolumn,10pt,superscriptaddress,notitlepage,longbibliography]{revtex4-2}
\usepackage[utf8]{inputenc}

\usepackage{bm}

\usepackage{hyperref}
 \hypersetup{
     colorlinks=true,
     linkcolor=blue,
     filecolor=blue,
     citecolor = magenta,      
     urlcolor=red,
     }

\usepackage{xcolor}
\usepackage{graphicx} % For including images

\usepackage{amsthm,amsmath,amssymb,amsfonts,bbm}% Math symbols
\usepackage{braket}
\usepackage{bbold}

\graphicspath{{Figures/}} 
\usepackage{pythonhighlight}

\begin{document}

\title{Theory and simulations of 
few-photon Fock state pulses strongly interacting 
with a single qubit in a waveguide: exact population dynamics and time-dependent spectra}

\author{Sofia Arranz Regidor}
\email{18sar4@queensu.ca}
\affiliation{Department of Physics,
Engineering Physics and Astronomy, Queen's University, Kingston, Ontario, Canada, K7L 3N6}
\author{Andreas Knorr}
\affiliation{Institut für Theoretische Physik, Nichtlineare Optik und Quantenelektronik, Technische Universität Berlin, Berlin, 10623, Germany}
\author{Stephen Hughes}
\affiliation{Department of Physics,
Engineering Physics and Astronomy, Queen's University, Kingston, Ontario, Canada, K7L 3N6}
\email{shughes@queensu.ca}

\date{\today}

\begin{abstract} 

We present a detailed quantum theory and simulations of a few-photon
Fock state pulse interacting with
a two-level system (TLS) in a 
waveguide. For a
 rectangular pulse shape, 
 we present an exact temporal scattering theory for the waveguide-QED system to derive analytical expressions  for the 
 TLS population,
 for 
  1-photon and 2-photon pulses,
 for both chiral and symmetric 
 emitters.
 We also derive the stationary (long time) and time-dependent spectra 
 for 1 photon excitation, and show how these differ at a fundamental level when connecting to TLS population effects.
 %(which would be missing in a regime of weak excitation). 
 Numerically, we also present matrix product state (MPS) simulations, which allow us to compute more general photon correlation functions for arbitrary
 quantum pulses, and we use this approach to also show results for Gaussian quantum pulses, and to confirm the accuracy of our analytical results.
 In addition,
 we show how significant population TLS effects also occur for pulses relatively long compared to the radiative decay time (showing that a weak excitation approximation cannot be made), and investigate the population signatures, nonlinear features and dynamical behavior as a function of pulse length.
These detailed theoretical  results extend and complement our related Letter results 
[Arranz Regidor et al., unpublished, 2024].
 
\end{abstract}

\maketitle

\section{Introduction}

Waveguides can couple to localized quantum emitters in ways that are not possible with cavities, opening up rich
light-matter coupling regimes~\cite{RevModPhys.95.015002,PhysRevA.82.063816,Witthaut_2010,PhysRevLett.106.053601,PhysRevA.101.023807,PhysRevA.106.013714,Shen:05,LeJeannic2022,PhysRevA.83.063828,PhysRevA.110.L031703}.
In a quantum optics picture,
such systems facilitate ``waveguide-QED,'' which can be realized in various material systems, including atoms, molecules, quantum dots,
and superconducting qubits, and these schemes use various waveguide environments~\cite{Trschmann2019,10.1063/1.5117888,PhysRevLett.101.113903,Mirhosseini2019,PhysRevLett.115.163603,PhysRevX.2.011014,PhysRevLett.131.103602,PhysRevA.92.063836,PhysRevX.7.031024,PhysRevLett.120.140404,PhysRevLett.131.033606,PhysRevResearch.2.043213}. 

With superconducting waveguides (transmission lines, which use  microwave photons), qubits can be
arranged as ``giant-atoms,'' where multiple qubits can be collectively coupled with great precision at various spatial points, typically separated in such a way as to cause either destructive or constructive interference~\cite{Kannan2020,PRXQuantum.4.030326,PhysRevResearch.6.013279,Wang_2022}. In atomic systems, 
 laser-cooled atoms can also be trapped
 in nanofibers, evanescently coupled 
 to the propagating fiber modes.
 From a practical perspective,
 semiconductor quantum dots 
 (grown within semiconductor waveguides)
 behave as ideal quantum emitters,
 emitting optical photons on-chip with
 very large waveguide beta factors~\cite{PhysRevB.75.205437,PhysRevLett.113.093603,Paesani2019} (i.e., almost
 all of the emitted photons are into the waveguide mode of interest).
Such waveguides also offer rich design 
flexibility, and, for example, one can easily tune the chirality of the emitter coupling~\cite{leFeber2015,PhysRevLett.115.153901}, opening up other areas of study such as chiral networks~\cite{Sllner2015,doi:10.1126/sciadv.aaw0297,PhysRevResearch.4.023082,Liu:22,PhysRevLett.117.240501,PhysRevX.10.031011}.

From a theoretical perspective, various approaches have been used to describe 
waveguide-QED systems, often with various levels of approximation. For example, with a weak excitation approximation (WEA), usually the problem can be solved within the domain of linear response, for example, using Green function techniques, allowing easy access to the polariton states, even for collective multi-emitter systems~\cite{PhysRevA.95.033818}. 
However, such approximations cannot be done
in the ultrastrong 
coupling 
regime~\cite{frisk_kockum_ultrastrong_2019,forn-diaz_ultrastrong_2019}, where 
%also 
unique quantum waveguide-QED effects appear
from not making a rotating wave approximation\cite{PhysRevLett.111.243602,PhysRevA.98.043816,PhysRevLett.113.263604,PhysRevA.104.053701}.
%PhysRevA.81.042311,PhysRevA.92.063830}.
For the regimes of interest in this paper, we will safely assume the rotating wave approximation (since we are interested in a single emitter whose waveguide couping rate is much smaller than the transition frequency).

 At the quantum field level, other simplified approaches also exist, for example, 
 instantaneous master equations, and
2nd-order Born-Markov solutions where the waveguide can be traced out~\cite{Chang_2012,Mirhosseini2019}. 
% \sh{add some of the Chuang stuff on atomic mirrors for example}. 
Unfortunately, such approaches are not 
suited to look at some of the unique quantum light-matter entanglement that can occur when one treats the waveguide photons exactly, 
in a regime 
where the photons and emitters interact strongly. 

One of the simplest waveguide-QED systems is a single qubit coupled to a waveguide, which already admits various non-trivial solutions. Fan and coworkers
have developed powerful scattering theories to model such systems, usually specialized to model the long-time
output states with respect to the input states, allowing for a powerful description of 
the transmitted and reflected quantum fields, while fully accounting for the
TLS~\cite{PhysRevA.82.063821,PhysRevLett.98.153003,PhysRevA.76.062709,Rephaeli2012FewPhotonSC}. 
In a waveguide-QED
single TLS system, when the waveguide mode interacts strongly with the TLS, the often used WEA 
can drastically fail, even when one
uses temporally short quantum fields at the level of one photon (a single Fock state). Indeed, it has been shown that
one can fully invert a TLS when the appropriate temporal excitation is employed~\cite{PhysRevA.82.033804}. Despite this, the WEA is still commonly employed in many scattering
theories, and it is not generally well known when such an approximation works in terms of the excitation fields. It is also not 
well appreciated if the single excitation, which can still yield significant population transfer and photon-matter entanglement, shows any spectral signatures, as one still expects
a linear response from a single photon
excitation, which is usually well described at the level of a simple classical harmonic oscillator. 

More direct time-dependent modelling
of few photon scattering can be modelled
by solving the time-dependent Schr\"odinger equation~\cite{Nysteen2015,Chen_2011}, or solving the relevant Heisenberg equations of motion~\cite{gardiner_zoller_2010,PhysRevA.106.023708}.
For easier access to the multi-photons, time retardation effects and various causality-modified correlation effects, matrix product states (MPS) is a powerful tensor network approach to solve the entire waveguide-QED system in a numerically exact way~\cite{iblisdir_matrix_2007,Guimond_2017,PhysRevA.103.033704,vanderstraeten_tangent-space_2019,orus_practical_2014,yang_matrix_2018,PhysRevX.10.031011}, in principle only limited by numerical step sizes, etc, which can easily be checked for good numerical convergence. 

In this work, we complement 
our related Letter paper~\cite{sofia2024letter},
by presenting a detailed theoretical study
of how a few-photon pulse interacts with a single TLS in a waveguide. We first present a real-space time-dependent solution to obtain analytical results for both 1-photon and 2-photon (Fock state) pulses, for the TLS population.
We also describe our MPS approach, and use both these methods to connect to the stationary as well as the time-dependent spectra, presenting a selection of results for both a symmetric TLS and a chiral TLS. 

One of our interesting findings is that the population dynamics of a
1-photon-pulse excitation completely cancels out in the transmitted spectra, {\it but} it remains an observable for the time-dependent spectra, which directly 
relates to similar observables that were measured recently with classical pulses in the regime of semiconductor quantum dot cavity systems~\cite{Liu2024,PhysRevLett.132.053602}.
We also show how the 2-photon results yield significantly different stationary spectra, as well as dynamic (time-dependent) spectra. 
Finally, we study how the peak TLS populations depend on the pulse duration, and present solutions for both top-hat pulse shapes as well as Gaussian pulses.

The rest of our article is arranged as follows.
First, in Sec.~\ref{sec:1photonpop}, we
derive an analytical solution for the TLS population using a rectangular (top-hat) temporal quantum pulse containing one photon, for both chiral and symmetrical quantum emitters. We then extend this approach in Sec.~\ref{sec:2photonpop} to a 2-photon quantum pulse with the same temporal shape. 

In Sec.~\ref{sec:4stationaryS}, we derive an analytical solution for the stationary spectrum (i.e., the long time limit) with a one-photon excitation pulse, again for both the symmetrical and chiral emitter case. Here, we see how all the TLS population effects that we observed in the previous sections cancel out, and in the chiral case we simply recover a spectrum that is identical to the input pulse. In order to further investigate this effect
and go beyond the long time limit result, in Sec.~\ref{sec:timedepS}, we derive an analytical solution for the chiral 
{\it time-dependent spectrum}. We start by deriving the transmitted mean photon population in Sec.~\ref{subsec:photonpop}. Then, in Sec.~\ref{subsec:correl}, we derive the transmitted first-order photon correlation functions, knowing that in the limit when $\tau=0$, these simply recover the mean photon population previously calculated. Here, we also verify our results by calculating the same observable using MPS, and show excellent agreement. Next, Sec.~\ref{subsec:Swt} shows the relation between the first-order photon correlations, and the time-dependent spectrum and spectral intensity.

In Sec.~\ref{sec:MPS}, we introduce the MPS theoretical approach we use that can be used to calculate 1-photon and 2-photon quantum pulses with any given temporal shape.

Population results (for the TLS and transmitted fields) are shown in Sec.~\ref{sec:popresults}, where we start by calculating the populations of a symmetrical TLS emitter and compare them with the chiral results shown in Ref.~\cite{sofia2024letter} (Sec.~\ref{subsec:sympopresults});
we then study the influence of pulse duration on the TLS populations in Sec.~\ref{subsec:pulseduration}, where we observe that populations are considerable even for relatively long pulses, still showing results that clearly deviate from any WEA results. Indeed, we observe that we do not recover the 
WEA results until
the pulse durations are on the order of $500 \gamma^{-1}$, where
$\gamma$ is the total TLS decay rate
into the waveguide.

In Sec.~\ref{sec:spectralresults}, we show example analytical results for the time-dependent spectra and spectral intensity for a 1-photon rectangular quantum pulse with a chiral emitter, and confirm that the results agree with the calculations performed using MPS, shown in Ref.~\cite{sofia2024letter}. In addition, we use MPS to show the symmetrical TLS case with rectangular pulses containing 1 and 2 photons.
For chiral TLSs, we also show the time-dependent spectra
for 1-photon and 2-photon pulses, for several longer 
pulse durations ($\gamma t_p= 10$ and $\gamma t_p= 50$), demonstrating pulse break up effects
for intermediate pulse lengths
($\gamma t_p= 10$).

Finally, in Sec.~\ref{sec:gaussian}, we further exploit MPS 
to study Gaussian temporal pulses, 
by calculating similar observables, populations and time-dependent spectral observables, and showing results for both 1 and 2 photons. 
Our conclusions are presented in Sec.~\ref{sec:conclusions}.

\section{Analytical Solutions for the Two-Level-System Population with a One-Photon (Fock State) Pulse}
\label{sec:1photonpop}

We consider an infinite waveguide, that supports a target forward and backward waveguide mode below the light line, and is thus lossless. This could be in the form of a semiconductor nanowire, a circuit-QED
waveguide, or a photonic crystal waveguide---which allows lots of flexibility in how one couples
to artificial atoms such as
quantum dots 
\cite{PhysRevX.2.011014,Sllner2015,JalaliMehrabad:20,RevModPhys.87.347,Trschmann2019,Mnaymneh2019,https://doi.org/10.1002/lpor.200810081,doi:10.1021/acs.nanolett.0c00607}.

Since we are dealing with forward 
(e.g., to the right) and backward (to the left) propagating modes for waveguides, we can derive
 space-time solutions
for the (quasi) one-dimensional (1D) envelopes
of the waveguide mode 
operators,
which satisfy the following propagation equation:
\begin{equation}
(\partial_t \pm v_g \partial_z) A^\dagger_{R/L}(z,t) =i\omega_0A^\dagger_{R/L}(z,t)  ,
\end{equation}
where $R$ and $L$ label `right' and `left' propagating fields, 
and we have 
expanded the waveguide dispersion solution as
$\omega_k = \omega_0 +v_g (k-k_0)$
($k>0$)
or 
$\omega_k = \omega_0 -v_g (k+k_0)$
($k<0$), i.e., we 
assume linear dispersion
at some frequency of interest,
$\omega_0$, which is below the light line.
The mode is assumed to be lossless.

Defining
$A(z,t) = a(z,t) e^{-i\omega_0 t}$,
we can also write in the rotating frame,
\begin{equation}
(\partial_t \pm v_g \partial_z) a^\dagger_{R/L}(z,t)=0,
\label{dazt_noint}
\end{equation}
which are the 1D envelopes for the
{\it slowly-varying} waveguide mode operators.
The boson operators satisfy the fundamental commutation relations,
\begin{equation}
[a_{R/L}^{\phantom \dagger}(z,t),a^{\dagger}_{R/L}(z',t)] = \delta(z-z'),
\label{com_a}
\end{equation}
where $a^\dagger(z,t)$
and $a(z,t)$
have units of ${\rm 1/\sqrt{m}}$.
These operators
describe the creation of a right- and left-moving state at position $z$,
respectively, and allow us to introduce a real space picture for the Hamiltonian description below.

For the TLS-waveguide
system, we start with the following total Hamiltonian,
\begin{equation}
    H = H_{\rm wg} + H_{\rm TLS} + H_{\rm int},
\end{equation}
where the free (uncoupled) waveguide term is
\begin{equation}
\begin{split}
H_{\rm wg}
&= \hbar\int dz
\Big[ (a_R^\dagger(z,t)(\omega_0-iv_g\partial_z )
a_R(z,t) \\
&+ (a_L^\dagger(z,t)(\omega_0+iv_g\partial_z )
a_L^\dagger(z,t)\Big], 
\end{split}
\end{equation}
and the free TLS term, in the Heisenberg picture (where operators are time-dependent
in general),
is
\begin{equation}
    H_{\rm TLS}= \hbar \delta \sigma^+(t) \sigma^-(t),
\end{equation}
where  $\delta = \omega_a - \omega_0
= \omega_a-\omega_p$ is the detuning between the 
TLS, with a resonant frequency $\omega_a$, and the 
waveguide mode pulse, with the main frequency $\omega_p$;
thus, for on resonance, $\delta =0$.

For the TLS-field interaction term,
after using a dipole and rotating wave approximation,
we have
\begin{equation}
\begin{split}
     &H_{\rm int} = i\hbar\int dz 
\Big[ \big( g_{R}^* \sigma^+(t)  a_R (z,t) \delta(z-z_0) \\
     & +
g_{L}^* \sigma^+(t)   a_L (z,t) \delta(z-z_0) \big) -{\rm {\rm H.c.}}\Big]
     ,     
     \label{Hintint}
\end{split}
\end{equation}
which, for a single TLS
at $z_0$, yields
\begin{equation}
\begin{split}
     &H_{\rm int} = i\hbar
\Big[ \big( g_{R}^* \sigma^+(t) a_R(z_0,t)   
      +
g_{L}^* \sigma^+(t) a_L (z_0,t)   \big) -{\rm {\rm H.c.}}\Big]
     ,     
     \label{Hint}
\end{split}
\end{equation}
where we introduced the photon-TLS coupling term
$g_{R/L}$ (with right and left coupling indices), 
for an atom at position
${\bf r}_0(x_0,y_0,z_0)$.
This waveguide photon-TLS coupling
term is defined from
\begin{align}
g_{R} =g_{R}(k_0) =  \sqrt{a_0} \sqrt{\frac{\omega_0}{2 \hbar \epsilon_0 }}
{\bf d} \cdot {\bf f}_{k_0}({\bf r}_0) 
 =  d \sqrt{\frac{a_0\omega_0}{2\epsilon_0 \hbar V^R_{\rm eff}  }}
,  \\
g_{L}=g_{L}(k_0) =  \sqrt{a_0} \sqrt{\frac{\omega_0}{2\epsilon_0  \hbar}}
{\bf d} \cdot {\bf f}^*_{k_0}({\bf r}_0)
 =  d  \sqrt{\frac{a_0 \omega_0}{2\epsilon_0   \hbar V^L_{\rm eff}}}
,
\end{align}
where $V^{L/R}_{\rm eff} = V^{L/R}_{\rm eff}({\bf r}_0)$ is an effective mode volume (per unit cell),
${\bf d}$ is the dipole moment, and  $|{\bf f}_k|^2$ has units of inverse volume, with the normalization
$\int dr^3 \epsilon({\bf r}) |{\bf f}_k({\bf r})|^2=1$~\cite{PhysRevB.75.205437}. Note that we write $g_{R/L}=g_{R/L}(k_0)$ for brevity, 
and these coupling strengths
have units of $\left[ {\rm \sqrt{m}/s} \right]$.
Here $a_0$ is the pitch in the case of a photonic crystal waveguide, or a length associated with periodic boundary conditions.
We can also relate this to the 
TLS population decay rate,
$\gamma_{R/L} = |g_{R/L}|^2/v_g$, so that
\begin{equation}
\gamma_{R/L} \equiv \gamma_{R/L}({\bf r}_0) = \frac{d^2 a_0 \omega_0}{2 v_g V^{R/L}_{\rm eff}({\bf r}_0) \hbar \epsilon_0},
\end{equation}
in agreement with known results for photonic crystal waveguides
\cite{PhysRevB.75.205437,PhysRevLett.115.153901}.

The transmitted and reflected fields can be calculated from the corresponding equations of motion for the photon operators, 
\begin{equation}
\begin{split}
    \frac{da_{R/L}(z,t)}{dt} &= \frac{i}{\hbar} \left[ H, a_{R/L}(z,t)\right] \\
    &=  \frac{i}{\hbar} \left[H_{\rm wg} + H_{\rm int}, a_{R/L}(z,t)\right].
\end{split}
\end{equation}
Using the result of the unperturbed  field from Eq.~\eqref{dazt_noint}, 
together with
Eq.~\eqref{Hint} and Eq.~\eqref{com_a},
we obtain
\begin{equation}
\begin{split}
     &(\partial_t - v_g \partial_z) a_{R}(z,t)= - \frac{g_R}{v_g} \sigma^-(t)\delta(z-z_0), \\
     &(\partial_t + v_g \partial_z) a_{L}(z,t)= - \frac{g_L}{v_g} \sigma^-(t)\delta(z_0-z),
     \end{split}
\end{equation}
and after integrating,
\begin{equation}
\begin{split}
     a_{R}(z,t) &= a_{\rm in,R}\left(t-\frac{z}{v_g}\right)\\
     & - \frac{g_R}{v_g} \sigma^-\left(t-\frac{z-z_0}{v_g}\right)\theta(z-z_0), \\
     a_{L}(z,t)&=a_{\rm in ,L}\left(t+\frac{z}{v_g}\right)\\ 
     &- \frac{g_L}{v_g} \sigma^-\left(t-\frac{z+z_0}{v_g}\right)\theta(z_0-z).
\end{split}
\end{equation}
Note that the input fields contain noise 
fields (or free fields) as well as any considered
quantum source. 

Now, considering a right input field operator,
and neglecting vacuum noise
terms as they will not contribute
in normal ordering), 
$a_{\rm in,R}(t-z/{v_g})=a_{\rm in}(t-z/{v_g})$ and $a_{\rm in,L}(t+z/{v_g})=0$, then the transmitted field is
\begin{equation}
\begin{split}
a_R(z,t) &= a_{\rm in}\left(t-\frac{z}{v_g}\right) -\frac{g_R}{{v_g}} \sigma^-\left(t-\frac{z-z_0}{v_g}\right)\theta(z-z_0), 
\label{generalfieldR}
\end{split}
\end{equation}
while the reflected field (to the left of the TLS) is
\begin{equation}
\begin{split}
a_L(z,t) &= -\frac{g_L}{{v_g}} \sigma^-\left(t-\frac{z_0-z}{v_g}\right)\theta(z_0-z), 
\label{generalfieldL}
\end{split}
\end{equation}
It is important to note again that $a(z,t)$ has units of $[1/\sqrt{\rm m}]$ 
and, consequently, $a_{\rm in}$ has the same units.

Let us now assume 
$z_0=0$, with a right (forward) propagating input field, and using $\theta(0) = \frac{1}{2}$, then 
%
%\begin{equation}
\begin{align}    
     a_R(0,t) &= a_{\rm in}(t) - \frac{g_{R}}{2v_g} \sigma^-(t),  \  z=0 \nonumber \\
     a_R(z,t) &= a_{\rm in}\left(t-\frac{z}{v_g}\right) - \frac{g_{R}}{{v_g}} \sigma^-\left(t-\frac{z}{v_g}\right) ,  \  z>0   
    \label{generalfieldR0}
\end{align}
and
\begin{align}    
     a_L(0,t) &= - \frac{g_{L}}{2{v_g}} \sigma^-(t),  \  z=0 \nonumber \\
     a_L(z,t) &= -\frac{g_{L}}{{v_g}} \sigma^-\left(t+\frac{z}{v_g}\right) ,  \ z<0. 
    \label{generalfieldL0}
\end{align}

For any temporal input pulse shape
$f(t)$ (which modulates the input quantum state), normalized with
\begin{equation}
\int dt |f(t)|^2=1,
\end{equation}
we can define
the input
operator as
$a_{\rm in}(t) =  a_0 \beta_1 f(t)$,
with $a^\dagger_0\ket{0}=1$.
The temporal pulse has a center frequency
$\omega_p = \omega_0$, and propagates freely in the waveguide in the absence of a TLS. The pulse bandwidth is then determined
by the quantum pulse time duration and temporal profile.

For a 1-photon pulse, 
we need to satisfy
\begin{align}
 &\int_{-\infty}^{\infty} dz \braket{a_{\rm in}^\dagger (t_z) a_{\rm in} (t_z)}
=  \nonumber \\
&\int_{-\infty}^{\infty} dz \braket{a_{\rm in}^\dagger (t-z/v_g) a_{\rm in} (t-z/v_g)} = 1,
\end{align}
which enables us to obtain  
$\beta_1$ for a 1-photon input state.
Note also that
\begin{equation}
\int_{-\infty}^{\infty} \braket{a_0^\dagger a_0 }|f(t)|^2 dz = \int_{-\infty}^{\infty} |f(t)|^2 dz = v_g. 
\end{equation}
Thus, to obtain a normalized pulse containing one photon, $\beta_1=-1/\sqrt{v_g}$, and the input boson operator  is 
\begin{equation}
a_{\rm in}(t)= -a_0 \frac{f(t)}{\sqrt{v_g}},
\end{equation}
which depends also the normalized pulse shape $f(t)$, and the annihilation operator $a_0$, which annihilates a photon in the 
waveguide, with the relation:  $a_0 \ket{0} =0$.

\subsection{Symmetrical Emitter (Two-Level System) Solution}

To obtain the expectation value of the TLS population, we seek to solve the appropriate differential equations of motion in the Heisenberg representation. 
We first
consider the case of symmetrical TLS coupling. The equation of motion
for $\sigma^-(t)$ is obtained from:
% %
\begin{equation}
\begin{split}
    &\frac{d \sigma^- (t)}{dt} = \frac{i}{\hbar} \left[ H(t), \sigma^-(t) \right] 
    = i \delta \sigma^-(t) \\
    &-g_R^{*}  \sigma^+(t) a_{\rm R}(0,t) \sigma^-(t) -g_L^{*}  \sigma^+(t) a_{\rm L}(0,t) \sigma^-(t) \\
    & + g_R^{*}  \sigma^-(t) \sigma^+(t)  a_{\rm R}(0,t) 
     + g_L^{*}  \sigma^-(t) \sigma^+(t)  a_{\rm L}(0,t) \\
     & = i \delta \sigma^-(t)-\left(\frac{\gamma_R}{2} +\frac{\gamma_L}{2}\right) \sigma^-(t) -g_R^* a_{\rm in}(t) \sigma^z (t),
\end{split}
    \label{dsigma}
\end{equation}
 with
$\gamma_R=\gamma_L=\gamma /2 $ and $g_R = g_L =g_0$, with $g_0$ real,
 Equation~\eqref{dsigma} gives
\begin{equation}
    \frac{d \sigma^- (t)}{dt} =  i \delta \sigma^- (t)  -\frac{\gamma}{2} \sigma^-(t) - g_0 a_{\rm in}(t) \sigma^z(t). 
    \label{dsigma2}
\end{equation}

Next, we derive the  differential equation for the TLS population,
\begin{equation}
\begin{split}
 &\frac{d \sigma^+(t)\sigma^-(t)}{dt} = \frac{i}{\hbar}[H,\sigma^+(t)\sigma^-(t)] = 
  g_R a_{\rm R}^\dagger(0,t)  \sigma^-(t) \\
 &+ g_L a_{\rm L}^\dagger(0,t) \sigma^-(t)  
 + g_R^*  \sigma^+(t) a_{\rm R}(0,t)  
 + g_L^*  \sigma^+(t) a_{\rm L}(0,t) \\
& = g_R a_{\rm in}^\dagger (t) \sigma^-(t) + g_R^* \sigma^+(t) a_{\rm in}(t) -\left( \gamma_R + \gamma_L\right)\sigma^+(t) \sigma^-(t) \\
&= 
-\gamma \sigma^+(t) \sigma^-(t)
+\left( g_{0}
a_{\rm in}^\dagger(t) \sigma^-(t) +
{\rm {\rm H.c.}} \right),
\end{split}
\label{sigma+-}
\end{equation}
which shows explicitly the
radiative emission rate,
$\gamma$, in the absence
of any quantum input pulse.

Now, if we apply the differential equation to an input pulse containing one photon, we obtain
\begin{widetext}
\begin{equation}
\begin{split}
    &\frac{d}{dt} \bra{1,g} \sigma^+ (t) \sigma^- (t) \ket{1,g} = 
    -\gamma \bra{1,g} \sigma^+(t) \sigma^-(t)  \ket{1,g} 
+\left[ \bra{1,g} g_{0}
a_{\rm in}^\dagger(t) \sigma^-(t) +
{\rm {\rm H.c.}} \ket{1,g} \right]\\
&=-\gamma \bra{1,g} \sigma^+(t) \sigma^-(t)  \ket{1,g}
- \sqrt{\frac{\gamma}{2}}\Big[ f^*(t)    
\bra{0,g}
 \sigma^-(t)  \ket{1,g}  
 + f(t)    
\bra{1,g}
 \sigma^+(t)  \ket{0,g} \Big], 
\end{split}
\label{sigma+-1p}
\end{equation}
and
\begin{equation}
\begin{split}
    &\frac{d}{dt} \bra{0,g} \sigma^- (t) \ket{1,g} = \left(i \delta - \frac{\gamma}{2} \right)\bra{0,g} \sigma^-(t)\ket{1,g} -g_0 \bra{0,g} a_{\rm in}(t) \left( 2\sigma^+(t) \sigma^-(t) - \mathbb{1} \right)\ket{1,g} \\
    &=  \left(i \delta - \frac{\gamma}{2} \right) \bra{0,g} \sigma^-(t)\ket{1,g} - \sqrt{\gamma/2} f(t) \left[1- 2\bra{0,g} \sigma^+(t) \sigma^-(t) \ket{0,g}\right]
    .
    % \label{sigmaminus1p} 
\end{split}
\end{equation}
Considering the resonant case,
i.e., $\delta=0$, then we have
\begin{equation}
\begin{split}
    &\frac{d}{dt} \bra{0,g} \sigma^- (t) \ket{1,g} =  
   - \frac{\gamma}{2} \bra{0,g} \sigma^-(t)\ket{1,g} - \sqrt{\gamma/2} f(t) \left[1- 2\bra{0,g} \sigma^+(t) \sigma^-(t) \ket{0,g}\right]
    .
    \label{sigmaminus1p} 
\end{split}
\end{equation}

This general approach can be extended to pulses with a higher number of photons ($n$)~\cite{PhysRevA.106.023708},
\begin{equation}
\begin{split}
    \frac{d}{dt} \bra{n,g}\sigma^+(t) \sigma^-(t)\ket{n,g}&=-\gamma \bra{n,g}\sigma^+(t) \sigma^-(t)\ket{n,g} 
    -\sqrt{\frac{n \gamma}{2}} \left[ f^*(t) \bra{n-1,g} \sigma^-(t)\ket{n,g} +
    {\rm {\rm H.c.}} \right],  
\end{split}
\label{general}
\end{equation}
where $n$ represents the photon number and $f(t)$ is the pulse shape. 
To solve Eq.~\eqref{general}, we also need the more general coherence equation of motion
\begin{equation}
\begin{split}
    &\frac{d}{dt} \bra{n-1,g} \sigma^-(t)\ket{n,g}=-\frac{\gamma}{2} \bra{n-1,g}\sigma^-(t)\ket{n,g} 
    -\sqrt{\frac{n \gamma}{2}} f(t) \left[1 -2\bra{n-1,g} \sigma^+(t) \sigma^-(t)\ket{n-1,g} \right].  
\end{split}
\label{sigmaminus}    
\end{equation}

For the case of a rectangular pulse, of length $t_p$, containing one photon,
the pulse {\it envelope} is defined from
\begin{align}
f_{\rm rect}(t) ={\rm rect} \left ( \frac{t}{t_p} \right) &=
0, \ \ t > t_p \nonumber \\
{\rm rect} \left (\frac{t}{t_p} \right) &=
\frac{1}{\sqrt{t_p}} \ \ 0 < t \leq t_p, 
\end{align}
and Eq.~\eqref{general} transforms to
\begin{equation}
\begin{split}
    &\frac{d}{dt} \bra{1,g}\sigma^+(t) \sigma^-(t)\ket{1,g}=-\gamma \bra{1,g}\sigma^+(t) \sigma^-(t)\ket{1,g} 
    -2\sqrt{\frac{\gamma}{2 t_p}}  \bra{0,g} \sigma^-(t)\ket{1,g} , \  t  \leq t_p  \\  
    &\frac{d}{dt} \bra{1,g}\sigma^+(t) \sigma^-(t)\ket{1,g}=-\gamma \bra{1,g}\sigma^+(t) \sigma^-(t)\ket{1,g} , \ t > t_p.
\end{split}    
\label{1phdiff}
\end{equation}
\end{widetext}

Solving for Eq.~\eqref{sigmaminus1p}
first, yields
\begin{equation}
\begin{split}
    &\frac{d}{dt} \bra{0,g}\sigma^-(t)\ket{1,g}= 
    -\frac{\gamma}{2} \bra{0,g}\sigma^-(t)\ket{1,g} - \sqrt{\frac{\gamma}{2 t_p}},  
\end{split}    
\end{equation}
which, after integrating, becomes
\begin{equation}
     \bra{0,g}\sigma^-(t)\ket{1,g} = \sqrt{\frac{2}{\gamma t_p}} \left[ e^{-\gamma t/2} -1 \right].
\end{equation}
Using this solution in Eq.~\eqref{1phdiff} gives
\begin{align}
%\begin{split}
    &\frac{d}{dt} \bra{1,g}\sigma^+(t) \sigma^-(t)\ket{1,g}= \nonumber \\
    &-\gamma \bra{1,g}\sigma^+(t) \sigma^-(t)\ket{1,g} 
    -\frac{2}{t_p} \left[ e^{-\gamma t/2} -1 \right] , \ t \leq t_p  \label{1phdifff1}  \\  
    &\frac{d}{dt} \bra{1,g}\sigma^+(t) \sigma^-(t)\ket{1,g}= \nonumber \\
    &-\gamma \bra{1,g}\sigma^+(t) \sigma^-(t)\ket{1,g} , \ t > t_p.
%\end{split}    
\label{1phdifff2}
\end{align}

Finally, integrating Eqs.~\eqref{1phdifff1} and ~\eqref{1phdifff2}, 
 gives the TLS population solution
 as a function of time
\begin{equation}
\begin{split}
    &\bra{1,g}\sigma^+(t) \sigma^-(t)\ket{1,g}=\frac{2}{\gamma t_p} \left[ e^{-\gamma t/2} -1 \right]^2 , \ t \leq t_p,  \\  
    &\bra{1,g}\sigma^+(t) \sigma^-(t)\ket{1,g}=\frac{2}{\gamma t_p} \left[ e^{\gamma t_p/2} -1 \right]^2 e^{-\gamma t} 
    , \ t>t_p.
\end{split}    
\label{1phsol0}
\end{equation}

\subsection{Chiral Emitter Solution}

In the case of having a chiral TLS system, where $\gamma_R=\gamma$ and $\gamma_L=0$ (which is the one studied in Ref.~\cite{sofia2024letter}, our companion Letter), the relevant interaction equations of motion are now

\begin{equation}
\begin{split}
 &\frac{d \sigma^+(t)\sigma^-(t)}{dt} 
=  
-\gamma \sigma^+(t) \sigma^-(t)
+  \left( g_{R}
a_{\rm in}^\dagger(t) \sigma^-(t) +
{\rm {\rm H.c.}} \right),
 \end{split}
\end{equation}
and
\begin{equation}
    \begin{split}
       & \frac{d \sigma^-(t)}{dt}= 
       \left (i\delta -\frac{\gamma}{2} \right)\sigma^-(t) - g_R
       a_{\rm in} \sigma^z(t).  
    \end{split}
\end{equation}
Considering again the system on resonance, then
\begin{equation}
    \begin{split}
       & \frac{d \sigma^-(t)}{dt} = 
       -\frac{\gamma}{2} \sigma^-(t) -  g_R
       a_{\rm in} \sigma^z(t) . 
    \end{split}
\end{equation}

Now using this with the one-photon rectangular pulse, and  considering that in the chiral case, ${|g_R|}^2 / v_g = \gamma_R =\gamma$, (for $t < t_p$, i.e., while the pulse is being applied), we obtain
\begin{equation}
    \begin{split}
       & \bra{0,g}\frac{d \sigma^-(t)}{dt} \ket{1,g} 
       = -\frac{\gamma}{2}\bra{0,g} \sigma^-(t)\ket{1,g} -\sqrt{ \frac{\gamma}{t_p}}.
       \label{sigma-chiral}
    \end{split}
\end{equation}
Integrating this equation gives
\begin{equation}
    \bra{0,g}\sigma^-(t)\ket{1,g} = \frac{2}{\sqrt{\gamma t_p}} \left[ 
 e^{-\gamma t/2} -1 \right],
\end{equation}
and the time-derivative for the population is derived to be
\begin{equation}
\begin{split}
     &\frac{d \sigma^+(t)\sigma^-(t)}{dt} = \\
     &-\gamma \sigma^+(t)\sigma^-(t) -  \left( g_R a_{\rm in}^\dagger(t) \sigma^-(t) +{\rm H.c.}  \right),    
\end{split}
\label{sigma+-chiral}
\end{equation}
and thus
\begin{equation}
\begin{split}
     &\bra{1,g}\frac{d \sigma^+(t)\sigma^-(t)}{dt} \ket{1,g} =\\
 &= -\gamma \bra{1,g}\sigma^+(t)\sigma^-(t) \ket{1,g} - \frac{4}{t_p}\left[ 
 e^{-\gamma t/2} -1 \right]. 
\end{split}
\end{equation}

Subsequently, the relevant differential equations 
of motion in this case are
as follows:
\begin{align}
%\begin{split}
    &\frac{d}{dt} \bra{1,g}\sigma^+(t) \sigma^-(t)\ket{1,g}= \nonumber \\
    &-\gamma \bra{1,g}\sigma^+(t) \sigma^-(t)\ket{1,g} 
    -\frac{4}{t_p} \left[ e^{-\gamma t/2} -1 \right] ,\ t \leq t_p, 
    \label{1phdiff3_a}
    \\  
    &\frac{d}{dt} \bra{1,g}\sigma^+(t) \sigma^-(t)\ket{1,g}= \nonumber \\
    &-\gamma \bra{1,g}\sigma^+(t) \sigma^-(t)\ket{1,g} ,\  t > t_p.
%\end{split}    
\label{1phdiff3}
\end{align}

Integrating Eqs.~\eqref{1phdiff3_a}-\eqref{1phdiff3} then gives
the desired population dynamics
for the chiral TLS,
\begin{equation}
\begin{split}
    &\bra{1,g}\sigma^+(t) \sigma^-(t)\ket{1,g}=\frac{4}{\gamma t_p} \left[ e^{-\gamma t/2} -1 \right]^2, \ t \leq t_p,  \\   
    &\bra{1,g}\sigma^+(t) \sigma^-(t)\ket{1,g}=\frac{4}{\gamma t_p} \left[ e^{\gamma t_p/2} -1 \right]^2 e^{-\gamma t}, \ t>t_p,
\end{split}    
\label{1phsol}
\end{equation}
which yields precisely double the population of the symmetrical case
[Eq.~\eqref{1phsol0}], as might be expected (since the coupling strength of the chiral-field interaction is double the symmetric case).

\section{Analytical Solution for the Two-Level-System Population using a Two-Photon Pulse}
\label{sec:2photonpop}

\subsection{Symmetrical Emitter Population Dynamics}
\label{subsec:sympopulation}

Using the same top-hat pulse profile as before, we now consider
a 2-photon excitation pulse.
While $t \leq t_p$, Eqs.~\eqref{general} and \eqref{sigmaminus} transform to
\begin{equation}
\begin{split}
    &\frac{d}{dt} \bra{2,g}\sigma^+(t) \sigma^-(t)\ket{2,g}= -\gamma \bra{2,g}\sigma^+(t) \sigma^-(t)\ket{2,g} \\
    &
    -\sqrt{ \frac{\gamma}{t_p}} \left[ \bra{1,g} \sigma^-(t)\ket{2,g} +
    {\rm {\rm H.c.}} \right],  
\end{split}
\label{2phdiff}
\end{equation}
and
\begin{equation}
\begin{split}
    &\frac{d}{dt} \bra{1,g} \sigma^-(t)\ket{2,g}=-\frac{\gamma}{2} \bra{1,g}\sigma^-(t)\ket{2,g} \\
    &-\sqrt{ \frac{\gamma}{t_p}} \left[1 -2\bra{1,g} \sigma^+(t) \sigma^-(t)\ket{1,g} \right],  
\end{split}   
\label{eq:1sigma-2}
\end{equation}
where the last term contains the 1-photon solution, derived explicitly in the previous 
subsection. Using Eq.~\eqref{1phsol},
then 
\begin{equation}
\begin{split}
    &\frac{d}{dt} \bra{1,g} \sigma^-(t)\ket{2,g}=-\frac{\gamma}{2} \bra{1,g}\sigma^-(t)\ket{2,g}  \\
    &-\sqrt{ \frac{\gamma}{t_p}} \left[1 - \frac{4}{\gamma t_p} \left( e^{-\gamma t/2} -1 \right)^2 \right],        
\end{split}
    \label{2phsigmaminus} 
\end{equation}
and after integrating 
Eq.~\eqref{2phsigmaminus},
we obtain
\begin{equation}
\begin{split}
    &\bra{1,g} \sigma^-(t)\ket{2,g}= \frac{1}{\sqrt{\gamma t_p}} \Big[ -\frac{4}{\gamma t_p} e^{-\gamma t} \\
    &+ \left( \frac{4t}{t_p} + 1 \right) e^{-\gamma t/2} + \left( \frac{4}{\gamma t_p} -1 \right)
  \Big].    
\end{split}
\end{equation}

We can use this result to solve Eq.~\eqref{2phdiff}, so that:
\begin{equation}
\begin{split}
    &\frac{d}{dt} \bra{2,g}\sigma^+(t) \sigma^-(t)\ket{2,g}=-\gamma \bra{2,g}\sigma^+(t) \sigma^-(t)\ket{2,g} \\
    &-\frac{4}{t_p} \left[ -\frac{4}{\gamma t_p} e^{-\gamma t} + \left( \frac{4t}{t_p} + 1 \right) e^{-\gamma t/2} + \left( \frac{4}{\gamma t_p} -1 \right)
    \right],    
\end{split}
\end{equation}
and then the TLS population, while $t \leq t_p$, is
\begin{equation}
\begin{split}
     &\bra{2,g}\sigma^+(t) \sigma^-(t)\ket{2,g}= \frac{4}{t_p} \Big[ \left( \frac{4t}{t_p} +\frac{20}{\gamma t_p} +1 \right) e^{-\gamma t} \\
     &+ \left( \frac{8t}{t_p} - \frac{16}{\gamma t_p} -2 \right) e^{-\gamma t/2} + \left( -\frac{4}{\gamma t_p} +1 \right)
    \Big].
\end{split}
    \label{part12ph}
\end{equation}

For $t > t_p$, Eq.~\eqref{general} yields
\begin{equation}
    \frac{d}{dt} \bra{2,g}\sigma^+(t) \sigma^-(t)\ket{2,g}=-\gamma \bra{2,g}\sigma^+(t) \sigma^-(t)\ket{2,g}. 
\end{equation}
Integrating and ensuring continuity at $t=t_p$, we obtain
\begin{equation}
\begin{split}
    &\bra{2,g}\sigma^+(t) \sigma^-(t)\ket{2,g} = \frac{4}{t_p} \Big[ \left( -\frac{4t}{t_p} +1 \right) e^{\gamma t_p} \\
    & + \left( -\frac{16}{\gamma t_p} + 6 \right) e^{\gamma t_p/2} + \left( \frac{20}{\gamma t_p} +5 \right)
    \Big] e^{-\gamma t}.       
\end{split}
    \label{part22ph}
\end{equation}

Finally, combining Eqs.~\eqref{part12ph} and \eqref{part22ph}, we get the
complete TLS population solution for the 2-photon top-hat pulse:
\begin{equation}
\begin{split}
    &\bra{2,g}\sigma^+(t) \sigma^-(t)\ket{2,g}= 
    \frac{4}{t_p} \Bigg[ \left( \frac{4t}{t_p} +\frac{20}{\gamma t_p} +1 \right) e^{-\gamma t} \\
    &+ \left( \frac{8t}{t_p} - \frac{16}{\gamma t_p} -2 \right) e^{-\gamma t/2} + \left( -\frac{4}{\gamma t_p} +1 \right)
    \Bigg] , \ t \leq t_p
\\
    &\bra{2,g}\sigma^+(t) \sigma^-(t)\ket{2,g}= 
    \frac{4}{t_p} \Bigg[ \left( -\frac{4t}{t_p} +1 \right) e^{\gamma t_p} \\ 
    &+ \left( -\frac{16}{\gamma t_p} + 6 \right) e^{\gamma t_p/2} + \left( \frac{20}{\gamma t_p} +5 \right)
    \Bigg] e^{-\gamma t}  , \ t > t_p.
\end{split}
\end{equation}

\subsection{Chiral Emitter Population Dynamics}

The population equation of motion in a chiral emitter system transforms to
\begin{equation}
\begin{split}
    &\frac{d}{dt} \bra{2,g}\sigma^+(t) \sigma^-(t)\ket{2,g}=-\gamma \bra{2,g}\sigma^+(t) \sigma^-(t)\ket{2,g} \\
    &-2\sqrt{\frac{\gamma}{t_p}} \left[ \bra{1,g} \sigma^-(t)\ket{2,g} +
    {\rm {\rm H.c.}} \right].  
\end{split}
\label{2phdiffch}
\end{equation}
Using the chiral solution for the 1-photon pulse,
Eq.~\eqref{eq:1sigma-2} gives
\begin{equation}
\begin{split}
    \frac{d}{dt} \bra{1,g} \sigma^-(t)\ket{2,g} &=-\frac{\gamma}{2} \bra{1,g}\sigma^-(t)\ket{2,g} \\
    &-\sqrt{ \frac{\gamma}{t_p}} \left[1 - \frac{8}{\gamma t_p} \left( e^{-\gamma t/2} -1 \right)^2 \right].
    \end{split}
    \label{2phsigmaminus2} 
\end{equation}
Integrating this, using  in Eq.~\eqref{2phdiffch} and proceeding in the same manner as in the symmetrical case, we obtain the 2-photon
TLS population solution for the chiral system,
\begin{equation}
\begin{split}
    &\bra{2,g}\sigma^+(t) \sigma^-(t)\ket{2,g}= 
    \frac{8}{t_p} \Bigg[ \left( \frac{8t}{t_p} +\frac{40}{\gamma t_p} +1 \right) e^{-\gamma t} \\
    &+ \left( \frac{16t}{t_p} - \frac{32}{\gamma t_p} -2 \right) e^{-\gamma t/2} + \left( -\frac{8}{\gamma t_p} +1 \right)
    \Bigg], \ t \leq t_p
\\
    &\bra{2,g}\sigma^+(t) \sigma^-(t)\ket{2,g}= 
    \frac{8}{t_p} \Bigg[ \left( -\frac{8t}{t_p} +1 \right) e^{\gamma t_p} \\ 
    &+ \left( -\frac{32}{\gamma t_p} + 14 \right) e^{\gamma t_p/2} + \left( \frac{40}{\gamma t_p} +9 \right)
    \Bigg] e^{-\gamma t}  , \ t > t_p.
\end{split}
\end{equation}

%%%%%%%%%%%%%%%%%%%%%%%%%%%%%%%%%%%%%%%%%%%%%%%%%%%%%%%%%%%%%%%%%

\section{Analytical Solution for the One-Photon Stationary Spectra}
\label{sec:4stationaryS}

\subsection{One-photon stationary spectra for the 
symmetrical emitter}

We next calculate the transmitted  stationary (long time) spectrum, $S_{\rm sym}(\omega)=S_{\rm sym}(\omega,t \rightarrow \infty)$, from
% \sh{fix $g_R$}
%
%\begin{widetext}
\begin{widetext}
\begin{equation}
\begin{split}
 S_{\rm sym}(\omega) &= v_g \braket{ a^\dagger_R(\omega)  a_R(\omega)} =
    v_g\left< \left( a_{\rm in}^\dagger(\omega) -\frac{g_R}{v_g} \sigma^+(\omega) \right) \left( a_{\rm in}(\omega)-\frac{g_R}{v_g} \sigma^-(\omega) \right) \right> \\
    &= v_g \left< a_{\rm in}^\dagger(\omega)a_{\rm in}(\omega)\right> -
    g_R\left< a_{\rm in}^\dagger(\omega)\sigma^-(\omega)\right> -g_R\left< \sigma^+(\omega)a_{\rm in}(\omega)\right> +  \frac{{|g_R|}^2}{v_g} \left< \sigma^+(\omega)\sigma^-(\omega)\right>. 
    \label{spectrum}
\end{split}
\end{equation}
\end{widetext}
The first term of Eq.~\eqref{spectrum} is
\begin{equation}
     v_g \left< a_{\rm in}^\dagger(\omega)a_{\rm in}(\omega)\right> = |f(\omega)|^2,
\end{equation}
which is just the spectrum of the incidence pulse.
For the symmetrical coupling case, where $g_R=g_0$, the second term is
\begin{equation}
    g_0 \left<  a_{\rm in}^\dagger(\omega)\sigma^-(\omega)\right> =
    \frac{-g_{0}}{\sqrt{v_g}} f^*(\omega) \bra{0,g}  \sigma^-(\omega)\ket{1,g}.
    \label{term2}
\end{equation}

In order to solve $\bra{0,g} \sigma^-(\omega)\ket{1,g}$, we need to use the equation of motion for $\sigma^-$ [Eq.~\eqref{dsigma2}], and perform a Fourier transform to obtain:
\begin{equation}
    i\omega \sigma^-(\omega) =   i \delta \sigma^- (\omega)  -\frac{\gamma}{2} \sigma^-(\omega) -  g_0 \mathcal{F}  \left[ a_{\rm in}(t)  \sigma^z(t) \right],
\end{equation}
and thus
\begin{equation}
    \sigma^-(\omega) =  \frac{-g_0}{i(\omega -\delta)+\frac{\gamma}{2}} \mathcal{F}  \left[ a_{\rm in}(t)  \sigma^z(t) \right].
\end{equation}
Applying this to Eq.~\eqref{term2}, gives
\begin{equation}
\begin{split}
    &g_0 \left<  a_{\rm in}^\dagger(\omega)\sigma^-(\omega)\right> \\
    &= \frac{-\gamma/2}{i(\omega -\delta)+\frac{\gamma}{2}} \left| f(\omega)\right|^2 \mathcal{F} \bra{0,g}    \sigma^z(t)  \ket{0,g}.
    \label{term2f}
\end{split}
\end{equation}
To calculate the right hand side, 
we use
\begin{equation}
   \bra{0,g}  \sigma^z(t) \ket{0,g}=  \bra{0,g} 2 \sigma^+(t) \sigma^-(t) - \mathbb{1} \ket{0,g} = 
- 1. 
\end{equation}
Inserting this result in  Eq~\eqref{term2f}, gives
\begin{equation}
    g_0 \left<  a_{\rm in}^\dagger(\omega)\sigma^-(\omega)\right> = 
    \frac{\gamma/2}{i(\omega-\delta)+\frac{\gamma}{2}} \left| f(\omega)\right|^2. 
    \label{term2F}
\end{equation}

The next term of Eq.~\eqref{spectrum} is the simply the conjugate of Eq.~\eqref{term2F},
\begin{equation}
    g_0 \left< \sigma^+(\omega)  a_{\rm in}(\omega) \right> = 
    \frac{\gamma/2}{-i(\omega-\delta)+\frac{\gamma}{2}} \left| f(\omega)\right|^2. 
    % \label{term2F}
\end{equation}

Finally, the last term of Eq.~\eqref{spectrum} can be solved using the previous results, and we obtain
\begin{equation}
\begin{split}
    &\frac{|g_0|^2}{v_g}  \left< \sigma^+(\omega) \sigma^-(\omega) \right> 
    = \frac{ \gamma^2/4 \left| f(\omega)\right|^2}{(\omega -\delta)^2+\gamma^2/4}. 
\end{split}    
\end{equation}
Adding up all the previous terms, we obtain 
\begin{equation}
\begin{split}
    S_{\rm sym}(\omega) 
    = |f(\omega)|^2 \left(1 -\frac{\gamma^2/4}{(\omega -\delta)^2+\gamma^2/4}\right),
    \label{eq:S_QM1}
\end{split} 
\end{equation}
\newline
where the second term represents a drop on the transmitted spectrum due to reflection from the TLS.

This is exactly the
same result
as a bosonic emitter, 
namely a harmonic oscillator model for the atom
or qubit,
which also coincides with the result from a WEA, 
{for this function}.
However, clearly the terminology of a ``weak excitation'' is not appropriate here, as the atom can indeed be efficiently excited; it just happens to yield the same stationary spectrum result as a bosonic atom in the strict one quantum subspace. Indeed, we could also simply have replaced $\sigma_z = -\mathbb{1}$ much earlier in the derivation, and identical final results will be obtained for the 1-photon case, but such an approximation would get the wrong dynamical result, discussed in more detail later.

Note to obtain Eq.~\eqref{eq:S_QM1}, we have exploited the
convolution theorem: $\mathcal{F} (h*g) = \mathcal{F}(h) \mathcal{F}(g)$,
and $h*g = \int_{-\infty}^{\infty} h(t') g(t-t') dt'$.
In this case, $h=\sigma^+(t) \sigma^-(t) f(t)$ and $g=\sigma^+(t-t') \sigma^-(t-t') f^*(t-t')$, and 
\begin{widetext}
\begin{equation}
\begin{split}
     h*g &=
     \int_{-\infty}^{\infty} 
    \sigma^+(t') \sigma^-(t') f(t') \sigma^+(t-t')\sigma^-(t-t') f^*(t-t') dt'\\
    &= \int_{-\infty}^{\infty} \sigma^+(t')\sigma^-(t') f(t') \left[  \sigma^+(\omega) \sigma^-(\omega)f^*(\omega) e^{i\omega(t-t')} d\omega \right] dt' \\
    &= \int_{-\infty}^{\infty}  \sigma^+(\omega)  \sigma^-(\omega)  f(\omega) e^{i\omega t'}  \sigma^+(\omega) \sigma^-(\omega)  f^*(\omega) e^{i\omega(t-t')} d\omega  dt'\\
    &=  \int_{-\infty}^{\infty} \left(\sigma^+(\omega) \sigma^-(\omega) \right)^2 |f(\omega)|^2 e^{i\omega t} d\omega = \left(\sigma^+(t)\sigma^-(t)\right)^2 |f(t)|^2 = \sigma^+(t)\sigma^-(t) |f(t)|^2.
\end{split} 
\end{equation}
\end{widetext}
% %

%---------------------------------------------------------------

\subsection{
One-photon stationary spectra for the  chiral emitter}

As before, 
we can calculate the stationary spectrum, $S_{\rm chir}(\omega)$,  from Eq.~\eqref{spectrum},
where the first term is again
\begin{equation}
    v_g \left< a_{\rm in}^\dagger(\omega)a_{\rm in}(\omega)\right> = |f(\omega)|^2,
\end{equation}
and the equation of motion for $\sigma^-$ is
\begin{equation}
    \frac{d \sigma^- (t)}{dt} = i \delta \sigma^- (t)  -\frac{\gamma}{2} \sigma^-(t) - g_R a_{\rm in}(t)  \sigma^z(t) 
    \label{dsigma2ch}.
\end{equation}
If we perform a Fourier transform to Eq.~\eqref{dsigma2ch},
\begin{equation}
    i\omega \sigma^-(\omega) =   i \delta \sigma^- (\omega)  -\frac{\gamma}{2} \sigma^-(\omega) -g_R \mathcal{F}  \left( a_{\rm in}(t)  \sigma^z(t) \right),
\end{equation}
then
\begin{equation}
    \sigma^-(\omega) =  \frac{-g_{R}}{i(\omega -\delta)+\frac{\gamma}{2}} \mathcal{F}  \left( a_{\rm in}(t)  \sigma^z(t) \right),
\end{equation}
and
\begin{equation}
\begin{split}
    g_R \left<  a_{\rm in}^\dagger(\omega)\sigma^-(\omega)\right> &=
    \frac{\gamma}{i(\omega -\delta)+\frac{\gamma}{2}} \left| f(\omega)\right|^2 .
    % \label{term2f}
\end{split}
\end{equation}

The third term takes the form,
\begin{equation}
    g_R \left< \sigma^+(\omega)  a_{\rm in}(\omega) \right> = 
    \frac{\gamma}{-i(\omega-\delta)+\frac{\gamma}{2}} \left| f(\omega)\right|^2, 
    % \label{term2F}
\end{equation}
and the last term is
\begin{equation}
\begin{split}
    \frac{{|g_R|}^2}{v_g}  \left< \sigma^+(\omega) \sigma^-(\omega) \right> &= 
     \frac{ \gamma^2 \left| f(\omega)\right|^2}{(\omega -\delta)^2+\gamma^2/4}. 
\end{split}    
\end{equation}

Adding up all the previous terms, we simply get the trivial result
\begin{equation}
\begin{split}
    S_{\rm chir}(\omega) & 
    = |f(\omega)|^2 ,
    \label{eq:S_QM}
\end{split} 
\end{equation}
and the transmitted spectrum is identical to the input spectrum in this chiral case where the emitter radiates only to the right
(at least for the stationary 
spectrum case).

It is important to stress that the cancellation of the final three terms 
{\it only occurs for the time-integrated spectra} as we show below, since the population
dynamics of the TLS indeed affect the time-dependent spectra, even for a 1-photon pulse. The chiral emitter results are shown in Ref.~\cite{sofia2024letter}, where the long-time limit spectra for 1-photon pulses are calculated for different pulse lengths, and in all cases they show agreement with Eq.~\eqref{eq:S_QM}.

\section{Solution for the chiral time-dependent spectrum}
\label{sec:timedepS}

Here we will first derive the population terms and then the 
{\it time-dependent} spectrum for the field after interacting with the 
quantum chiral emitter. Results
for a non-chiral emitter can be derived in a similar way.

\vspace{0.3cm}

\subsection{Derivation of the transmitted photon mean population}
\label{subsec:photonpop}

In general, the transmitted and reflected flux can be calculated from
\begin{equation}
    n_{R/L}^{\rm out}(t) = v_g \braket{a_{R/L}^\dagger (t) a_{R/L}(t)}, 
    \label{btbt}
\end{equation}
where $R$ and $L$, as before, refer to 
the right (transmitted) and left (reflected) photons, respectively.

We start by deriving the chiral solution for the time-dependent transmitted photon flux,
which is needed for the initial condition of the two-time correlation functions,
\begin{widetext}
\begin{equation}
\begin{split}
 v_g  \braket{ a_R^\dagger(t)  a_R(t)} &= v_g 
    \left< \left( a_{\rm in}^\dagger(t) -\frac{g_R}{v_g} \sigma^+(t) \right) \left( a_{\rm in}(t) - \frac{g_R}{v_g} \sigma^-(t) \right) \right> \\
    &= v_g  \left< a_{\rm in}^\dagger(t)a_{\rm in}(t)\right> -
    {g_R} \left< a_{\rm in}^\dagger(t)\sigma^-(t)\right> -g_R \left< \sigma^+(t)a_{\rm in}(t)\right> +  \frac{{|g_R|^2}}{v_g} \left< \sigma^+(t)\sigma^-(t)\right>. 
    \label{timecorrtau0}
\end{split}
\end{equation}
\end{widetext}
The first term of Eq.~\eqref{timecorrtau0} is
\begin{equation}
    v_g \left< a_{\rm in}^\dagger(t)a_{\rm in}(t)\right> = |f(t)|^2,
\end{equation}
while the second term is
\begin{equation}
    g_R \left< a_{\rm in}^\dagger(t)\sigma^-(t)\right> = -\frac{g_R}{\sqrt{v_g}} 
f^*(t) \bra{0,g} \sigma^-(t) \ket{1,g}.
\label{term2ttau0}
\end{equation}

The coherence term evolves as
\begin{equation}
\begin{split}
    &\bra{0,g} \frac{d \sigma^-(t)}{dt} \ket{1,g} = \\
    & \left(i \delta - \gamma/2 \right) \bra{0,g} \sigma^- (t)  \ket{1,g} - \frac{g_R}{\sqrt{v_g}} f(t),
    \end{split}
\end{equation}
and after integrating,
\begin{equation}
\begin{split}
   & \bra{0,g} \sigma^-(t) \ket{1,g} = \\
    &-\frac{g_R}{\sqrt{v_g}} e^{(i\delta-\gamma/2)t} \int_{-\infty}^t  f(t') e^{(-i\delta+\gamma/2)t'} dt'. 
    \end{split}
    \label{sigma-tau0}
\end{equation}
Introducing this in Eq.~\eqref{term2ttau0}, gives
\begin{equation}
\begin{split}
    &g_R \left< a_{\rm in}^\dagger(t)\sigma^-(t)\right> = \\
    &\frac{{|g_R|}^2}{v_g} 
f^*(t)  e^{(i\delta-\gamma/2)(t)} \int_{-\infty}^{t}  f(t') e^{(-i\delta+\gamma/2)t'} dt'.
\end{split}
\end{equation}
While for the third term, we obtain
\begin{equation}
\begin{split}
   & g_R \left< \sigma^+(t)a_{\rm in}(t)\right> = \\
   &-\frac{{|g_R}|^2}{v_g} 
f(t)  e^{(-i\delta-\gamma/2)t} \int_{-\infty}^t  f(t') e^{(i\delta+\gamma/2)t'} dt'.
\end{split}
\end{equation}

Finally, to calculate the last term in Eq.~\eqref{timecorrtau0},
 we can use Eq.~\eqref{sigma+-1p}, 
and considering the chiral case,
then
\begin{equation}
\begin{split}
    &\bra{1,g}\frac{d \sigma^+(t) \sigma^-(t)}{dt} \ket{1,g} 
= -\gamma \bra{1,g} \sigma^+(t) \sigma^-(t)  \ket{1,g} \\
&+\gamma e^{-\gamma t/2}  \Big[ f^*(t)   e^{i\delta t} \int_{-\infty}^{t}  f(t') e^{(-i\delta+\gamma/2)t'} dt' \\
&+ f(t)  e^{-i\delta t} \int_{-\infty}^t  f(t') e^{(i\delta+\gamma/2)t'} dt'  \Big].
\end{split}
\end{equation}
Assuming the resonant case ($\delta=0$), then
\begin{equation}
\begin{split}
    &\bra{1,g}\frac{d \sigma^+(t) \sigma^-(t)}{dt} \ket{1,g} 
=  -\gamma \bra{1,g} \sigma^+(t) \sigma^-(t)  \ket{1,g} \\
&+\gamma e^{-\gamma t/2}  \left(f^*(t) + f(t) \right)  \int_{-\infty}^{t}  f(t') e^{(\gamma/2)t'} dt' .
\end{split}
\end{equation}

To further check that this general solution is correct, we can calculate the TLS population for a rectangular pulse (since we worked this out earlier),
\begin{equation}
\begin{split}
    &\bra{1,g}\frac{d \sigma^+(t) \sigma^-(t)}{dt} \ket{1,g} 
\\
&=  -\gamma \bra{1,g} \sigma^+(t) \sigma^-(t)  \ket{1,g}
- \frac{4}{t_p}  \left[ e^{-\gamma t/2} - 1 \right],
\end{split}
\end{equation}
and after integrating,
\begin{equation}
    \bra{1,g} \sigma^+(t) \sigma^-(t)  \ket{1,g} = \frac{4}{\gamma t_p}   \left[ e^{-\gamma t/2} - 1 \right]^2,
\end{equation}
which perfectly agrees with the result calculated in Eq.~\eqref{1phdiff3}.

The more general integrated expression (in the resonant chiral case), for arbitrary pulse shapes, is
\begin{widetext}
\begin{equation}
\begin{split}
     \bra{1,g} \sigma^+(t) \sigma^-(t)  \ket{1,g} &=
     \gamma e^{-\gamma t} \int_{-\infty}^t dt'  e^{-\gamma t'/2}  \Big(f^*(t') 
     %\\
     %&
     + f(t') \Big)  \int_{-\infty}^{t'}  f(t'') e^{\gamma t'' /2} dt''.       
\end{split}
\end{equation}

Finally putting everything together, we obtain the transmitted 
photon population:
%\begin{widetext}
\begin{equation}
\begin{split}
v_g  \braket{ a_R^\dagger(t)  a_R(t)}& = 
  |f(t)|^2 +  \gamma
f^*(t)  e^{-\gamma t/2} \int_{-\infty}^{t}  f(t') e^{\gamma t'/2} dt' - \gamma f(t)  e^{-\gamma t/2} \int_{-\infty}^t  f^*(t') e^{\gamma t'/2} dt'\\
&  + \gamma^2 e^{-\gamma t} \int_{-\infty}^t dt'  e^{-\gamma t'/2}  \left(f^*(t') + f(t') \right)  \int_{-\infty}^{t'}  f(t'') e^{\gamma t'' /2} dt''. 
\end{split}
\end{equation}
\end{widetext}
We  observe here that, in general, the
TLS interaction terms do not cancel out, even for the non-resonant case ($\delta \neq 0$). Thus (apart from the case of relatively very long pulses, quantified later) a WEA would not make sense here,
and produce the wrong result, even if it happens to yield the same stationary spectrum.

The transmitted flux
for the rectangular pulse can be calculated by considering the four terms in Eq.~\eqref{timecorrtau0}.
For the first term in this case,
\begin{equation}
\begin{split}
    &v_g \bra{1,g} a^\dagger_{\rm in}(t) a_{\rm in}(t)  \ket{1,g} = \frac{1}{t_p} , \ t \leq t_p \\
    &v_g \bra{1,g} a^\dagger_{\rm in}(t) a_{\rm in}(t)  \ket{1,g} = 0 , \ t > t_p,
\end{split}
\end{equation}
while for  second term,
\begin{equation}
\begin{split}
    &g_R \left< a_{\rm in}^\dagger(t)\sigma^-(t)\right> = \frac{-2}{t_p} \left[ e^{-\gamma t/2} -1 \right], \ t \leq t_p \\
    &g_R\left< a_{\rm in}^\dagger(t)\sigma^-(t)\right> = 0 , \ t > t_p. 
\end{split}
\end{equation}
For the third term,
% \sh{fix all the $g_R$ terms}
\begin{equation}
\begin{split}
    &g_R \left< \sigma^+(t)a_{\rm in}(t)\right>  = \frac{2}{t_p} \left[1- e^{-\gamma t/2}  \right], \ t \leq t_p \\
    &g_R\left< \sigma^+(t)a_{\rm in}(t)\right>  = 0 , \ t > t_p,
\end{split}
\end{equation}
and for the last term, we can use the solution of the TLS population, so that
\begin{equation}
\begin{split}
    \frac{|g_R|^2}{v_g}  &\bra{1,g} \sigma^+(t) \sigma^-(t)  \ket{1,g}  = \\
    &\frac{4}{t_p} \left[ e^{-\gamma t/2} -1 \right]^2 , \ t \leq t_p \\
    \frac{|g_R|^2}{v_g}  & \bra{1,g} \sigma^+(t) \sigma^-(t)  \ket{1,g}  = \\
    &\frac{4}{ t_p} \left[ e^{\gamma t_p/2} -1 \right]^2 e^{-\gamma t} , \ t > t_p. 
\end{split}
\end{equation}
\vspace{0.5cm}

Putting everything together, we obtain the
desired solution for the transmitted
photon flux for a chiral TLS interaction:
\begin{equation}
\begin{split}
v_g  \braket{ a_R^\dagger(t)  a_R(t)} &= \frac{1}{t_p} 
 + \frac{4}{t_p} \left[ e^{-\gamma t/2} -1 \right] \\
 &+ \frac{4}{t_p} \left[ e^{-\gamma t/2} -1 \right]^2 , \ t \leq t_p \\
 v_g  \braket{\hat a_R^\dagger(t) \hat a_R(t)} &= \frac{4}{ t_p} \left[ e^{\gamma t_p/2} -1 \right]^2 e^{-\gamma t}
  , \ t > t_p.
\label{eq:bdagbt}
\end{split}
\end{equation}
We note that for $t>t_p$, the only term contributing is the TLS population one.
The symmetric system can be derived in a similar way, which yields both transmitted and reflected fields.

\subsection{Derivation of the transmitted first-order photon correlation function}
\label{subsec:correl}

Considering the same chiral system as above,
 we now wish to calculate $\braket{a_R^\dagger(t)  a_R(t+\tau)}$,
since this is needed to connect to the time-dependent 
spectra. This two-time correlation 
function is obtained from 
\vspace{0.5cm}
\begin{widetext}
\begin{align}
 &v_g \braket{ a_R^\dagger(t)  a_R(t+\tau)} =
    v_g \left< \left( a_{\rm in}^\dagger(t) -\frac{g_R}{v_g} \sigma^+(t) \right) \left( a_{\rm in}(t+\tau) -\frac{g_R}{v_g} \sigma^-(t+\tau) \right) \right> 
    \nonumber \\
    &= v_g  \left< a_{\rm in}^\dagger(t)a_{\rm in}(t+\tau)\right> -
    g_R \left< a_{\rm in}^\dagger(t)\sigma^-(t+\tau)\right> -g_R\left< \sigma^+(t)a_{\rm in}(t+\tau)\right> +  \frac{{|g_R|}^2}{v_g} \left< \sigma^+(t)\sigma^-(t+\tau)\right> \nonumber
    \\
    &\equiv C_1(t,t+\tau)
    + C_2(t,t+\tau)
    + C_3(t,t+\tau)
    + C_4(t,t+\tau),
    \label{timecorr}
\end{align}
\end{widetext}
which we have split into four parts for analysis purposes.

The first term of Eq.~\eqref{timecorr} is
\begin{equation}
   v_g  \left<  a_{\rm in}^\dagger(t) a_{\rm in}(t+\tau)\right> 
    = f^*(t)f(t+\tau),
\end{equation}
and the second term is
\begin{equation}
    -g_R \left<  a_{\rm in}^\dagger(t)\sigma^-(t+\tau)\right> = 
    \sqrt{\gamma}
f^*(t) \bra{0,g} \sigma^-(t+\tau) \ket{1,g}.
\label{term2t}
\end{equation}

To solve these, we can first use Eq.~\eqref{sigma-chiral} for any general pulse shape:
\begin{equation}
\begin{split}
     \bra{0,g}\frac{d \sigma^-(t)}{dt} \ket{1,g} 
        =  -\frac{\gamma}{2} \bra{0,g}\sigma^-(t) \ket{1,g} -\sqrt{\gamma} f(t),
\end{split}
\end{equation}
which gives a formal solution,
\begin{equation}
\begin{split}
    &\bra{0,g} \sigma^-(t) \ket{1,g} \\
    &= -\sqrt{\gamma} e^{(i\delta-\gamma/2)t} \int_{-\infty}^t  f(t') e^{(-i\delta+\gamma/2)t'} dt'. 
\label{sigma-(t)}
\end{split}
\end{equation}
Introducing this in Eq.~\eqref{term2t}, gives
\begin{equation}
\begin{split}
    &- g_R \left< a_{\rm in}^\dagger(t)\sigma^-(t+\tau)\right> = \\
    &-\gamma
f^*(t)  e^{(i\delta-\gamma/2)(t+\tau)} \int_{-\infty}^{(t+\tau)}  f(t') e^{(-i\delta+\gamma/2)t'} dt',
\end{split}
\end{equation}
which will depend on the pulse profile, $f(t)$, used in each case.

The third term in Eq.~\eqref{timecorr} can be derived in a similar way,
\begin{equation}
\begin{split}
   & g_R \left< \sigma^+(t)a_{\rm in}(t+\tau)\right> = \\
   &\gamma
f(t+\tau)  e^{(-i\delta-\gamma/2)t} \int_{-\infty}^t  f^*(t') e^{(i\delta+\gamma/2)t'} dt'.
\end{split}
\end{equation}

\vspace{0.5cm}
\begin{widetext}
Finally, to calculate the last term of Eq.~\eqref{timecorr}, we need to use the quantum regression theorem, 
\begin{equation}
\begin{split}
    &\frac{d}{d\tau} \left< \sigma^+(t) \sigma^-(t+\tau) \right> 
    =
    -\gamma  \left< \sigma^+(t) \sigma^-(t+\tau) \right> - \sqrt{\gamma} \left[f(t + \tau )  \left< \sigma^+(t) \right>
 \right],
 \end{split}
\end{equation}
where one uses $\left< \sigma^+(t) \sigma^-(t) \right>$ as the initial condition, so clearly the 
time-dependent TLS population plays a significant role here.

For a rectangular pulse, the previous terms can be derived analytically, %become
\begin{equation}
\begin{split}
    &C_1(t,t+\tau)= \frac{1}{t_p} , \  0<t \leq t_p \ \ and \ \ 0<t+\tau \leq t_p \\
    &C_1(t,t+\tau) = 0  , \ t \leq t_p \ \ and  \ \ t+\tau > t_p \\
    &C_1(t,t+\tau) = 0  , \ t>t_p.
\end{split}
\label{eqC1}
\end{equation}
The second term, in this case, gives
\begin{equation}
\begin{split}
    &C_2(t,t+\tau)  = \frac{2}{t_p} \left( e^{-\gamma (t+\tau) /2} -1 \right) , \ 0<t \leq t_p \ \ and \ \ 0<t+\tau \leq t_p \\
    &C_2(t,t+\tau)   = \frac{2}{t_p} \left( 1-e^{\gamma t_p /2}  \right) e^{-\gamma (t+\tau) /2} , \ 0<t \leq t_p \ \ and \ \ t+\tau>t_p \\
    &C_2(t,t+\tau)   = 0 , \ t>t_p.
\end{split}
\end{equation}
The third term becomes
\begin{equation}
\begin{split}
    & C_3(t,t+\tau)  = \frac{2}{t_p} \left( e^{-\gamma t/2} - 1 \right), \ 0<t \leq t_p \ \ and \ \ 0<t+\tau \leq t_p
    \\
    &C_3(t,t+\tau) = 0  , \ 0<t \leq t_p \ \ and \ \ t+\tau>t_p 
    \\
    &C_3(t,t+\tau) = 0 , \ t > t_p
    .
\end{split}
\end{equation}

\begin{figure}[t]
\centering
\includegraphics[width=0.7\columnwidth]{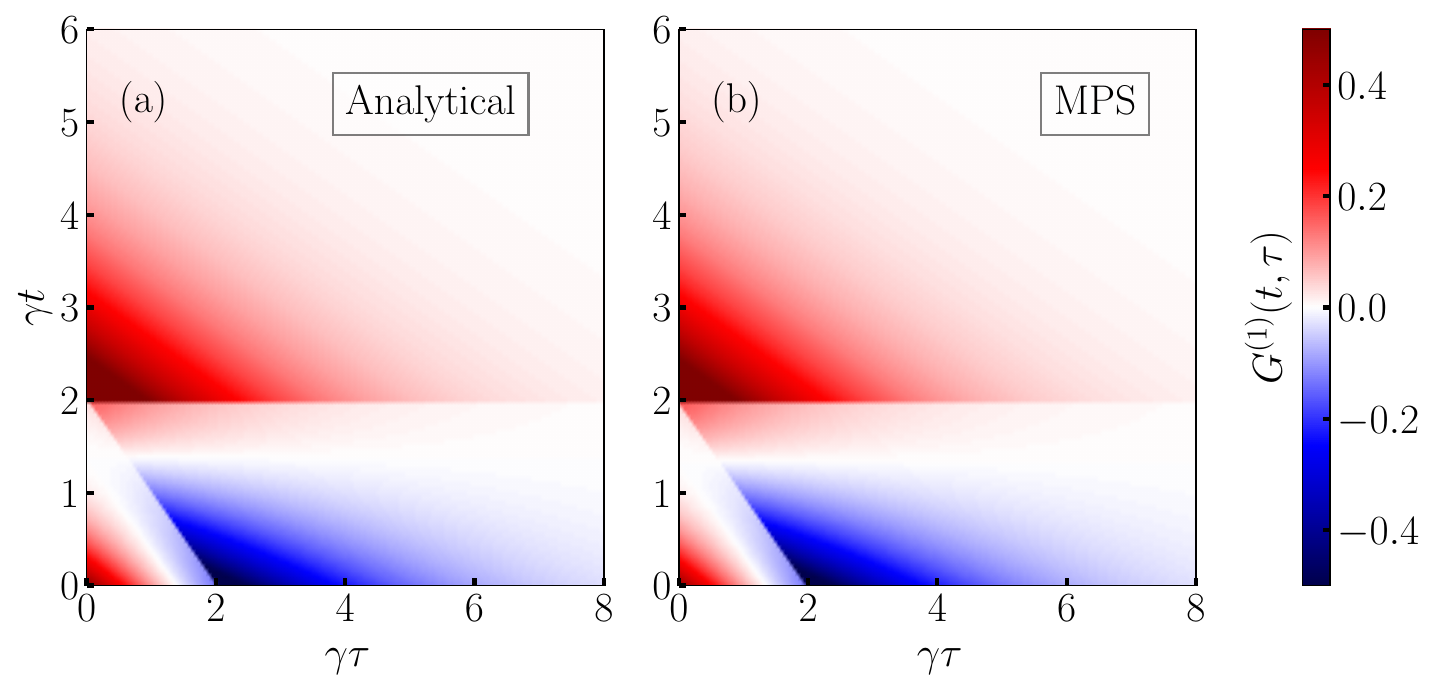}
\caption{ {\bf Chiral TLS results.} First order correlation function ($G^{(1)}(t,\tau)$) for a chiral TLS system with a rectangular 1-photon quantum pulse with a pulse duration of $\gamma t_p =2$, solved analytically in (a) using Eq.~\eqref{timecorr},
and using MPS in (b).
}
\label{fig:correl_analytical}
\end{figure}

Finally, the fourth term is,
\begin{equation}
\begin{split}
    & C_4(t,t+\tau) = \frac{4}{t_p} \left( 
    1-  e^{-\gamma t/2} + e^{-\gamma (t+\tau/2)} - e^{-\gamma (t+\tau)/2}
    \right) , \ 0<t \leq t_p \ \ and \ \ 0<t+\tau \leq t_p
    \\
    &C_4(t,t+\tau) = \frac{4}{t_p} e^{-\gamma \tau/2} \left( 
    -  e^{\gamma (t_p/2 -t)} + e^{\gamma (t_p -t)/2} + e^{-\gamma t} -e^{-\gamma t/2}
    \right)
    , \ 0<t \leq t_p \ \ and \ \ t+\tau > t_p \\
     &C_4(t,t+\tau)= \frac{4}{t_p} \left( e^{\gamma t_p/2} -1 \right)^2 e^{-\gamma (t +\tau/2)} , \ t > t_p.
\end{split}
\end{equation}    
\end{widetext}
We highlight that this last term is the only term contributing to the correlation function when $t > t_p$, showing a clear 
influence from the
TLS population effects.  
We have also confirmed that the above analytical expressions agree perfectly with those obtained numerically using MPS (Sec.~\ref{sec:MPS}), as shown in  Fig.~\ref{fig:correl_analytical}.

\subsection{Time dependent spectra and spectral intensity}
\label{subsec:Swt}

In a rotating wave approximation, the time-dependent spectrum is obtained from the two-time first-order correlation functions calculated in Sec.~\ref{subsec:correl},
using
\begin{equation}
	S(\omega,t) = \text{Re}\left[ \int_0^{t} dt' \int_0^{t-t'} \!\! d\tau v_g \langle a_R^{\dagger}(t')a_R(t'+\tau)\rangle e^{i\Delta_{\omega {\rm p}}\tau} \right], 
 \label{eq:Swt}
\end{equation}
where $\Delta_{\omega {\rm p}} = \omega - \omega_{\rm p}$.

The time-dependent spectral intensity can also be calculated from the same two-time correlation functions:
\begin{equation}
 I(\omega, t) = 
     {\rm Re} \left[ \int_0^{\infty} \! d\tau v_g \braket{a_R^{\dagger} (t) a_R(t+\tau)}  e^{i \Delta _{\rm \omega {\rm p}}\tau}    \right].
    \label{eqI}
\end{equation}
Here, it is important to note that the integration of Eq.~\eqref{eqI} over time,
$ I(\omega)  = \int_0^{\infty}  I(\omega, t) dt $, agrees with the stationary spectra, i.e., $I(\omega) = S(\omega)$~\cite{Liu2024}, so the time-integrated spectral intensity is equivalent to the long-time spectra, even though $S(\omega,t)$ and $I(\omega,t)$ contain very different dynamical signatures.

\section{Matrix Product States approach}
\label{sec:MPS}

In order to extend the analytical results previously calculated, we have also made use of MPS, which is an exact numerical approach based on tensor network theories~\cite{orus_practical_2014,yang_matrix_2018}.

Using the MPS approach, the system, which contains both the TLS and the waveguide, is discretized in time. The state is decomposed in a tensor product by performing Schmidt decompositions (or singular value decompositions). Each of the tensors receives the name of `bin', with the TLS as the `TLS bin' and the photonic part as the `time bins'~\cite{PhysRevResearch.3.023030,PhysRevLett.116.093601}. 

The quantum pulse is contained in the photonic part of the initial state~\cite{Guimond_2017}. If we have an initial state,
\begin{equation}
    \ket{\psi_0} = \ket{i}_s \otimes \ket{\phi_0},
\end{equation}
where $\ket{i}_s$ corresponds to the TLS part and $\ket{\phi_0}$ to the photonic part, where each time bin contains a tensor product of the right and left channels at the corresponding time step. Thus, a 1-photon pulse will be introduced from
\begin{equation}
        \ket{\phi_0} = b_{\rm in}^\dagger \ket{0,...,0} ,  
\end{equation}
where
\begin{equation}
    b_{\rm in}^\dagger = \int dt \ f(t) b_\alpha^\dagger(t),
    \label{bin}
\end{equation}
where $f(t)$ is normalized as before, and 
\begin{equation}
    b_\alpha(t)=\frac{1}{\sqrt{2\pi}} \int d\omega b_\alpha(\omega) e^{-i(\omega - \omega_{0})t},
    \label{continuous-time-ops}
\end{equation}
is the quantum noise creation operator in the time domain with the commutation relations $\left[ 
 b_{\alpha}(t), b_{\alpha'}(t')\right] = \delta_{\alpha, \alpha'} \delta(t-t')$ where $\alpha = L,R$ correspond to the left and right channel, respectively~\cite{gardiner_zoller_2010}. In this case, for a right propagating pulse, the input pulse is introduced in the right channel,
 \begin{equation}
    b_{\rm in}^\dagger = \int dt \ f(t) b_R^\dagger(t).
\end{equation}

Writing this in the MPS description,
\begin{widetext}
\begin{equation}
\begin{split}
 \ket{\phi_0}_N &= \sum_{k=1}^m f_k \Delta B_k^\dagger /\sqrt{\Delta t} \ket{0,...,0} = 
     % \ket{\phi_0}_N &=
     \sum_{k=1}^m \Delta B_k^\dagger / \sqrt{\Delta }t\ket{0,...,0}_{1,...,m} \otimes \ket{0,...,0}_{m+1,...,N} \\
    &=
   \sum_{k=1}^m A_{a_1}^{(i_1)}A_{a_1,a_2}^{(i_2)}...A_{a_{m-2},a_{m-1}}^{(i_{m-1})}A_{a_{m-1}}^{(i_{m})}\ket{0,...,0}_{1,...,m} \otimes \ket{0,...,0}_{m+1,...,N},
\end{split}
\end{equation}
\end{widetext}
where $f_k$ is the discretized version of $f(t)$ which contains the shape of the pulse at each time bin,
\begin{equation}
    \Delta B_k^\dagger = \int_{t_k}^{t_{k+1}} dt' b_R^\dagger(t')
\end{equation}
is the quantum noise creation matrix product operator (MPO), with the commutator $\left[ \Delta B_k, \Delta B_{k'}^\dagger \right] = \Delta t \delta_{k,k'}$, and $\Delta t$ is a time step. It is important to note that $\Delta B_k$ has units of $\left[\sqrt{s}\right]$, hence,  to calculate the photon flux as in Eq.~\eqref{btbt} with the MPS approach,
\begin{equation}
    n_{R/L}^{\rm out} (t_k)=  \braket{\Delta B_k^\dagger  \Delta B_k} /\left(\Delta t\right)^2.
\end{equation}

Considering this, $A_k^{(i_k)}$  can be written as
follows:
\begin{itemize}
    \item if $k=1$:
\begin{equation}
    A_1^{(0)}=
    \begin{pmatrix}
        1 & 0
    \end{pmatrix}
    , \ A_1^{(1)}=
    \begin{pmatrix}
        0 & f_1
    \end{pmatrix},
\end{equation}
\item if $1 < k < m$:
\begin{equation}
A_k^{(0)}=
    \begin{pmatrix}
         1 & 0 \\
         0 & 1
    \end{pmatrix}, \ A_k^{(1)}=
    \begin{pmatrix}
        0 &  f_k \\
        0 & 0
    \end{pmatrix},
\end{equation}
\item if $k=m$:
\begin{equation}
A_m^{(0)}=
    \begin{pmatrix}
        0 \\
        1
    \end{pmatrix}, \ A_m^{(1)}=
    \begin{pmatrix}
        f_m \\
        0
    \end{pmatrix},
\end{equation}
\end{itemize}
where, in this case, each $i_k$ has 2 dimensions corresponding to 0 and 1 excitations. 

For a 2-photon pulse, we have
\begin{equation}
\begin{split}
    &\ket{\phi_0}_m = \frac{1}{\Delta t \sqrt{2}}  \\
    &\left(\sum_{k=1}^m \left(f_k \Delta B_k^\dagger\right)^2 + 2 \sum_{k=1}^m  \sum_{l=k+1}^m f_k \Delta B_k^\dagger f_l \Delta a_L^\dagger \right) \ket{0,...,0},     
\end{split}
\end{equation}  
and the tensors now contain another dimension to count for the two excitations:
\begin{widetext}
\begin{itemize}
    \item if $k=1$:
\begin{equation}
    A_1^{(0)}=
    \begin{pmatrix}
        1 & 0 & 0
    \end{pmatrix}
    , \ A_1^{(1)}=
    \begin{pmatrix}
        0 & \sqrt{2} f_1 & 0
    \end{pmatrix}
    , \ A_1^{(2)}=
    \begin{pmatrix}
        0 & 0 & \sqrt{2} f_1
    \end{pmatrix}
    ,
\end{equation}
\item if $1 < k < m$:
\begin{equation}
A_k^{(0)}=
    \begin{pmatrix}
         1 & 0 & 0\\
         0 & 1 & 0 \\
         0 & 0 & 1
    \end{pmatrix}, \ A_k^{(1)}=
    \begin{pmatrix}
        0 & \sqrt{2} f_k & 0\\
        0 & 0 & \sqrt{2} f_k \\
        0 & 0 & 0
    \end{pmatrix},
     \ A_k^{(2)}=
    \begin{pmatrix}
        0 & 0 & \sqrt{2} f_k \\
        0 & 0 & 0 \\
        0 & 0 & 0
    \end{pmatrix},
\end{equation}
\item if $k=m$:
\begin{equation}
A_m^{(0)}=
    \begin{pmatrix}
        0 \\
        0 \\
        1
    \end{pmatrix}, \ A_m^{(1)}=
    \begin{pmatrix}
        0 \\
        \sqrt{2}f_m \\
        0
    \end{pmatrix},
    \ A_m^{(2)}=
    \begin{pmatrix}
        0 \\
        0 \\
        \sqrt{2}f_m
    \end{pmatrix}.
\end{equation}
\end{itemize}

\begin{figure}[h]
\centering
\includegraphics[width=\textwidth]{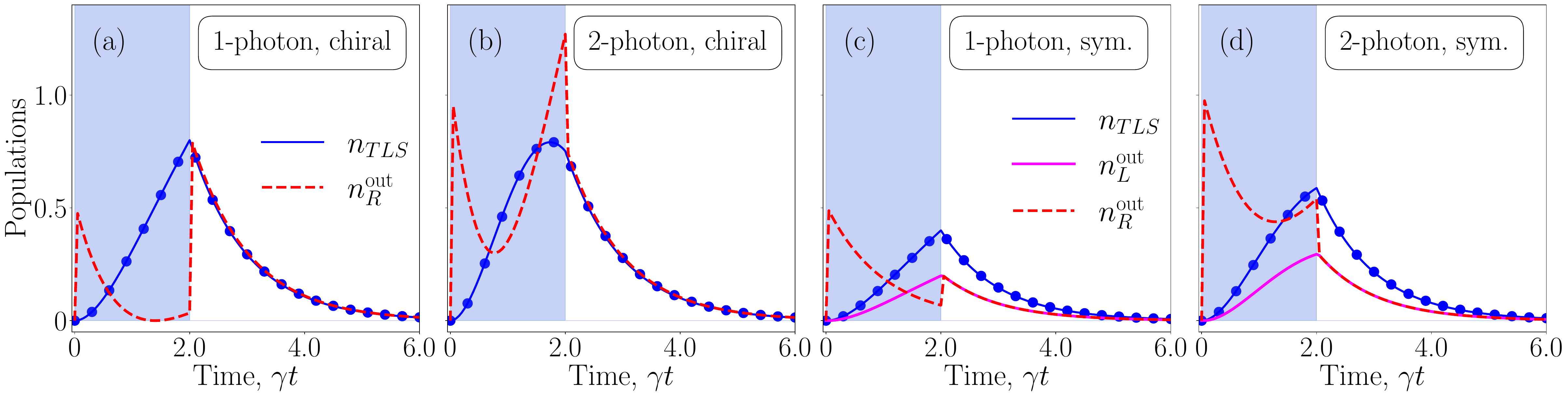}
\caption{
(a,b) {\bf Chiral TLS results.} Quantum dynamics for a chiral single TLS-waveguide system, using an incident rectangular quantum pulse (with pulse duration $t_p =2\gamma^{-1}$, cyan shaded region) containing: (a) 1 photon, and (b) 2 photons. Both panels show the TLS population (solid blue), with the corresponding analytical solution (blue circles), and the normalized photon population of the right (dashed red) (c,d). {\bf Symmetric TLS results.} Population dynamics for a symmetrical single TLS-waveguide system, using an incident rectangular quantum pulse (with the same duration) containing: (c) 1 photon, and (d) 2 photons. In this case, in the symmetric cases (c,d), the left (magenta) photon flux is also shown.
}
\label{fig:2sym}
\end{figure}
\end{widetext}

This is then transformed to have the required canonical shape,
where we have the Orthogonality Center (OC) in the present time bin and the rest of the pulse time bins are left normalized, in order to combine them with the TLS bin, and solve the time dynamics as usual. This method can be easily extended to pulses containing higher numbers of photons.

\section{Photon Population and Two-Level-System Population Results}
\label{sec:popresults}

\subsection{Symmetrical emitter}
\label{subsec:sympopresults}

In our related Letter, Ref.~\onlinecite{sofia2024letter}, we showed specific results for the chiral TLS. For completeness, here we also show them along with the results for the symmetric TLS in Fig.~\ref{fig:2sym}. 
For these simulations, we consider an incident pulse with a width $\gamma t_p = 2$, containing either 1 photon [Figs.~\ref{fig:2sym}(a,c)] or 2 photons [Figs.~\ref{fig:2sym}(b,d)], which interacts with a chiral TLS with $\gamma_R = \gamma $, $\gamma_L =0$ in Figs.~\ref{fig:2sym} (a,b),  and with a symmetrical TLS with $\gamma_R = \gamma_L = \gamma /2$ in Figs.~\ref{fig:2sym} (c,d). 

The observables are calculated using MPS (solid lines), and the TLS population is compared with the analytical solution (blue circles) derived in Sec.~\ref{sec:1photonpop} and \ref{sec:2photonpop}, which agree perfectly.

First, in the 1-photon solution, it can be seen that the TLS population of the symmetrical emitter is exactly half  the population observed with a chiral TLS. This agrees with the analytical results from the previous sections. However, the 2-photon solution deviates from this, due to the photon non-linear interactions. Additionally, the transmitted and reflected flux are shown in dashed red and solid magenta, respectively,
where the nonlinearities of the 2-photon can again be observed. 

The photon flux from Eq.~\eqref{btbt} can be integrated in time, yielding
\begin{equation}
    N_{R/L}^{\rm out} (t) =  \int_0^t dt' v_g \braket{a_{R/L}^\dagger(t') a_{R/L} (t')}.
\end{equation}
When this quantity is 
 added to the TLS population, we recover the number of quantum excitations in the system.
For example, for a chiral emitter, this is  
\begin{equation}
N_{\rm total}^{\rm chiral}(t) = N_R^{\rm out} (t) +n_{TLS} (t),
\end{equation}
and for a symmetrical case we now need to consider both the transmitted and reflected flux, so that
\begin{equation}
N^{\rm sym}_{\rm total}(t) = N_R^{\rm out} (t) +N_L^{\rm out} (t) +n_{TLS} (t),
\end{equation}
where 
$n_{TLS} (t)$ is the instantaneous TLS population.
In both cases, $N^{\rm sym/chiral}_{\rm total}$$(t) = 1$ for a 1-photon pulse, and 
$N^{\rm sym/chiral}_{\rm total}$$(t) = 2$ for the 2-photon one. 
\\
\begin{widetext}
\begin{figure*}[t]
\centering
\includegraphics[width=0.8\textwidth]{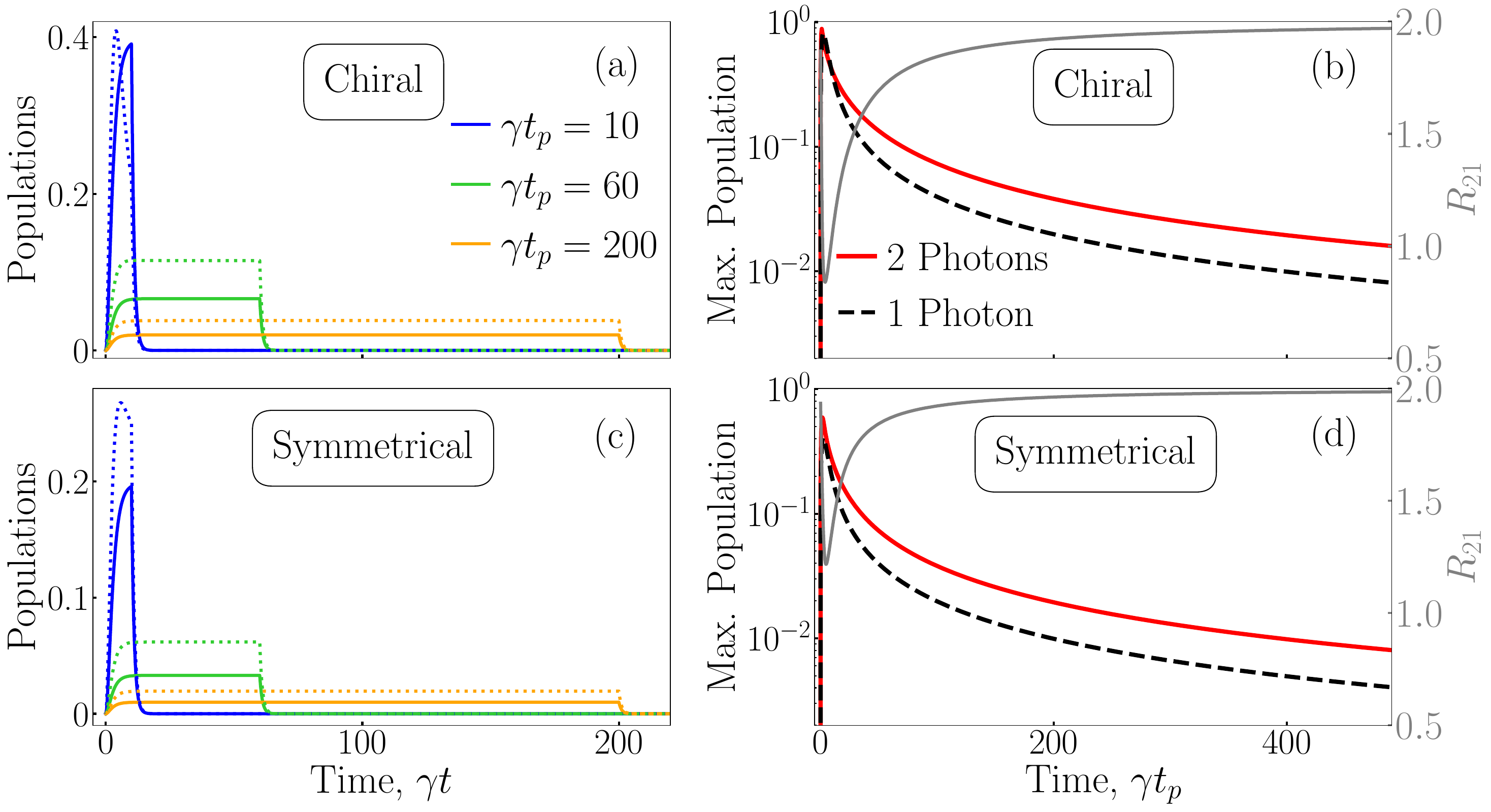}
\caption{ (a,b) {\bf Chiral TLS results.} (a) Quantum pulse excited TLS populations for 
    pulse lengths of $\gamma t_p = 10$ (blue), $\gamma t_p = 60$ (green) and $\gamma t_p = 200$ (orange) for a 1-photon pulse (solid line) and a 2-photon pulse (dotted lines). (b) Maximum value of the TLS population versus pulse length for a 1-photon pulse (dashed black) and a 2-photon pulse (red). The ratio between the two and one-photon solutions ($R_{21}$) is shown in the grey line.
    (c,d) {\bf Symmetric TLS results.} (c) Quantum pulse excited TLS populations for the same
    pulse lengths. (d) Maximum value of the TLS population versus pulse length. 
    }
\label{fig:longpop}
\end{figure*}

\end{widetext}

\subsection{Influence of pulse duration}
\label{subsec:pulseduration}

To better understand the time dynamics as a function of pulse duration, in Fig~\ref{fig:longpop} we study the
TLS population for different pulse lengths for 1-  and 2-
photon pulses.

In 
Figs~\ref{fig:longpop}(a,c), the TLS populations are shown for 1-photon pulses (solid lines) and 2-photon pulses (dotted lines) with three different pulse lengths: $\gamma t_p = 10$ (blue), $\gamma t_p = 60$ (green) and $\gamma t_p = 200$ (orange). Figure~\ref{fig:longpop}(a) shows the chiral solution while Fig.~\ref{fig:longpop}(c) is the symmetrical solution. We observe a clear deviation from the WEA approximation in every solution with significant population dynamics in all the results.
Indeed, the excitation of the TLS is still considerable even for pulses as long as $\gamma t_p = 200$. In addition, we see how the population
nonlinearites between the 1-photon and 2-photon solutions appear 
more in the shorter pulses, where they both have highly different dynamics, but when the pulses get long enough, they reach a plateau, and the two-photon solution tends to just double the single photon one.

In order to see the
trend of population ratios more clearly, in Figs.~\ref{fig:longpop}(b,d), the maximum values of the TLS population are shown versus the pulse duration for the 1-photon pulses (solid red lines) and 2-photon pulses (dashed black lines), where again Fig.~\ref{fig:longpop}(b) is the chiral solution and Fig.~\ref{fig:longpop}(d) is the symmetrical one. Additionally, the ratio between both solutions are computed from:
\begin{equation}
        R_{21} = {\rm max}[n^{(2)}_{TLS}]/ {\rm max}[n^{(1)}_{TLS}].
    \end{equation}

Here, we highlight two important
effects:
(i) it can be seen that the TLS population does not tend to zero until remarkably long pulses ($\gamma t_p \approx 500$) even in the 1-photon solution, showing that the WEA does not hold for
these longer pulses;
and (ii),  the ratio between the maximum population for the 2-photon solution and the 1-photon only tends to two only for sufficiently long pulses, which indicates that there are nonlinearities in the 2-photon solution 
even for these relatively long pulses. This is important as often a WEA is assumed in models of quantum pulses interacting with    TLSs in waveguides, and our theory allows one to quantify if this is valid or not.
    
\begin{figure*}[t]
\centering
\includegraphics[width=0.75\textwidth]{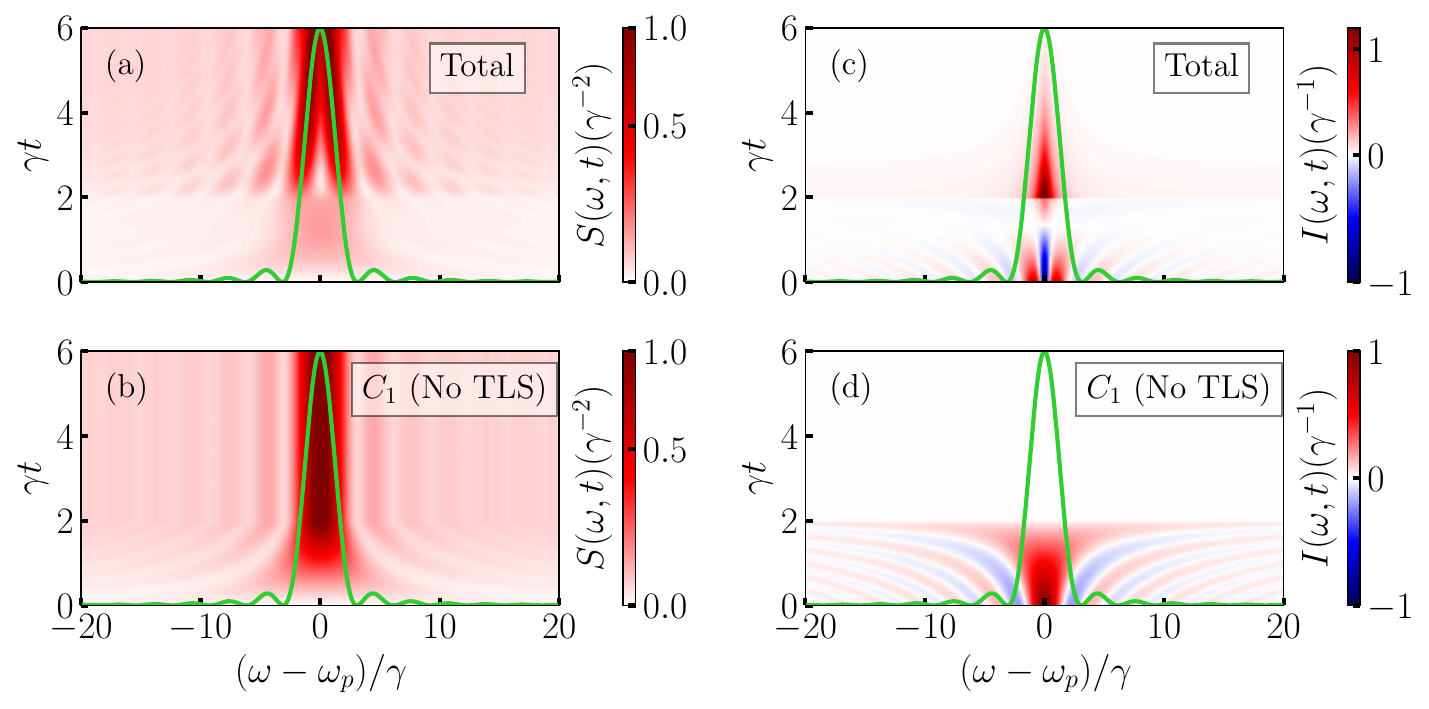}
\caption{
{\bf Chiral TLS, analytical results, 
with $\bm{ \gamma t_p=2}$.}
Analytical calculations for the transmitted time-dependent spectra and frequency-dependent flux. (a,b) Time-dependent spectrum, $S(\omega,t)$ computed from Eq.~\eqref{eq:Swt},
 for a 1-photon quantum pulse with a pulse length $\gamma t_p=2$ interacting with a chiral TLS, considering all the terms of the correlation function [Eq.~\eqref{timecorr}] (a) and only the first term $C_1(t,t+\tau)$ [Eq.~\eqref{eqC1}] (b).  (c,d) Time-dependent spectral intensity, $I(\omega,t)$ for the same cases, where (c) is the full solution and (d) is the solution considering only $C_1(t,t+\tau)$ [Eq.~\eqref{eqC1}] (i.e., no TLS interaction).
 The corresponding long time limit $S(\omega)$, is plotted in the four cases in solid green. 
}
\label{fig:analytical}
\vspace{0.2cm}
\includegraphics[width=\textwidth]{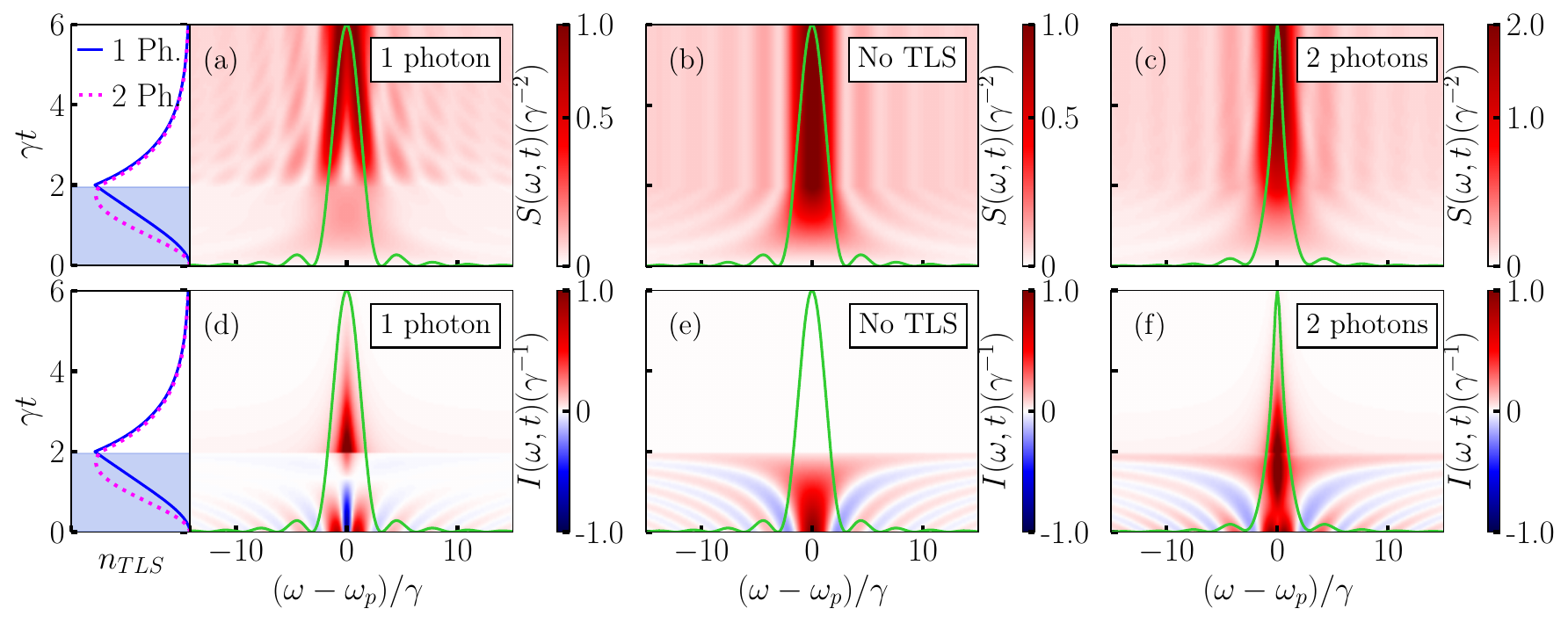}
\caption{{\bf Chiral TLS results, numerical results, with $\bm{ \gamma t_p=2}$.}
Similar results to Fig.~\ref{fig:analytical},
but using MPS, and 
now also showing the populations versus time, with 1-photon and 2-photon pulses.
The 2-photon spectra are shown in 
(c,d).
This graph is duplicated from
the results
in \cite{sofia2024letter}, but is shown here 
as also a reference to compare with the
symmetric TLS results shown in the next graph.
}
\label{fig:new_fig5}
\end{figure*}

\begin{figure*}[t]
\centering
\includegraphics[width=\textwidth]{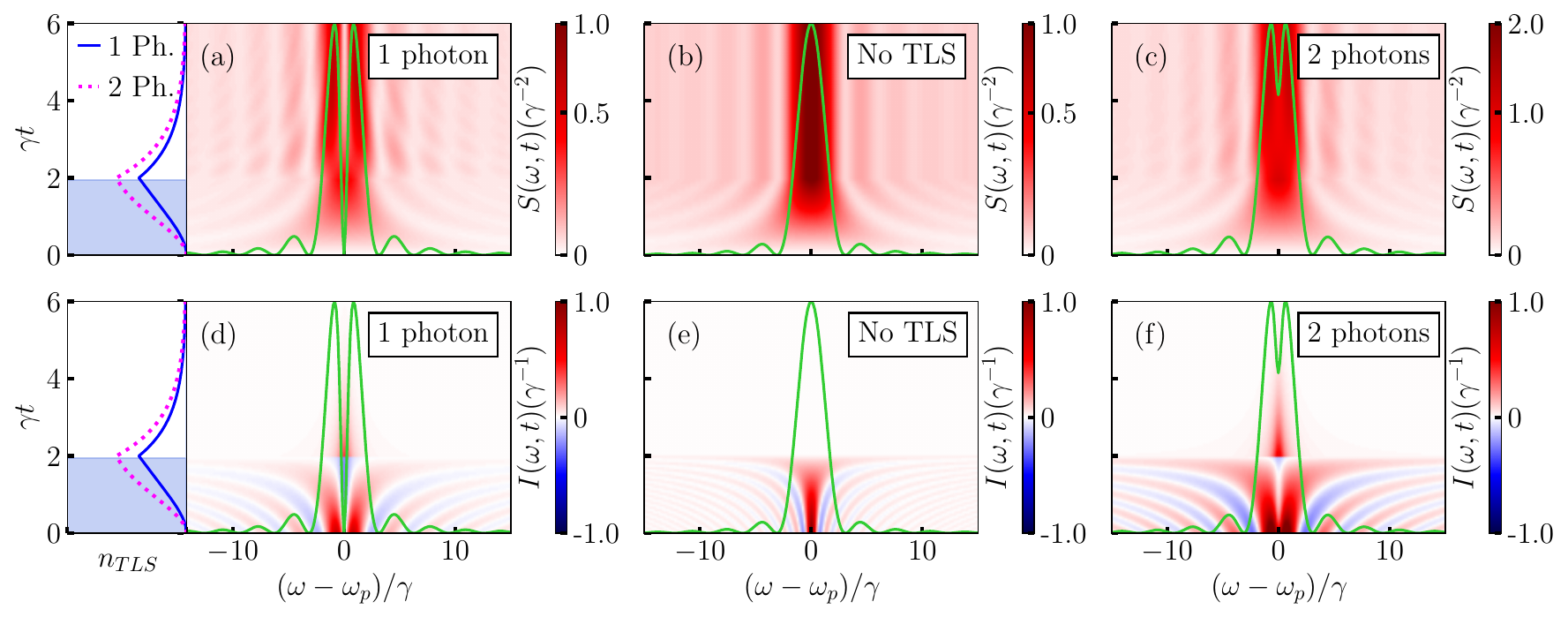}
\caption{{\bf Symmetric TLS results, with $\bm{ \gamma t_p=2}$.} (a,b,c) Time-dependent spectrum, $S(\omega,t)$, of (a) a 1-photon and (b) a 2-photon pulse with a pulse length $\gamma t_p=2$ interacting with a symmetrical TLS, and (c) a 1-photon pulse of the same length without the TLS interaction. (d,e,f) Time-dependent spectral intensity, $I(\omega,t)$, of (d) a 1-photon and (e) a 2-photon quantum pulse with a pulse length $\gamma t_p=2$ interacting with a symmetrical TLS, and (f) a 1-photon pulse of the same length without the TLS interaction. The corresponding long time limit $S(\omega)$, is plotted in the six cases in green. On the left, the TLS population ($n_{TLS}$) is plotted for both the interaction with a 1-photon pulse (solid black line) and a 2-photon pulse (dashed magenta line) and the pulse is shown with a blue shade.
}
\label{fig:symtimedepS}
\end{figure*}

\begin{figure*}[t]
\centering
\includegraphics[width=\textwidth]{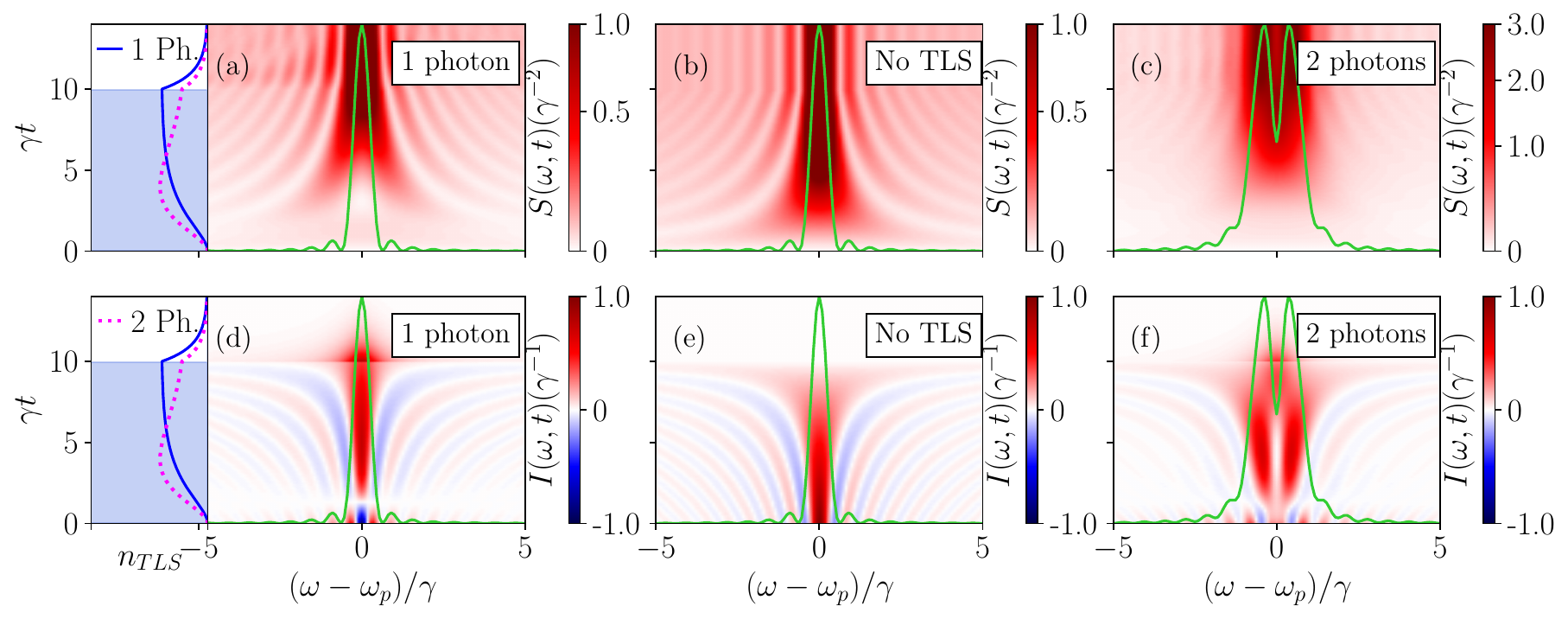}
\caption{{\bf Chiral TLS results, with $\bm{ \gamma t_p=10}$.} (a,b,c) Time-dependent spectrum, $S(\omega,t)$, of (a) a 1-photon and (b) a 2-photon pulse interacting with a chiral TLS, and (c) a 1-photon pulse of the same length without the TLS interaction. (d,e,f) Time-dependent spectral intensity, $I(\omega,t)$ for the same cases as in (a,b,c) respectively. The corresponding long time limit $S(\omega)$, is plotted in the six cases in green. On the left, the TLS population ($n_{TLS}$) is plotted for both the interaction with a 1-photon pulse (solid blue line) and a 2-photon pulse (dashed magenta line) and the pulse is shown with a blue shade.
}
\label{fig:timedepStp10}
\end{figure*}

\begin{figure*}[t]
\centering
\includegraphics[width=\textwidth]{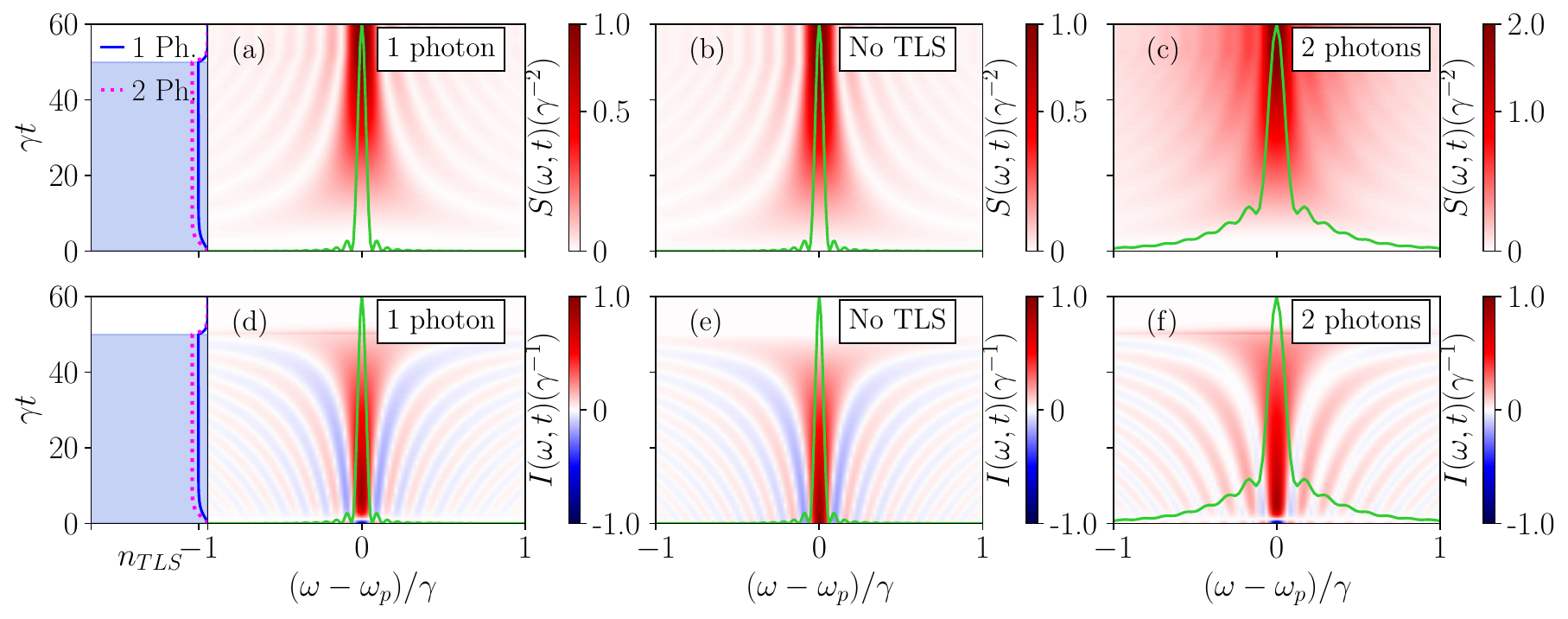}
\caption{{\bf Chiral TLS results, with $\bm{ \gamma t_p=50}$.} (a,b,c) Time-dependent spectrum, $S(\omega,t)$, of (a) a 1-photon and (b) a 2-photon pulse interacting with a chiral TLS, and (c) a 1-photon pulse of the same length without the TLS interaction. (d,e,f) Time-dependent spectral intensity, $I(\omega,t)$ for the same cases as in (a,b,c) respectively. The corresponding long time limit $S(\omega)$, is plotted in the six cases in green. On the left, the TLS population ($n_{TLS}$) is plotted for both the interaction with a 1-photon pulse (solid blue line) and a 2-photon pulse (dashed magenta line) and the pulse is shown with a blue shade.}
\label{fig:timedepStp50}
\end{figure*}

\section{Stationary and Time-Dependent Spectral Results}
\label{sec:spectralresults}

For the limit of 1-photon excitation, 
we found the  interesting
result that the time-dependent spectrum evolves to exactly the same spectrum as the input pulse when working with a chiral system. 
To better understand this, consider the
earlier derived 
long-time spectrum, $S(\omega)$, for a chiral pulse
[Eq.~\eqref{eq:S_QM}], 
    \begin{equation}
    \begin{split}
        S(\omega)= |f(\omega)|^2,
    \end{split} 
    \end{equation}
which is a consequence of the fact that the population and TLS effects completely cancel out.  

Clearly, this is different in the
time-dependent case 
[$S(\omega,t)$], when time
$t$ is not in the long time limit. For instance, if we study the $\braket{ a_R^\dagger(t)  a_R(t+\tau)}$ in the region where $t>t_p$, the only term remaining is fourth term of Eq.~\eqref{timecorr}, $C_4(t,t+\tau)$, therefore dominating the dynamics in this region,
\begin{equation}
     v_g  \braket{ a_R^\dagger(t)  a_R(t+\tau)} 
      = \frac{4}{t_p} \left( e^{\gamma t_p/2} -1 \right)^2 e^{-\gamma (t +\tau/2)},
\end{equation}
which requires as the initial condition, 
\begin{equation}
     v_g  \braket{ a_R^\dagger(t)  a_R(t)} 
     = \frac{4}{ t_p} \left[ e^{\gamma t_p/2} -1 \right]^2 e^{-\gamma t}. 
     \label{ardagar}
\end{equation}
Equation~\eqref{ardagar} resembles closely Eq.~\eqref{1phsol}, i.e. the population dynamics, 
showing that TLS population effects directly affect both $S(\omega,t)$
and $I(\omega,t)$.

Figure~\ref{fig:analytical} shows the analytical solution of the time-dependent spectrum, $S(\omega,t)$, of a 1-photon rectangular quantum pulse with $\gamma t_p=2$
interacting with a chiral TLS. In 
Fig.~\ref{fig:analytical}(a) we can see the full solution and in Fig.~\ref{fig:analytical}(b), the solution when we do not consider population effects (only considering the first term of the first order correlation function, $C_1(t,t+\tau)$). First of all, we can see again the agreement between the analytical results and the ones calculated using MPS (see Ref.~\cite{sofia2024letter}). In addition, these results show the importance of considering the population dynamics when calculating the time-dependent spectrum. The solution with only the first term of the analytical solution, $C_1(t,t+\tau)$, gives the same results as when not considering the TLS when solving the problem.

This is additionally shown in Fig.~\ref{fig:analytical}(c,d) where the spectral intensity is calculated for the same cases. Again, from this result, we can clearly see the contribution of the population dynamics.

Note that unphysical solutions are
 obtained if we simply neglect the
 population term ($C_4$) in computing the spectra, where
 the sum of $C_1+C_2+C_3$ yield negative spectra with  qualitatively different 
 features.

These analytical results
agree  agree perfectly with the numerical results shown in
Ref.~\onlinecite{sofia2024letter},
which are reproduced
in Fig.~\ref{fig:new_fig5}, where we also show the 2-photon spectra (in part to compare with the symmetric TLS results discussed next).

Clearly, 
this population behavior is not exclusive to the chiral solution, and in Fig.~\ref{fig:symtimedepS}, we show the time-dependent spectral results when considering a symmetrical emitter. The spectrum $S(\omega ,t)$ and the spectral intensity $I(\omega ,t)$ are calculated using MPS, where we use again a rectangular pulse of pulse length $\gamma t_p = 2$ containing one photon Fig.~\ref{fig:symtimedepS}(a,d). For comparison, the same observables are calculated without TLS interaction in Fig.~\ref{fig:symtimedepS}(b,e). We observe how, in the stationary spectra, where we recover the harmonic oscillator solution. However, the population effects are evident in both the dynamical spectrum and the spectral intensity, with results that highly differ from the case with no TLS. For instance, in the spectral intensity [Fig.~\ref{fig:symtimedepS}(d)], we observe positive values at times longer than the pulse length (for $\gamma t > 2$), while these effects are (obviously) not visible when there is not a TLS interacting with the pulse [Fig.~\ref{fig:symtimedepS}(e)].

Further, we have also calculated the solution for a 2-photon pulse with the same pulse length ($\gamma t_p = 2$) interacting with the symmetrical emitter [Fig.~\ref{fig:symtimedepS}(c,f)], where we again observe similar population effects. However, in this case we can also observe the population effects in the stationary spectrum (represented by the green solid line).

\subsection*{Influence of pulse duration on the transmitted spectra}

As previously seen in Sec.~\ref{subsec:pulseduration}, the TLS population dynamics depend on the pulse duration, with results that deviate from the WEA approximation and visible nonlinear effects in the case of 2-photon pulse.

We next examine the role of using a longer pulse
on the emitted spectra.
In 
Fig.~\ref{fig:timedepStp10},  the spectra and spectral intensity are calculated for a chiral TLS interacting with a rectangular pulse of length $\gamma t_p = 10$, containing 1 photon Fig.~\ref{fig:timedepStp10} (a,d) and 2 photons Fig.~\ref{fig:timedepStp10} (c,f). The pulse results with no interaction are shown in Fig.~\ref{fig:timedepStp10}(b,e). Here, we observe how the TLS effects can still be appreciated in both cases, with solutions that deviate from the cases with no TLS interaction. In the 1-photon solutions, we observe how the presence of the TLS effects start to become more subtle. However, in the 2-photon cases, the nonlinearities are still very present, where we can even see a stronger break up of the pulse in frequency. 
Similar pulse break up effects have been found in
\cite{PhysRevX.10.031011} using
ensembles of chiral emitters.

In Fig.~\ref{fig:timedepStp50}, we show results similar to the previous ones, but with an even longer pulse ($\gamma t_p = 50$). In this case, we observe solutions that are much closer to the case with no TLS. 
Specifically, in the one-photon solutions, some small dynamical effects can still be seen when the pulse turns on and off, with the populations reaching a quasi- steady-state during most of the pulse. The two-photon solution still has some significant deviations from the case without the TLS interaction, but shows no visible pulse break up effects (caused by TLS saturation).

These results confirm that the dynamical effects seen in the spectral results are directly related to the TLS population effects and that they persist as long as there is a reasonable TLS population.

\begin{figure}[t]
\centering
\includegraphics[width=0.8\columnwidth]{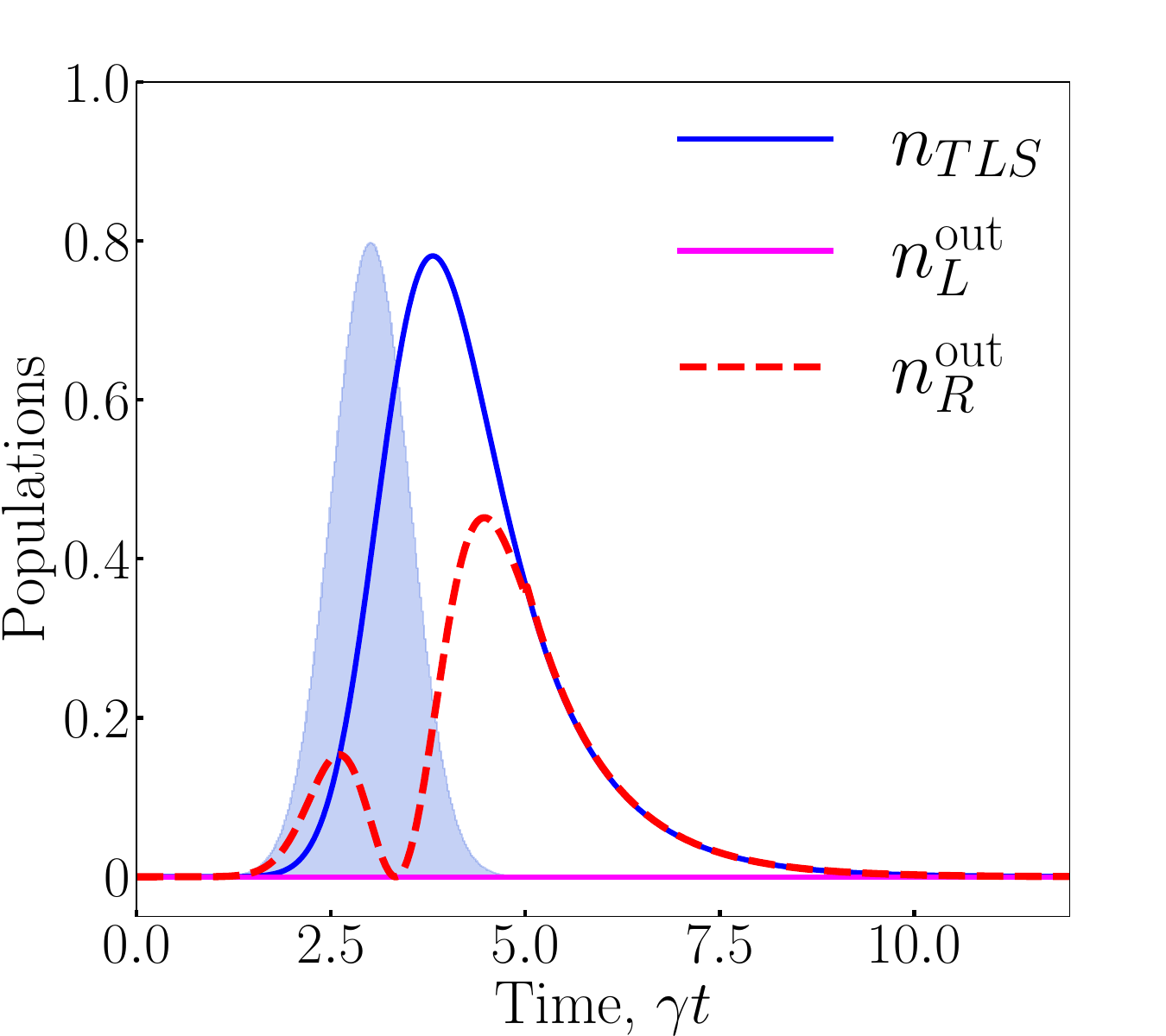}
\caption{{\bf Chiral TLS results with a 
1-photon Gaussian pulse, with $\bm{ \gamma t_p=1}$.} Quantum dynamics for a chiral single TLS-waveguide system, using an incident Gaussian quantum pulse (cyan shaded region) containing 1 photon. The figure shows the TLS population (solid blue), and the normalized photon population of the right (dashed red) and left (magenta) moving photons.
}
\label{fig:gaussianpop}
\end{figure}

\section{Gaussian pulses}
\label{sec:gaussian}

\begin{figure*}[t]
\centering
\includegraphics[width=\textwidth]{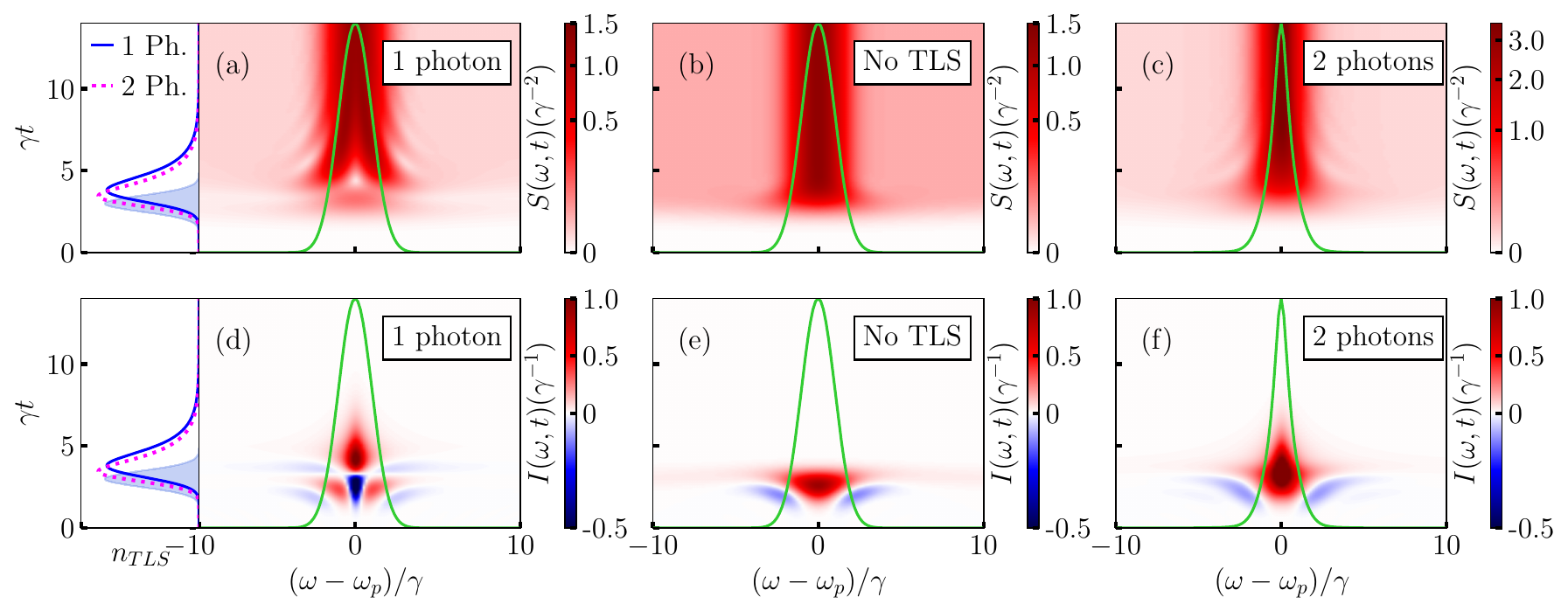}
\caption{{\bf Chiral TLS results for a Gaussian pulse, with $\bm{ \gamma t_p=1}$.} (a,b,c) Time-dependent spectrum, $S(\omega,t)$, of (a) a 1-photon and (b) a 2-photon pulse with a pulse width of $\gamma t_p=1$ interacting with a chiral TLS, and (c) a 1-photon pulse of the same shape without the TLS interaction. (d,e,f) Time-dependent spectral intensity, $I(\omega,t)$, of (d) a 1-photon and (e) a 2-photon quantum pulse with a pulse width $\gamma t_p=1$ interacting with a chiral TLS, and (f) a 1-photon pulse of the same length without the TLS interaction. The corresponding long time limit $S(\omega)$, is plotted in the six cases in green. On the left, the TLS population ($n_{TLS}$) is plotted for both the interaction with a 1-photon pulse (solid blue curve) and a 2-photon pulse (dashed magenta curve) and the pulse is shown with a blue shade. 
}
\label{fig:gaussianspec}
\end{figure*}

For completeness, we next show results
for few-photon excitation with Gaussian pulses, since the MPS results can use arbitrary waveforms. In this case, the temporal pulse has the following normalized shape,
\begin{equation}
    f(t) = \frac{1}{\sqrt{2\pi} t_p} 
    \exp\left \{-\frac{(t-t_c)^2}{2t_p^2}\right \},
\end{equation}
where $t_c$ is the center of the pulse and $t_p$ is its width.
Figure~\ref{fig:gaussianpop}
shows the various populations,
for a 1-photon pulse, with a width of $\gamma t_p=1$ and centered at $\gamma t_c=3$, interacting with a chiral emitter. Here again, the shaded blue area represents the incident pulse, the solid blue line is the TLS population, and the red dashed line represents the normalized transmitted photon population. Interestingly, there is an instant that the photon population is zero in this case (just after the center of the pulse). Additionally, the left photon population  stay at zero at all times since this is a chiral system. 

The corresponding
spectral functions
$S(\omega,t)$
and $I(\omega,t)$ are shown in
Fig.~\ref{fig:gaussianspec}. To better appreciate the influence of population effects on the dynamical spectrum, in Fig.~\ref{fig:gaussianspec},
again 
we show both the population decay and the dynamic spectra side by side. As also seen in Fig.~\ref{fig:gaussianpop}, one can clearly observe in  Fig.~\ref{fig:gaussianspec}(a) the effects of the TLS population 
dynamics, which causes a spectral hole and additional interferences. This can be appreciated by comparing these solutions with the corresponding $S(\omega,t)$
 without the TLS, shown in 
 Fig.~\ref{fig:gaussianspec}(b).  However, as in the case of the rectangular pulse, in the long time limit, all the population effects get cancelled and we recover the same stationary spectra in both cases (green solid lines).
 
 In addition, rich features are observed in the spectral intensity $I(\omega,t)$ in Fig.~\ref{fig:gaussianspec}(d), where again, by comparison with the case without the TLS [Fig.~\ref{fig:gaussianspec}(e)], we see additional interferences during the pulse, and intensity positive values after the pulse while the TLS is decaying, that cannot be observed without this interaction. Nevertheless, if we integrate the spectral intensity in both cases, we again recover the same solution which agrees with the stationary spectra (green solid lines).

Finally, Figs.~\ref{fig:gaussianspec}(c) and (f) show the same spectral observables, $S(\omega,t)$ and $I(\omega,t)$ respectively, when using a similar Gaussian pulse but now containing 2 photons. Here we see again the TLS dynamical effects in both cases, and, in agreement with what we saw with rectangular pulses, the stationary spectrum now shows the nonlinear excitation, with a narrower width and a more pronounced central peak.

These results show further details about how population dynamics influence the $S(\omega, t)$ spectral plots, and also which ones are due to the specific shape of the pulse used as well (e.g., showing convolution effects with a sinc function or a Gaussian function). Note also, that an analytical solution for the 1-photon excitation is possible in terms of the error function (or via a numerical integration), but may not be so easy to obtain for the two-photon case.

\section{Conclusions}
\label{sec:conclusions}

We have presented a detailed theoretical analysis of few-photon quantum pulses interacting with a single TLS in a waveguide, using both chiral and symmetric TLS emitters. This 
complements and adds substantial details and extensions to the results shown in our related 
Letter~\cite{sofia2024letter}.

For the theory,
we have shown two approaches to compute the TLS population information, visible both in the time dynamics and in certain spectral observables. We started by deriving the analytical solution of the TLS population by solving the corresponding equations of motion in the Heisenberg picture. We have presented the general differential equations for a pulse containing 1 photon, and then solved for a rectangular pulse. Here, we have also shown both solutions with a symmetrical emitter or a chiral emitter
that couples only to the right photon channel, and then we have repeated the same procedure for a rectangular pulse containing 2 photons. 

Next, we have used the same method to derive the analytical solution for the 1-photon stationary spectra
(i.e., in the long time limit). Here, we have seen how all the population effects that were observed in the time dynamics perfectly cancel out, and we recover the same solution with or without the interaction with the TLS (for this observable),
for a chiral emitter. For a symmetric emitter, we obtain the same result as a WEA, essentially recovering elastic scattering theory.

In order to better understand these effects, we then introduced a derivation for a time-dependent spectrum, starting with the derivation of the transmitted photon population, and then the transmitted first-order photon correlation functions, and we have shown how these relate to the time-dependent spectra and spectral intensity.

Our analytical results are benchmarked, checked,  and extended by using a second theoretical method, based on matrix product states (MPS), which we have introduced after the analytical derivations. This is a tensor network approach which both helped us to verify the analytical solutions, and extend our simulations to more efficient calculations with pulses of different shapes and containing more photons.

After presenting our models, solutions with a rectangular incident pulse are shown, starting with the population results, where a comparison between symmetrical and chiral emitters is shown, for both 1 and 2-photon pulses. In the general cases of short pulses, we have observed high values of TLS populations making clear a deviation from the weak excitation limit. Furthermore, this has been extended with a study of the maximum TLS population as a function of pulse length, and we have seen how the weak excitation approximation is not valid until very long pulses, on the order of $\gamma t \sim 500$. 
We have also observed in these cases how nonlinear effects are visible in the 2-photon cases until similar lengths.
In addition, we showed how the time-dependent spectral profiles depend on the length of the pulse, where eventually the population effects visible for 1-photon pulses become less and less important,
and we showed the clear difference between 1-photon and 2-photon pulses.

The spectral results have shown us how population effects influence both the dynamical spectra and the spectral intensity, even though they completely cancel out in the 1-photon stationary solutions. We have also seen that this is not the case when considering a 2-photon pulse, where the nonlinear effects are also seen in the stationary limit,
yielding a narrowing of the central pulse peak for a chiral TLS and pulse break up for
a symmetric TLS.

Finally, we have calculated similar observables for a Gaussian incident pulse, both containing 1 and 2 photons, interacting with a chiral TLS. In all the cases we observe similar behaviors as in the case of the rectangular pulse (which permits simpler analytical solutions). These
results confirm that the findings in this paper are not particular to the shape of the pulse, and can be extended to other pulse shapes
as well.

\vspace{0.5cm}

\acknowledgements

This work was supported by the Natural Sciences and Engineering Research Council of Canada (NSERC) [Discovery and Quantum Alliance Grants],
the National Research Council of Canada (NRC),
the Canadian Foundation for Innovation (CFI), Queen's University, Canada, and the 
Alexander von Humboldt Foundation 
through a Humboldt Award. 
We also thank Kisa Barkemeyer,
Jacob Ewaniuk, and Nir Rotenberg
for valuable contributions and discussions.

\vspace{0.5cm}

\bibliography{references_all}

\begin{thebibliography}{70}%
\makeatletter
\providecommand \@ifxundefined [1]{%
 \@ifx{#1\undefined}
}%
\providecommand \@ifnum [1]{%
 \ifnum #1\expandafter \@firstoftwo
 \else \expandafter \@secondoftwo
 \fi
}%
\providecommand \@ifx [1]{%
 \ifx #1\expandafter \@firstoftwo
 \else \expandafter \@secondoftwo
 \fi
}%
\providecommand \natexlab [1]{#1}%
\providecommand \enquote  [1]{``#1''}%
\providecommand \bibnamefont  [1]{#1}%
\providecommand \bibfnamefont [1]{#1}%
\providecommand \citenamefont [1]{#1}%
\providecommand \href@noop [0]{\@secondoftwo}%
\providecommand \href [0]{\begingroup \@sanitize@url \@href}%
\providecommand \@href[1]{\@@startlink{#1}\@@href}%
\providecommand \@@href[1]{\endgroup#1\@@endlink}%
\providecommand \@sanitize@url [0]{\catcode `\\12\catcode `\$12\catcode
  `\&12\catcode `\#12\catcode `\^12\catcode `\_12\catcode `\%12\relax}%
\providecommand \@@startlink[1]{}%
\providecommand \@@endlink[0]{}%
\providecommand \url  [0]{\begingroup\@sanitize@url \@url }%
\providecommand \@url [1]{\endgroup\@href {#1}{\urlprefix }}%
\providecommand \urlprefix  [0]{URL }%
\providecommand \Eprint [0]{\href }%
\providecommand \doibase [0]{https://doi.org/}%
\providecommand \selectlanguage [0]{\@gobble}%
\providecommand \bibinfo  [0]{\@secondoftwo}%
\providecommand \bibfield  [0]{\@secondoftwo}%
\providecommand \translation [1]{[#1]}%
\providecommand \BibitemOpen [0]{}%
\providecommand \bibitemStop [0]{}%
\providecommand \bibitemNoStop [0]{.\EOS\space}%
\providecommand \EOS [0]{\spacefactor3000\relax}%
\providecommand \BibitemShut  [1]{\csname bibitem#1\endcsname}%
\let\auto@bib@innerbib\@empty
%</preamble>
\bibitem [{\citenamefont {Sheremet}\ \emph {et~al.}(2023)\citenamefont
  {Sheremet}, \citenamefont {Petrov}, \citenamefont {Iorsh}, \citenamefont
  {Poshakinskiy},\ and\ \citenamefont {Poddubny}}]{RevModPhys.95.015002}%
  \BibitemOpen
  \bibfield  {author} {\bibinfo {author} {\bibfnamefont {A.~S.}\ \bibnamefont
  {Sheremet}}, \bibinfo {author} {\bibfnamefont {M.~I.}\ \bibnamefont
  {Petrov}}, \bibinfo {author} {\bibfnamefont {I.~V.}\ \bibnamefont {Iorsh}},
  \bibinfo {author} {\bibfnamefont {A.~V.}\ \bibnamefont {Poshakinskiy}},\ and\
  \bibinfo {author} {\bibfnamefont {A.~N.}\ \bibnamefont {Poddubny}},\
  }\bibfield  {title} {\bibinfo {title} {Waveguide quantum electrodynamics:
  Collective radiance and photon-photon correlations},\ }\href
  {https://doi.org/10.1103/RevModPhys.95.015002} {\bibfield  {journal}
  {\bibinfo  {journal} {Rev. Mod. Phys.}\ }\textbf {\bibinfo {volume} {95}},\
  \bibinfo {pages} {015002} (\bibinfo {year} {2023})}\BibitemShut {NoStop}%
\bibitem [{\citenamefont {Zheng}\ \emph {et~al.}(2010)\citenamefont {Zheng},
  \citenamefont {Gauthier},\ and\ \citenamefont
  {Baranger}}]{PhysRevA.82.063816}%
  \BibitemOpen
  \bibfield  {author} {\bibinfo {author} {\bibfnamefont {H.}~\bibnamefont
  {Zheng}}, \bibinfo {author} {\bibfnamefont {D.~J.}\ \bibnamefont
  {Gauthier}},\ and\ \bibinfo {author} {\bibfnamefont {H.~U.}\ \bibnamefont
  {Baranger}},\ }\bibfield  {title} {\bibinfo {title} {Waveguide {QED}:
  Many-body bound-state effects in coherent and {Fock-state} scattering from a
  two-level system},\ }\href {https://doi.org/10.1103/PhysRevA.82.063816}
  {\bibfield  {journal} {\bibinfo  {journal} {Phys. Rev. A}\ }\textbf {\bibinfo
  {volume} {82}},\ \bibinfo {pages} {063816} (\bibinfo {year}
  {2010})}\BibitemShut {NoStop}%
\bibitem [{\citenamefont {Witthaut}\ and\ \citenamefont
  {Sørensen}(2010)}]{Witthaut_2010}%
  \BibitemOpen
  \bibfield  {author} {\bibinfo {author} {\bibfnamefont {D.}~\bibnamefont
  {Witthaut}}\ and\ \bibinfo {author} {\bibfnamefont {A.~S.}\ \bibnamefont
  {Sørensen}},\ }\bibfield  {title} {\bibinfo {title} {Photon scattering by a
  three-level emitter in a one-dimensional waveguide},\ }\href
  {https://doi.org/10.1088/1367-2630/12/4/043052} {\bibfield  {journal}
  {\bibinfo  {journal} {New Journal of Physics}\ }\textbf {\bibinfo {volume}
  {12}},\ \bibinfo {pages} {043052} (\bibinfo {year} {2010})}\BibitemShut
  {NoStop}%
\bibitem [{\citenamefont {Roy}(2011)}]{PhysRevLett.106.053601}%
  \BibitemOpen
  \bibfield  {author} {\bibinfo {author} {\bibfnamefont {D.}~\bibnamefont
  {Roy}},\ }\bibfield  {title} {\bibinfo {title} {Two-photon scattering by a
  driven three-level emitter in a one-dimensional waveguide and
  electromagnetically induced transparency},\ }\href
  {https://doi.org/10.1103/PhysRevLett.106.053601} {\bibfield  {journal}
  {\bibinfo  {journal} {Phys. Rev. Lett.}\ }\textbf {\bibinfo {volume} {106}},\
  \bibinfo {pages} {053601} (\bibinfo {year} {2011})}\BibitemShut {NoStop}%
\bibitem [{\citenamefont {Crowder}\ \emph {et~al.}(2020)\citenamefont
  {Crowder}, \citenamefont {Carmichael},\ and\ \citenamefont
  {Hughes}}]{PhysRevA.101.023807}%
  \BibitemOpen
  \bibfield  {author} {\bibinfo {author} {\bibfnamefont {G.}~\bibnamefont
  {Crowder}}, \bibinfo {author} {\bibfnamefont {J.~H.}\ \bibnamefont
  {Carmichael}},\ and\ \bibinfo {author} {\bibfnamefont {S.}~\bibnamefont
  {Hughes}},\ }\bibfield  {title} {\bibinfo {title} {Quantum trajectory theory
  of few-photon {cavity-{QED}} systems with a time-delayed coherent feedback},\
  }\href {https://doi.org/10.1103/PhysRevA.101.023807} {\bibfield  {journal}
  {\bibinfo  {journal} {Phys. Rev. A}\ }\textbf {\bibinfo {volume} {101}},\
  \bibinfo {pages} {023807} (\bibinfo {year} {2020})}\BibitemShut {NoStop}%
\bibitem [{\citenamefont {Crowder}\ \emph {et~al.}(2022)\citenamefont
  {Crowder}, \citenamefont {Ramunno},\ and\ \citenamefont
  {Hughes}}]{PhysRevA.106.013714}%
  \BibitemOpen
  \bibfield  {author} {\bibinfo {author} {\bibfnamefont {G.}~\bibnamefont
  {Crowder}}, \bibinfo {author} {\bibfnamefont {L.}~\bibnamefont {Ramunno}},\
  and\ \bibinfo {author} {\bibfnamefont {S.}~\bibnamefont {Hughes}},\
  }\bibfield  {title} {\bibinfo {title} {Quantum trajectory theory and
  simulations of nonlinear spectra and multiphoton effects in {waveguide-{QED}}
  systems with a time-delayed coherent feedback},\ }\href
  {https://doi.org/10.1103/PhysRevA.106.013714} {\bibfield  {journal} {\bibinfo
   {journal} {Phys. Rev. A}\ }\textbf {\bibinfo {volume} {106}},\ \bibinfo
  {pages} {013714} (\bibinfo {year} {2022})}\BibitemShut {NoStop}%
\bibitem [{\citenamefont {Shen}\ and\ \citenamefont {Fan}(2005)}]{Shen:05}%
  \BibitemOpen
  \bibfield  {author} {\bibinfo {author} {\bibfnamefont {J.~T.}\ \bibnamefont
  {Shen}}\ and\ \bibinfo {author} {\bibfnamefont {S.}~\bibnamefont {Fan}},\
  }\bibfield  {title} {\bibinfo {title} {Coherent photon transport from
  spontaneous emission in one-dimensional waveguides},\ }\href
  {https://doi.org/10.1364/OL.30.002001} {\bibfield  {journal} {\bibinfo
  {journal} {Opt. Lett.}\ }\textbf {\bibinfo {volume} {30}},\ \bibinfo {pages}
  {2001} (\bibinfo {year} {2005})}\BibitemShut {NoStop}%
\bibitem [{\citenamefont {Le~Jeannic}\ \emph {et~al.}(2022)\citenamefont
  {Le~Jeannic}, \citenamefont {Tiranov}, \citenamefont {Carolan}, \citenamefont
  {Ramos}, \citenamefont {Wang}, \citenamefont {Appel}, \citenamefont {Scholz},
  \citenamefont {Wieck}, \citenamefont {Ludwig}, \citenamefont {Rotenberg},
  \citenamefont {Midolo}, \citenamefont {García-Ripoll}, \citenamefont
  {Sørensen},\ and\ \citenamefont {Lodahl}}]{LeJeannic2022}%
  \BibitemOpen
  \bibfield  {author} {\bibinfo {author} {\bibfnamefont {H.}~\bibnamefont
  {Le~Jeannic}}, \bibinfo {author} {\bibfnamefont {A.}~\bibnamefont {Tiranov}},
  \bibinfo {author} {\bibfnamefont {J.}~\bibnamefont {Carolan}}, \bibinfo
  {author} {\bibfnamefont {T.}~\bibnamefont {Ramos}}, \bibinfo {author}
  {\bibfnamefont {Y.}~\bibnamefont {Wang}}, \bibinfo {author} {\bibfnamefont
  {M.~H.}\ \bibnamefont {Appel}}, \bibinfo {author} {\bibfnamefont
  {S.}~\bibnamefont {Scholz}}, \bibinfo {author} {\bibfnamefont {A.~D.}\
  \bibnamefont {Wieck}}, \bibinfo {author} {\bibfnamefont {A.}~\bibnamefont
  {Ludwig}}, \bibinfo {author} {\bibfnamefont {N.}~\bibnamefont {Rotenberg}},
  \bibinfo {author} {\bibfnamefont {L.}~\bibnamefont {Midolo}}, \bibinfo
  {author} {\bibfnamefont {J.~J.}\ \bibnamefont {García-Ripoll}}, \bibinfo
  {author} {\bibfnamefont {A.~S.}\ \bibnamefont {Sørensen}},\ and\ \bibinfo
  {author} {\bibfnamefont {P.}~\bibnamefont {Lodahl}},\ }\bibfield  {title}
  {\bibinfo {title} {Dynamical photon–photon interaction mediated by a
  quantum emitter},\ }\href {https://doi.org/10.1038/s41567-022-01720-x}
  {\bibfield  {journal} {\bibinfo  {journal} {Nature Physics}\ }\textbf
  {\bibinfo {volume} {18}},\ \bibinfo {pages} {1191–1195} (\bibinfo {year}
  {2022})}\BibitemShut {NoStop}%
\bibitem [{\citenamefont {Longo}\ \emph {et~al.}(2011)\citenamefont {Longo},
  \citenamefont {Schmitteckert},\ and\ \citenamefont
  {Busch}}]{PhysRevA.83.063828}%
  \BibitemOpen
  \bibfield  {author} {\bibinfo {author} {\bibfnamefont {P.}~\bibnamefont
  {Longo}}, \bibinfo {author} {\bibfnamefont {P.}~\bibnamefont
  {Schmitteckert}},\ and\ \bibinfo {author} {\bibfnamefont {K.}~\bibnamefont
  {Busch}},\ }\bibfield  {title} {\bibinfo {title} {Few-photon transport in
  low-dimensional systems},\ }\href
  {https://doi.org/10.1103/PhysRevA.83.063828} {\bibfield  {journal} {\bibinfo
  {journal} {Phys. Rev. A}\ }\textbf {\bibinfo {volume} {83}},\ \bibinfo
  {pages} {063828} (\bibinfo {year} {2011})}\BibitemShut {NoStop}%
\bibitem [{\citenamefont {Crowder}\ \emph {et~al.}(2024)\citenamefont
  {Crowder}, \citenamefont {Ramunno},\ and\ \citenamefont
  {Hughes}}]{PhysRevA.110.L031703}%
  \BibitemOpen
  \bibfield  {author} {\bibinfo {author} {\bibfnamefont {G.}~\bibnamefont
  {Crowder}}, \bibinfo {author} {\bibfnamefont {L.}~\bibnamefont {Ramunno}},\
  and\ \bibinfo {author} {\bibfnamefont {S.}~\bibnamefont {Hughes}},\
  }\bibfield  {title} {\bibinfo {title} {Improving on-demand
  single-photon-source coherence and indistinguishability through a
  time-delayed coherent feedback},\ }\href
  {https://doi.org/10.1103/PhysRevA.110.L031703} {\bibfield  {journal}
  {\bibinfo  {journal} {Phys. Rev. A}\ }\textbf {\bibinfo {volume} {110}},\
  \bibinfo {pages} {L031703} (\bibinfo {year} {2024})}\BibitemShut {NoStop}%
\bibitem [{\citenamefont {T\"{u}rschmann}\ \emph {et~al.}(2019)\citenamefont
  {T\"{u}rschmann}, \citenamefont {Jeannic}, \citenamefont {Simonsen},
  \citenamefont {Haakh}, \citenamefont {G\"{o}tzinger}, \citenamefont
  {Sandoghdar}, \citenamefont {Lodahl},\ and\ \citenamefont
  {Rotenberg}}]{Trschmann2019}%
  \BibitemOpen
  \bibfield  {author} {\bibinfo {author} {\bibfnamefont {P.}~\bibnamefont
  {T\"{u}rschmann}}, \bibinfo {author} {\bibfnamefont {H.~L.}\ \bibnamefont
  {Jeannic}}, \bibinfo {author} {\bibfnamefont {S.~F.}\ \bibnamefont
  {Simonsen}}, \bibinfo {author} {\bibfnamefont {H.~R.}\ \bibnamefont {Haakh}},
  \bibinfo {author} {\bibfnamefont {S.}~\bibnamefont {G\"{o}tzinger}}, \bibinfo
  {author} {\bibfnamefont {V.}~\bibnamefont {Sandoghdar}}, \bibinfo {author}
  {\bibfnamefont {P.}~\bibnamefont {Lodahl}},\ and\ \bibinfo {author}
  {\bibfnamefont {N.}~\bibnamefont {Rotenberg}},\ }\bibfield  {title} {\bibinfo
  {title} {Coherent nonlinear optics of quantum emitters in nanophotonic
  waveguides},\ }\href {https://doi.org/10.1515/nanoph-2019-0126} {\bibfield
  {journal} {\bibinfo  {journal} {Nanophotonics}\ }\textbf {\bibinfo {volume}
  {8}},\ \bibinfo {pages} {1641} (\bibinfo {year} {2019})}\BibitemShut
  {NoStop}%
\bibitem [{\citenamefont {Pregnolato}\ \emph {et~al.}(2020)\citenamefont
  {Pregnolato}, \citenamefont {Chu}, \citenamefont {Schröder}, \citenamefont
  {Schott}, \citenamefont {Wieck}, \citenamefont {Ludwig}, \citenamefont
  {Lodahl},\ and\ \citenamefont {Rotenberg}}]{10.1063/1.5117888}%
  \BibitemOpen
  \bibfield  {author} {\bibinfo {author} {\bibfnamefont {T.}~\bibnamefont
  {Pregnolato}}, \bibinfo {author} {\bibfnamefont {X.-L.}\ \bibnamefont {Chu}},
  \bibinfo {author} {\bibfnamefont {T.}~\bibnamefont {Schröder}}, \bibinfo
  {author} {\bibfnamefont {R.}~\bibnamefont {Schott}}, \bibinfo {author}
  {\bibfnamefont {A.~D.}\ \bibnamefont {Wieck}}, \bibinfo {author}
  {\bibfnamefont {A.}~\bibnamefont {Ludwig}}, \bibinfo {author} {\bibfnamefont
  {P.}~\bibnamefont {Lodahl}},\ and\ \bibinfo {author} {\bibfnamefont
  {N.}~\bibnamefont {Rotenberg}},\ }\bibfield  {title} {\bibinfo {title}
  {{Deterministic positioning of nanophotonic waveguides around single
  self-assembled quantum dots}},\ }\href {https://doi.org/10.1063/1.5117888}
  {\bibfield  {journal} {\bibinfo  {journal} {APL Photonics}\ }\textbf
  {\bibinfo {volume} {5}},\ \bibinfo {pages} {086101} (\bibinfo {year}
  {2020})}\BibitemShut {NoStop}%
\bibitem [{\citenamefont {Lund-Hansen}\ \emph {et~al.}(2008)\citenamefont
  {Lund-Hansen}, \citenamefont {Stobbe}, \citenamefont {Julsgaard},
  \citenamefont {Thyrrestrup}, \citenamefont {S\"unner}, \citenamefont {Kamp},
  \citenamefont {Forchel},\ and\ \citenamefont
  {Lodahl}}]{PhysRevLett.101.113903}%
  \BibitemOpen
  \bibfield  {author} {\bibinfo {author} {\bibfnamefont {T.}~\bibnamefont
  {Lund-Hansen}}, \bibinfo {author} {\bibfnamefont {S.}~\bibnamefont {Stobbe}},
  \bibinfo {author} {\bibfnamefont {B.}~\bibnamefont {Julsgaard}}, \bibinfo
  {author} {\bibfnamefont {H.}~\bibnamefont {Thyrrestrup}}, \bibinfo {author}
  {\bibfnamefont {T.}~\bibnamefont {S\"unner}}, \bibinfo {author}
  {\bibfnamefont {M.}~\bibnamefont {Kamp}}, \bibinfo {author} {\bibfnamefont
  {A.}~\bibnamefont {Forchel}},\ and\ \bibinfo {author} {\bibfnamefont
  {P.}~\bibnamefont {Lodahl}},\ }\bibfield  {title} {\bibinfo {title}
  {Experimental realization of highly efficient broadband coupling of single
  quantum dots to a photonic crystal waveguide},\ }\href
  {https://doi.org/10.1103/PhysRevLett.101.113903} {\bibfield  {journal}
  {\bibinfo  {journal} {Phys. Rev. Lett.}\ }\textbf {\bibinfo {volume} {101}},\
  \bibinfo {pages} {113903} (\bibinfo {year} {2008})}\BibitemShut {NoStop}%
\bibitem [{\citenamefont {Mirhosseini}\ \emph {et~al.}(2019)\citenamefont
  {Mirhosseini}, \citenamefont {Kim}, \citenamefont {Zhang}, \citenamefont
  {Sipahigil}, \citenamefont {Dieterle}, \citenamefont {Keller}, \citenamefont
  {Asenjo-Garcia}, \citenamefont {Chang},\ and\ \citenamefont
  {Painter}}]{Mirhosseini2019}%
  \BibitemOpen
  \bibfield  {author} {\bibinfo {author} {\bibfnamefont {M.}~\bibnamefont
  {Mirhosseini}}, \bibinfo {author} {\bibfnamefont {E.}~\bibnamefont {Kim}},
  \bibinfo {author} {\bibfnamefont {X.}~\bibnamefont {Zhang}}, \bibinfo
  {author} {\bibfnamefont {A.}~\bibnamefont {Sipahigil}}, \bibinfo {author}
  {\bibfnamefont {P.~B.}\ \bibnamefont {Dieterle}}, \bibinfo {author}
  {\bibfnamefont {A.~J.}\ \bibnamefont {Keller}}, \bibinfo {author}
  {\bibfnamefont {A.}~\bibnamefont {Asenjo-Garcia}}, \bibinfo {author}
  {\bibfnamefont {D.~E.}\ \bibnamefont {Chang}},\ and\ \bibinfo {author}
  {\bibfnamefont {O.}~\bibnamefont {Painter}},\ }\bibfield  {title} {\bibinfo
  {title} {Cavity quantum electrodynamics with atom-like mirrors},\ }\href
  {https://doi.org/10.1038/s41586-019-1196-1} {\bibfield  {journal} {\bibinfo
  {journal} {Nature}\ }\textbf {\bibinfo {volume} {569}},\ \bibinfo {pages}
  {692} (\bibinfo {year} {2019})}\BibitemShut {NoStop}%
\bibitem [{\citenamefont {Gonz\'alez-Tudela}\ \emph {et~al.}(2015)\citenamefont
  {Gonz\'alez-Tudela}, \citenamefont {Paulisch}, \citenamefont {Chang},
  \citenamefont {Kimble},\ and\ \citenamefont
  {Cirac}}]{PhysRevLett.115.163603}%
  \BibitemOpen
  \bibfield  {author} {\bibinfo {author} {\bibfnamefont {A.}~\bibnamefont
  {Gonz\'alez-Tudela}}, \bibinfo {author} {\bibfnamefont {V.}~\bibnamefont
  {Paulisch}}, \bibinfo {author} {\bibfnamefont {D.~E.}\ \bibnamefont {Chang}},
  \bibinfo {author} {\bibfnamefont {H.~J.}\ \bibnamefont {Kimble}},\ and\
  \bibinfo {author} {\bibfnamefont {J.~I.}\ \bibnamefont {Cirac}},\ }\bibfield
  {title} {\bibinfo {title} {Deterministic generation of arbitrary photonic
  states assisted by dissipation},\ }\href
  {https://doi.org/10.1103/PhysRevLett.115.163603} {\bibfield  {journal}
  {\bibinfo  {journal} {Phys. Rev. Lett.}\ }\textbf {\bibinfo {volume} {115}},\
  \bibinfo {pages} {163603} (\bibinfo {year} {2015})}\BibitemShut {NoStop}%
\bibitem [{\citenamefont {Laucht}\ \emph {et~al.}(2012)\citenamefont {Laucht},
  \citenamefont {P\"utz}, \citenamefont {G\"unthner}, \citenamefont {Hauke},
  \citenamefont {Saive}, \citenamefont {Fr\'ed\'erick}, \citenamefont
  {Bichler}, \citenamefont {Amann}, \citenamefont {Holleitner}, \citenamefont
  {Kaniber},\ and\ \citenamefont {Finley}}]{PhysRevX.2.011014}%
  \BibitemOpen
  \bibfield  {author} {\bibinfo {author} {\bibfnamefont {A.}~\bibnamefont
  {Laucht}}, \bibinfo {author} {\bibfnamefont {S.}~\bibnamefont {P\"utz}},
  \bibinfo {author} {\bibfnamefont {T.}~\bibnamefont {G\"unthner}}, \bibinfo
  {author} {\bibfnamefont {N.}~\bibnamefont {Hauke}}, \bibinfo {author}
  {\bibfnamefont {R.}~\bibnamefont {Saive}}, \bibinfo {author} {\bibfnamefont
  {S.}~\bibnamefont {Fr\'ed\'erick}}, \bibinfo {author} {\bibfnamefont
  {M.}~\bibnamefont {Bichler}}, \bibinfo {author} {\bibfnamefont {M.-C.}\
  \bibnamefont {Amann}}, \bibinfo {author} {\bibfnamefont {A.~W.}\ \bibnamefont
  {Holleitner}}, \bibinfo {author} {\bibfnamefont {M.}~\bibnamefont
  {Kaniber}},\ and\ \bibinfo {author} {\bibfnamefont {J.~J.}\ \bibnamefont
  {Finley}},\ }\bibfield  {title} {\bibinfo {title} {A waveguide-coupled
  on-chip single-photon source},\ }\href
  {https://doi.org/10.1103/PhysRevX.2.011014} {\bibfield  {journal} {\bibinfo
  {journal} {Phys. Rev. X}\ }\textbf {\bibinfo {volume} {2}},\ \bibinfo {pages}
  {011014} (\bibinfo {year} {2012})}\BibitemShut {NoStop}%
\bibitem [{\citenamefont {Nie}\ \emph {et~al.}(2023)\citenamefont {Nie},
  \citenamefont {Shi}, \citenamefont {Liu},\ and\ \citenamefont
  {Nori}}]{PhysRevLett.131.103602}%
  \BibitemOpen
  \bibfield  {author} {\bibinfo {author} {\bibfnamefont {W.}~\bibnamefont
  {Nie}}, \bibinfo {author} {\bibfnamefont {T.}~\bibnamefont {Shi}}, \bibinfo
  {author} {\bibfnamefont {Y.-x.}\ \bibnamefont {Liu}},\ and\ \bibinfo {author}
  {\bibfnamefont {F.}~\bibnamefont {Nori}},\ }\bibfield  {title} {\bibinfo
  {title} {{Non-Hermitian} waveguide cavity {QED} with tunable atomic
  mirrors},\ }\href {https://doi.org/10.1103/PhysRevLett.131.103602} {\bibfield
   {journal} {\bibinfo  {journal} {Phys. Rev. Lett.}\ }\textbf {\bibinfo
  {volume} {131}},\ \bibinfo {pages} {103602} (\bibinfo {year}
  {2023})}\BibitemShut {NoStop}%
\bibitem [{\citenamefont {Li}\ and\ \citenamefont
  {Wei}(2015)}]{PhysRevA.92.063836}%
  \BibitemOpen
  \bibfield  {author} {\bibinfo {author} {\bibfnamefont {X.}~\bibnamefont
  {Li}}\ and\ \bibinfo {author} {\bibfnamefont {L.~F.}\ \bibnamefont {Wei}},\
  }\bibfield  {title} {\bibinfo {title} {Designable single-photon quantum
  routings with atomic mirrors},\ }\href
  {https://doi.org/10.1103/PhysRevA.92.063836} {\bibfield  {journal} {\bibinfo
  {journal} {Phys. Rev. A}\ }\textbf {\bibinfo {volume} {92}},\ \bibinfo
  {pages} {063836} (\bibinfo {year} {2015})}\BibitemShut {NoStop}%
\bibitem [{\citenamefont {Asenjo-Garcia}\ \emph
  {et~al.}(2017{\natexlab{a}})\citenamefont {Asenjo-Garcia}, \citenamefont
  {Moreno-Cardoner}, \citenamefont {Albrecht}, \citenamefont {Kimble},\ and\
  \citenamefont {Chang}}]{PhysRevX.7.031024}%
  \BibitemOpen
  \bibfield  {author} {\bibinfo {author} {\bibfnamefont {A.}~\bibnamefont
  {Asenjo-Garcia}}, \bibinfo {author} {\bibfnamefont {M.}~\bibnamefont
  {Moreno-Cardoner}}, \bibinfo {author} {\bibfnamefont {A.}~\bibnamefont
  {Albrecht}}, \bibinfo {author} {\bibfnamefont {H.~J.}\ \bibnamefont
  {Kimble}},\ and\ \bibinfo {author} {\bibfnamefont {D.~E.}\ \bibnamefont
  {Chang}},\ }\bibfield  {title} {\bibinfo {title} {Exponential improvement in
  photon storage fidelities using subradiance and ``selective radiance'' in
  atomic arrays},\ }\href {https://doi.org/10.1103/PhysRevX.7.031024}
  {\bibfield  {journal} {\bibinfo  {journal} {Phys. Rev. X}\ }\textbf {\bibinfo
  {volume} {7}},\ \bibinfo {pages} {031024} (\bibinfo {year}
  {2017}{\natexlab{a}})}\BibitemShut {NoStop}%
\bibitem [{\citenamefont {Kockum}\ \emph {et~al.}(2018)\citenamefont {Kockum},
  \citenamefont {Johansson},\ and\ \citenamefont
  {Nori}}]{PhysRevLett.120.140404}%
  \BibitemOpen
  \bibfield  {author} {\bibinfo {author} {\bibfnamefont {A.~F.}\ \bibnamefont
  {Kockum}}, \bibinfo {author} {\bibfnamefont {G.}~\bibnamefont {Johansson}},\
  and\ \bibinfo {author} {\bibfnamefont {F.}~\bibnamefont {Nori}},\ }\bibfield
  {title} {\bibinfo {title} {Decoherence-free interaction between giant atoms
  in waveguide quantum electrodynamics},\ }\href
  {https://doi.org/10.1103/PhysRevLett.120.140404} {\bibfield  {journal}
  {\bibinfo  {journal} {Phys. Rev. Lett.}\ }\textbf {\bibinfo {volume} {120}},\
  \bibinfo {pages} {140404} (\bibinfo {year} {2018})}\BibitemShut {NoStop}%
\bibitem [{\citenamefont {Chu}\ \emph {et~al.}(2023)\citenamefont {Chu},
  \citenamefont {Papon}, \citenamefont {Bart}, \citenamefont {Wieck},
  \citenamefont {Ludwig}, \citenamefont {Midolo}, \citenamefont {Rotenberg},\
  and\ \citenamefont {Lodahl}}]{PhysRevLett.131.033606}%
  \BibitemOpen
  \bibfield  {author} {\bibinfo {author} {\bibfnamefont {X.-L.}\ \bibnamefont
  {Chu}}, \bibinfo {author} {\bibfnamefont {C.}~\bibnamefont {Papon}}, \bibinfo
  {author} {\bibfnamefont {N.}~\bibnamefont {Bart}}, \bibinfo {author}
  {\bibfnamefont {A.~D.}\ \bibnamefont {Wieck}}, \bibinfo {author}
  {\bibfnamefont {A.}~\bibnamefont {Ludwig}}, \bibinfo {author} {\bibfnamefont
  {L.}~\bibnamefont {Midolo}}, \bibinfo {author} {\bibfnamefont
  {N.}~\bibnamefont {Rotenberg}},\ and\ \bibinfo {author} {\bibfnamefont
  {P.}~\bibnamefont {Lodahl}},\ }\bibfield  {title} {\bibinfo {title}
  {Independent electrical control of two quantum dots coupled through a
  photonic-crystal waveguide},\ }\href
  {https://doi.org/10.1103/PhysRevLett.131.033606} {\bibfield  {journal}
  {\bibinfo  {journal} {Phys. Rev. Lett.}\ }\textbf {\bibinfo {volume} {131}},\
  \bibinfo {pages} {033606} (\bibinfo {year} {2023})}\BibitemShut {NoStop}%
\bibitem [{\citenamefont {Masson}\ and\ \citenamefont
  {Asenjo-Garcia}(2020)}]{PhysRevResearch.2.043213}%
  \BibitemOpen
  \bibfield  {author} {\bibinfo {author} {\bibfnamefont {S.~J.}\ \bibnamefont
  {Masson}}\ and\ \bibinfo {author} {\bibfnamefont {A.}~\bibnamefont
  {Asenjo-Garcia}},\ }\bibfield  {title} {\bibinfo {title} {Atomic-waveguide
  quantum electrodynamics},\ }\href
  {https://doi.org/10.1103/PhysRevResearch.2.043213} {\bibfield  {journal}
  {\bibinfo  {journal} {Phys. Rev. Res.}\ }\textbf {\bibinfo {volume} {2}},\
  \bibinfo {pages} {043213} (\bibinfo {year} {2020})}\BibitemShut {NoStop}%
\bibitem [{\citenamefont {Kannan}\ \emph {et~al.}(2020)\citenamefont {Kannan},
  \citenamefont {Ruckriegel}, \citenamefont {Campbell}, \citenamefont
  {Frisk~Kockum}, \citenamefont {Braum\"{u}ller}, \citenamefont {Kim},
  \citenamefont {Kjaergaard}, \citenamefont {Krantz}, \citenamefont {Melville},
  \citenamefont {Niedzielski}, \citenamefont {Veps\"{a}l\"{a}inen},
  \citenamefont {Winik}, \citenamefont {Yoder}, \citenamefont {Nori},
  \citenamefont {Orlando}, \citenamefont {Gustavsson},\ and\ \citenamefont
  {Oliver}}]{Kannan2020}%
  \BibitemOpen
  \bibfield  {author} {\bibinfo {author} {\bibfnamefont {B.}~\bibnamefont
  {Kannan}}, \bibinfo {author} {\bibfnamefont {M.~J.}\ \bibnamefont
  {Ruckriegel}}, \bibinfo {author} {\bibfnamefont {D.~L.}\ \bibnamefont
  {Campbell}}, \bibinfo {author} {\bibfnamefont {A.}~\bibnamefont
  {Frisk~Kockum}}, \bibinfo {author} {\bibfnamefont {J.}~\bibnamefont
  {Braum\"{u}ller}}, \bibinfo {author} {\bibfnamefont {D.~K.}\ \bibnamefont
  {Kim}}, \bibinfo {author} {\bibfnamefont {M.}~\bibnamefont {Kjaergaard}},
  \bibinfo {author} {\bibfnamefont {P.}~\bibnamefont {Krantz}}, \bibinfo
  {author} {\bibfnamefont {A.}~\bibnamefont {Melville}}, \bibinfo {author}
  {\bibfnamefont {B.~M.}\ \bibnamefont {Niedzielski}}, \bibinfo {author}
  {\bibfnamefont {A.}~\bibnamefont {Veps\"{a}l\"{a}inen}}, \bibinfo {author}
  {\bibfnamefont {R.}~\bibnamefont {Winik}}, \bibinfo {author} {\bibfnamefont
  {J.~L.}\ \bibnamefont {Yoder}}, \bibinfo {author} {\bibfnamefont
  {F.}~\bibnamefont {Nori}}, \bibinfo {author} {\bibfnamefont {T.~P.}\
  \bibnamefont {Orlando}}, \bibinfo {author} {\bibfnamefont {S.}~\bibnamefont
  {Gustavsson}},\ and\ \bibinfo {author} {\bibfnamefont {W.~D.}\ \bibnamefont
  {Oliver}},\ }\bibfield  {title} {\bibinfo {title} {Waveguide quantum
  electrodynamics with superconducting artificial giant atoms},\ }\href
  {https://doi.org/10.1038/s41586-020-2529-9} {\bibfield  {journal} {\bibinfo
  {journal} {Nature}\ }\textbf {\bibinfo {volume} {583}},\ \bibinfo {pages}
  {775–779} (\bibinfo {year} {2020})}\BibitemShut {NoStop}%
\bibitem [{\citenamefont {Alushi}\ \emph {et~al.}(2023)\citenamefont {Alushi},
  \citenamefont {Ramos}, \citenamefont {Garc\'{\i}a-Ripoll}, \citenamefont
  {Di~Candia},\ and\ \citenamefont {Felicetti}}]{PRXQuantum.4.030326}%
  \BibitemOpen
  \bibfield  {author} {\bibinfo {author} {\bibfnamefont {U.}~\bibnamefont
  {Alushi}}, \bibinfo {author} {\bibfnamefont {T.}~\bibnamefont {Ramos}},
  \bibinfo {author} {\bibfnamefont {J.~J.}\ \bibnamefont {Garc\'{\i}a-Ripoll}},
  \bibinfo {author} {\bibfnamefont {R.}~\bibnamefont {Di~Candia}},\ and\
  \bibinfo {author} {\bibfnamefont {S.}~\bibnamefont {Felicetti}},\ }\bibfield
  {title} {\bibinfo {title} {Waveguide {QED} with quadratic light-matter
  interactions},\ }\href {https://doi.org/10.1103/PRXQuantum.4.030326}
  {\bibfield  {journal} {\bibinfo  {journal} {PRX Quantum}\ }\textbf {\bibinfo
  {volume} {4}},\ \bibinfo {pages} {030326} (\bibinfo {year}
  {2023})}\BibitemShut {NoStop}%
\bibitem [{\citenamefont {Wang}\ \emph {et~al.}(2024)\citenamefont {Wang},
  \citenamefont {Zhu}, \citenamefont {Liu},\ and\ \citenamefont
  {Nori}}]{PhysRevResearch.6.013279}%
  \BibitemOpen
  \bibfield  {author} {\bibinfo {author} {\bibfnamefont {X.}~\bibnamefont
  {Wang}}, \bibinfo {author} {\bibfnamefont {H.-B.}\ \bibnamefont {Zhu}},
  \bibinfo {author} {\bibfnamefont {T.}~\bibnamefont {Liu}},\ and\ \bibinfo
  {author} {\bibfnamefont {F.}~\bibnamefont {Nori}},\ }\bibfield  {title}
  {\bibinfo {title} {Realizing quantum optics in structured environments with
  giant atoms},\ }\href {https://doi.org/10.1103/PhysRevResearch.6.013279}
  {\bibfield  {journal} {\bibinfo  {journal} {Phys. Rev. Res.}\ }\textbf
  {\bibinfo {volume} {6}},\ \bibinfo {pages} {013279} (\bibinfo {year}
  {2024})}\BibitemShut {NoStop}%
\bibitem [{\citenamefont {Wang}\ and\ \citenamefont {Li}(2022)}]{Wang_2022}%
  \BibitemOpen
  \bibfield  {author} {\bibinfo {author} {\bibfnamefont {X.}~\bibnamefont
  {Wang}}\ and\ \bibinfo {author} {\bibfnamefont {H.-R.}\ \bibnamefont {Li}},\
  }\bibfield  {title} {\bibinfo {title} {Chiral quantum network with giant
  atoms},\ }\href {https://doi.org/10.1088/2058-9565/ac6a04} {\bibfield
  {journal} {\bibinfo  {journal} {Quantum Science and Technology}\ }\textbf
  {\bibinfo {volume} {7}},\ \bibinfo {pages} {035007} (\bibinfo {year}
  {2022})}\BibitemShut {NoStop}%
\bibitem [{\citenamefont {Manga~Rao}\ and\ \citenamefont
  {Hughes}(2007)}]{PhysRevB.75.205437}%
  \BibitemOpen
  \bibfield  {author} {\bibinfo {author} {\bibfnamefont {V.~S.~C.}\
  \bibnamefont {Manga~Rao}}\ and\ \bibinfo {author} {\bibfnamefont
  {S.}~\bibnamefont {Hughes}},\ }\bibfield  {title} {\bibinfo {title} {Single
  quantum-dot {Purcell} factor and $\ensuremath{\beta}$ factor in a photonic
  crystal waveguide},\ }\href {https://doi.org/10.1103/PhysRevB.75.205437}
  {\bibfield  {journal} {\bibinfo  {journal} {Phys. Rev. B}\ }\textbf {\bibinfo
  {volume} {75}},\ \bibinfo {pages} {205437} (\bibinfo {year}
  {2007})}\BibitemShut {NoStop}%
\bibitem [{\citenamefont {Arcari}\ \emph {et~al.}(2014)\citenamefont {Arcari},
  \citenamefont {S\"ollner}, \citenamefont {Javadi}, \citenamefont
  {Lindskov~Hansen}, \citenamefont {Mahmoodian}, \citenamefont {Liu},
  \citenamefont {Thyrrestrup}, \citenamefont {Lee}, \citenamefont {Song},
  \citenamefont {Stobbe},\ and\ \citenamefont
  {Lodahl}}]{PhysRevLett.113.093603}%
  \BibitemOpen
  \bibfield  {author} {\bibinfo {author} {\bibfnamefont {M.}~\bibnamefont
  {Arcari}}, \bibinfo {author} {\bibfnamefont {I.}~\bibnamefont {S\"ollner}},
  \bibinfo {author} {\bibfnamefont {A.}~\bibnamefont {Javadi}}, \bibinfo
  {author} {\bibfnamefont {S.}~\bibnamefont {Lindskov~Hansen}}, \bibinfo
  {author} {\bibfnamefont {S.}~\bibnamefont {Mahmoodian}}, \bibinfo {author}
  {\bibfnamefont {J.}~\bibnamefont {Liu}}, \bibinfo {author} {\bibfnamefont
  {H.}~\bibnamefont {Thyrrestrup}}, \bibinfo {author} {\bibfnamefont {E.~H.}\
  \bibnamefont {Lee}}, \bibinfo {author} {\bibfnamefont {J.~D.}\ \bibnamefont
  {Song}}, \bibinfo {author} {\bibfnamefont {S.}~\bibnamefont {Stobbe}},\ and\
  \bibinfo {author} {\bibfnamefont {P.}~\bibnamefont {Lodahl}},\ }\bibfield
  {title} {\bibinfo {title} {Near-unity coupling efficiency of a quantum
  emitter to a photonic crystal waveguide},\ }\href
  {https://doi.org/10.1103/PhysRevLett.113.093603} {\bibfield  {journal}
  {\bibinfo  {journal} {Phys. Rev. Lett.}\ }\textbf {\bibinfo {volume} {113}},\
  \bibinfo {pages} {093603} (\bibinfo {year} {2014})}\BibitemShut {NoStop}%
\bibitem [{\citenamefont {Paesani}\ \emph {et~al.}(2019)\citenamefont
  {Paesani}, \citenamefont {Ding}, \citenamefont {Santagati}, \citenamefont
  {Chakhmakhchyan}, \citenamefont {Vigliar}, \citenamefont {Rottwitt},
  \citenamefont {Oxenløwe}, \citenamefont {Wang}, \citenamefont {Thompson},\
  and\ \citenamefont {Laing}}]{Paesani2019}%
  \BibitemOpen
  \bibfield  {author} {\bibinfo {author} {\bibfnamefont {S.}~\bibnamefont
  {Paesani}}, \bibinfo {author} {\bibfnamefont {Y.}~\bibnamefont {Ding}},
  \bibinfo {author} {\bibfnamefont {R.}~\bibnamefont {Santagati}}, \bibinfo
  {author} {\bibfnamefont {L.}~\bibnamefont {Chakhmakhchyan}}, \bibinfo
  {author} {\bibfnamefont {C.}~\bibnamefont {Vigliar}}, \bibinfo {author}
  {\bibfnamefont {K.}~\bibnamefont {Rottwitt}}, \bibinfo {author}
  {\bibfnamefont {L.~K.}\ \bibnamefont {Oxenløwe}}, \bibinfo {author}
  {\bibfnamefont {J.}~\bibnamefont {Wang}}, \bibinfo {author} {\bibfnamefont
  {M.~G.}\ \bibnamefont {Thompson}},\ and\ \bibinfo {author} {\bibfnamefont
  {A.}~\bibnamefont {Laing}},\ }\bibfield  {title} {\bibinfo {title}
  {Generation and sampling of quantum states of light in a silicon chip},\
  }\href {https://doi.org/10.1038/s41567-019-0567-8} {\bibfield  {journal}
  {\bibinfo  {journal} {Nature Physics}\ }\textbf {\bibinfo {volume} {15}},\
  \bibinfo {pages} {925–929} (\bibinfo {year} {2019})}\BibitemShut {NoStop}%
\bibitem [{\citenamefont {le~Feber}\ \emph {et~al.}(2015)\citenamefont
  {le~Feber}, \citenamefont {Rotenberg},\ and\ \citenamefont
  {Kuipers}}]{leFeber2015}%
  \BibitemOpen
  \bibfield  {author} {\bibinfo {author} {\bibfnamefont {B.}~\bibnamefont
  {le~Feber}}, \bibinfo {author} {\bibfnamefont {N.}~\bibnamefont
  {Rotenberg}},\ and\ \bibinfo {author} {\bibfnamefont {L.}~\bibnamefont
  {Kuipers}},\ }\bibfield  {title} {\bibinfo {title} {Nanophotonic control of
  circular dipole emission},\ }\bibfield  {journal} {\bibinfo  {journal}
  {Nature Communications}\ }\textbf {\bibinfo {volume} {6}},\ \href
  {https://doi.org/10.1038/ncomms7695} {10.1038/ncomms7695} (\bibinfo {year}
  {2015})\BibitemShut {NoStop}%
\bibitem [{\citenamefont {Young}\ \emph {et~al.}(2015)\citenamefont {Young},
  \citenamefont {Thijssen}, \citenamefont {Beggs}, \citenamefont
  {Androvitsaneas}, \citenamefont {Kuipers}, \citenamefont {Rarity},
  \citenamefont {Hughes},\ and\ \citenamefont
  {Oulton}}]{PhysRevLett.115.153901}%
  \BibitemOpen
  \bibfield  {author} {\bibinfo {author} {\bibfnamefont {A.~B.}\ \bibnamefont
  {Young}}, \bibinfo {author} {\bibfnamefont {A.~C.~T.}\ \bibnamefont
  {Thijssen}}, \bibinfo {author} {\bibfnamefont {D.~M.}\ \bibnamefont {Beggs}},
  \bibinfo {author} {\bibfnamefont {P.}~\bibnamefont {Androvitsaneas}},
  \bibinfo {author} {\bibfnamefont {L.}~\bibnamefont {Kuipers}}, \bibinfo
  {author} {\bibfnamefont {J.~G.}\ \bibnamefont {Rarity}}, \bibinfo {author}
  {\bibfnamefont {S.}~\bibnamefont {Hughes}},\ and\ \bibinfo {author}
  {\bibfnamefont {R.}~\bibnamefont {Oulton}},\ }\bibfield  {title} {\bibinfo
  {title} {Polarization engineering in photonic crystal waveguides for
  spin-photon entanglers},\ }\href
  {https://doi.org/10.1103/PhysRevLett.115.153901} {\bibfield  {journal}
  {\bibinfo  {journal} {Phys. Rev. Lett.}\ }\textbf {\bibinfo {volume} {115}},\
  \bibinfo {pages} {153901} (\bibinfo {year} {2015})}\BibitemShut {NoStop}%
\bibitem [{\citenamefont {S\"{o}llner}\ \emph {et~al.}(2015)\citenamefont
  {S\"{o}llner}, \citenamefont {Mahmoodian}, \citenamefont {Hansen},
  \citenamefont {Midolo}, \citenamefont {Javadi}, \citenamefont
  {Kir{\v{s}}ansk{\.{e}}}, \citenamefont {Pregnolato}, \citenamefont {El-Ella},
  \citenamefont {Lee}, \citenamefont {Song}, \citenamefont {Stobbe},\ and\
  \citenamefont {Lodahl}}]{Sllner2015}%
  \BibitemOpen
  \bibfield  {author} {\bibinfo {author} {\bibfnamefont {I.}~\bibnamefont
  {S\"{o}llner}}, \bibinfo {author} {\bibfnamefont {S.}~\bibnamefont
  {Mahmoodian}}, \bibinfo {author} {\bibfnamefont {S.~L.}\ \bibnamefont
  {Hansen}}, \bibinfo {author} {\bibfnamefont {L.}~\bibnamefont {Midolo}},
  \bibinfo {author} {\bibfnamefont {A.}~\bibnamefont {Javadi}}, \bibinfo
  {author} {\bibfnamefont {G.}~\bibnamefont {Kir{\v{s}}ansk{\.{e}}}}, \bibinfo
  {author} {\bibfnamefont {T.}~\bibnamefont {Pregnolato}}, \bibinfo {author}
  {\bibfnamefont {H.}~\bibnamefont {El-Ella}}, \bibinfo {author} {\bibfnamefont
  {E.~H.}\ \bibnamefont {Lee}}, \bibinfo {author} {\bibfnamefont {J.~D.}\
  \bibnamefont {Song}}, \bibinfo {author} {\bibfnamefont {S.}~\bibnamefont
  {Stobbe}},\ and\ \bibinfo {author} {\bibfnamefont {P.}~\bibnamefont
  {Lodahl}},\ }\bibfield  {title} {\bibinfo {title} {Deterministic
  photon{\textendash}emitter coupling in chiral photonic circuits},\ }\href
  {https://doi.org/10.1038/nnano.2015.159} {\bibfield  {journal} {\bibinfo
  {journal} {Nature Nanotechnology}\ }\textbf {\bibinfo {volume} {10}},\
  \bibinfo {pages} {775} (\bibinfo {year} {2015})}\BibitemShut {NoStop}%
\bibitem [{\citenamefont {Bello}\ \emph {et~al.}(2019)\citenamefont {Bello},
  \citenamefont {Platero}, \citenamefont {Cirac},\ and\ \citenamefont
  {González-Tudela}}]{doi:10.1126/sciadv.aaw0297}%
  \BibitemOpen
  \bibfield  {author} {\bibinfo {author} {\bibfnamefont {M.}~\bibnamefont
  {Bello}}, \bibinfo {author} {\bibfnamefont {G.}~\bibnamefont {Platero}},
  \bibinfo {author} {\bibfnamefont {J.~I.}\ \bibnamefont {Cirac}},\ and\
  \bibinfo {author} {\bibfnamefont {A.}~\bibnamefont {González-Tudela}},\
  }\bibfield  {title} {\bibinfo {title} {Unconventional quantum optics in
  topological waveguide {QED}},\ }\href
  {https://doi.org/10.1126/sciadv.aaw0297} {\bibfield  {journal} {\bibinfo
  {journal} {Science Advances}\ }\textbf {\bibinfo {volume} {5}},\ \bibinfo
  {pages} {eaaw0297} (\bibinfo {year} {2019})}\BibitemShut {NoStop}%
\bibitem [{\citenamefont {Hauff}\ \emph {et~al.}(2022)\citenamefont {Hauff},
  \citenamefont {Le~Jeannic}, \citenamefont {Lodahl}, \citenamefont {Hughes},\
  and\ \citenamefont {Rotenberg}}]{PhysRevResearch.4.023082}%
  \BibitemOpen
  \bibfield  {author} {\bibinfo {author} {\bibfnamefont {N.~V.}\ \bibnamefont
  {Hauff}}, \bibinfo {author} {\bibfnamefont {H.}~\bibnamefont {Le~Jeannic}},
  \bibinfo {author} {\bibfnamefont {P.}~\bibnamefont {Lodahl}}, \bibinfo
  {author} {\bibfnamefont {S.}~\bibnamefont {Hughes}},\ and\ \bibinfo {author}
  {\bibfnamefont {N.}~\bibnamefont {Rotenberg}},\ }\bibfield  {title} {\bibinfo
  {title} {Chiral quantum optics in broken-symmetry and topological photonic
  crystal waveguides},\ }\href
  {https://doi.org/10.1103/PhysRevResearch.4.023082} {\bibfield  {journal}
  {\bibinfo  {journal} {Phys. Rev. Res.}\ }\textbf {\bibinfo {volume} {4}},\
  \bibinfo {pages} {023082} (\bibinfo {year} {2022})}\BibitemShut {NoStop}%
\bibitem [{\citenamefont {Liu}\ \emph {et~al.}(2022)\citenamefont {Liu},
  \citenamefont {Wang}, \citenamefont {Wang}, \citenamefont {Ma},\ and\
  \citenamefont {Cheng}}]{Liu:22}%
  \BibitemOpen
  \bibfield  {author} {\bibinfo {author} {\bibfnamefont {N.}~\bibnamefont
  {Liu}}, \bibinfo {author} {\bibfnamefont {X.}~\bibnamefont {Wang}}, \bibinfo
  {author} {\bibfnamefont {X.}~\bibnamefont {Wang}}, \bibinfo {author}
  {\bibfnamefont {X.-S.}\ \bibnamefont {Ma}},\ and\ \bibinfo {author}
  {\bibfnamefont {M.-T.}\ \bibnamefont {Cheng}},\ }\bibfield  {title} {\bibinfo
  {title} {Tunable single photon nonreciprocal scattering based on giant
  atom-waveguide chiral couplings},\ }\href {https://doi.org/10.1364/OE.460255}
  {\bibfield  {journal} {\bibinfo  {journal} {Opt. Express}\ }\textbf {\bibinfo
  {volume} {30}},\ \bibinfo {pages} {23428} (\bibinfo {year}
  {2022})}\BibitemShut {NoStop}%
\bibitem [{\citenamefont {Mahmoodian}\ \emph {et~al.}(2016)\citenamefont
  {Mahmoodian}, \citenamefont {Lodahl},\ and\ \citenamefont
  {S\o{}rensen}}]{PhysRevLett.117.240501}%
  \BibitemOpen
  \bibfield  {author} {\bibinfo {author} {\bibfnamefont {S.}~\bibnamefont
  {Mahmoodian}}, \bibinfo {author} {\bibfnamefont {P.}~\bibnamefont {Lodahl}},\
  and\ \bibinfo {author} {\bibfnamefont {A.~S.}\ \bibnamefont {S\o{}rensen}},\
  }\bibfield  {title} {\bibinfo {title} {Quantum networks with
  chiral-light--matter interaction in waveguides},\ }\href
  {https://doi.org/10.1103/PhysRevLett.117.240501} {\bibfield  {journal}
  {\bibinfo  {journal} {Phys. Rev. Lett.}\ }\textbf {\bibinfo {volume} {117}},\
  \bibinfo {pages} {240501} (\bibinfo {year} {2016})}\BibitemShut {NoStop}%
\bibitem [{\citenamefont {Mahmoodian}\ \emph {et~al.}(2020)\citenamefont
  {Mahmoodian}, \citenamefont {Calaj\'o}, \citenamefont {Chang}, \citenamefont
  {Hammerer},\ and\ \citenamefont {S\o{}rensen}}]{PhysRevX.10.031011}%
  \BibitemOpen
  \bibfield  {author} {\bibinfo {author} {\bibfnamefont {S.}~\bibnamefont
  {Mahmoodian}}, \bibinfo {author} {\bibfnamefont {G.}~\bibnamefont
  {Calaj\'o}}, \bibinfo {author} {\bibfnamefont {D.~E.}\ \bibnamefont {Chang}},
  \bibinfo {author} {\bibfnamefont {K.}~\bibnamefont {Hammerer}},\ and\
  \bibinfo {author} {\bibfnamefont {A.~S.}\ \bibnamefont {S\o{}rensen}},\
  }\bibfield  {title} {\bibinfo {title} {Dynamics of many-body photon bound
  states in chiral waveguide {QED}},\ }\href
  {https://doi.org/10.1103/PhysRevX.10.031011} {\bibfield  {journal} {\bibinfo
  {journal} {Phys. Rev. X}\ }\textbf {\bibinfo {volume} {10}},\ \bibinfo
  {pages} {031011} (\bibinfo {year} {2020})}\BibitemShut {NoStop}%
\bibitem [{\citenamefont {Asenjo-Garcia}\ \emph
  {et~al.}(2017{\natexlab{b}})\citenamefont {Asenjo-Garcia}, \citenamefont
  {Hood}, \citenamefont {Chang},\ and\ \citenamefont
  {Kimble}}]{PhysRevA.95.033818}%
  \BibitemOpen
  \bibfield  {author} {\bibinfo {author} {\bibfnamefont {A.}~\bibnamefont
  {Asenjo-Garcia}}, \bibinfo {author} {\bibfnamefont {J.~D.}\ \bibnamefont
  {Hood}}, \bibinfo {author} {\bibfnamefont {D.~E.}\ \bibnamefont {Chang}},\
  and\ \bibinfo {author} {\bibfnamefont {H.~J.}\ \bibnamefont {Kimble}},\
  }\bibfield  {title} {\bibinfo {title} {Atom-light interactions in
  quasi-one-dimensional nanostructures: A green's-function perspective},\
  }\href {https://doi.org/10.1103/PhysRevA.95.033818} {\bibfield  {journal}
  {\bibinfo  {journal} {Phys. Rev. A}\ }\textbf {\bibinfo {volume} {95}},\
  \bibinfo {pages} {033818} (\bibinfo {year} {2017}{\natexlab{b}})}\BibitemShut
  {NoStop}%
\bibitem [{\citenamefont {Frisk~Kockum}\ \emph {et~al.}(2019)\citenamefont
  {Frisk~Kockum}, \citenamefont {Miranowicz}, \citenamefont {De~Liberato},
  \citenamefont {Savasta},\ and\ \citenamefont
  {Nori}}]{frisk_kockum_ultrastrong_2019}%
  \BibitemOpen
  \bibfield  {author} {\bibinfo {author} {\bibfnamefont {A.}~\bibnamefont
  {Frisk~Kockum}}, \bibinfo {author} {\bibfnamefont {A.}~\bibnamefont
  {Miranowicz}}, \bibinfo {author} {\bibfnamefont {S.}~\bibnamefont
  {De~Liberato}}, \bibinfo {author} {\bibfnamefont {S.}~\bibnamefont
  {Savasta}},\ and\ \bibinfo {author} {\bibfnamefont {F.}~\bibnamefont
  {Nori}},\ }\bibfield  {title} {\bibinfo {title} {Ultrastrong coupling between
  light and matter},\ }\href {https://doi.org/10.1038/s42254-018-0006-2}
  {\bibfield  {journal} {\bibinfo  {journal} {Nature Reviews Physics}\ }\textbf
  {\bibinfo {volume} {1}},\ \bibinfo {pages} {19} (\bibinfo {year}
  {2019})}\BibitemShut {NoStop}%
\bibitem [{\citenamefont {Forn-D{\'i}az}\ \emph {et~al.}(2019)\citenamefont
  {Forn-D{\'i}az}, \citenamefont {Lamata}, \citenamefont {Rico}, \citenamefont
  {Kono},\ and\ \citenamefont {Solano}}]{forn-diaz_ultrastrong_2019}%
  \BibitemOpen
  \bibfield  {author} {\bibinfo {author} {\bibfnamefont {P.}~\bibnamefont
  {Forn-D{\'i}az}}, \bibinfo {author} {\bibfnamefont {L.}~\bibnamefont
  {Lamata}}, \bibinfo {author} {\bibfnamefont {E.}~\bibnamefont {Rico}},
  \bibinfo {author} {\bibfnamefont {J.}~\bibnamefont {Kono}},\ and\ \bibinfo
  {author} {\bibfnamefont {E.}~\bibnamefont {Solano}},\ }\bibfield  {title}
  {\bibinfo {title} {Ultrastrong coupling regimes of light-matter
  interaction},\ }\href {https://doi.org/10.1103/RevModPhys.91.025005}
  {\bibfield  {journal} {\bibinfo  {journal} {Reviews of Modern Physics}\
  }\textbf {\bibinfo {volume} {91}},\ \bibinfo {pages} {025005} (\bibinfo
  {year} {2019})}\BibitemShut {NoStop}%
\bibitem [{\citenamefont {Peropadre}\ \emph {et~al.}(2013)\citenamefont
  {Peropadre}, \citenamefont {Zueco}, \citenamefont {Porras},\ and\
  \citenamefont {Garc\'{\i}a-Ripoll}}]{PhysRevLett.111.243602}%
  \BibitemOpen
  \bibfield  {author} {\bibinfo {author} {\bibfnamefont {B.}~\bibnamefont
  {Peropadre}}, \bibinfo {author} {\bibfnamefont {D.}~\bibnamefont {Zueco}},
  \bibinfo {author} {\bibfnamefont {D.}~\bibnamefont {Porras}},\ and\ \bibinfo
  {author} {\bibfnamefont {J.~J.}\ \bibnamefont {Garc\'{\i}a-Ripoll}},\
  }\bibfield  {title} {\bibinfo {title} {Nonequilibrium and nonperturbative
  dynamics of ultrastrong coupling in open lines},\ }\href
  {https://doi.org/10.1103/PhysRevLett.111.243602} {\bibfield  {journal}
  {\bibinfo  {journal} {Phys. Rev. Lett.}\ }\textbf {\bibinfo {volume} {111}},\
  \bibinfo {pages} {243602} (\bibinfo {year} {2013})}\BibitemShut {NoStop}%
\bibitem [{\citenamefont {Gheeraert}\ \emph {et~al.}(2018)\citenamefont
  {Gheeraert}, \citenamefont {Zhang}, \citenamefont {S\'epulcre}, \citenamefont
  {Bera}, \citenamefont {Roch}, \citenamefont {Baranger},\ and\ \citenamefont
  {Florens}}]{PhysRevA.98.043816}%
  \BibitemOpen
  \bibfield  {author} {\bibinfo {author} {\bibfnamefont {N.}~\bibnamefont
  {Gheeraert}}, \bibinfo {author} {\bibfnamefont {X.~H.~H.}\ \bibnamefont
  {Zhang}}, \bibinfo {author} {\bibfnamefont {T.}~\bibnamefont {S\'epulcre}},
  \bibinfo {author} {\bibfnamefont {S.}~\bibnamefont {Bera}}, \bibinfo {author}
  {\bibfnamefont {N.}~\bibnamefont {Roch}}, \bibinfo {author} {\bibfnamefont
  {H.~U.}\ \bibnamefont {Baranger}},\ and\ \bibinfo {author} {\bibfnamefont
  {S.}~\bibnamefont {Florens}},\ }\bibfield  {title} {\bibinfo {title}
  {Particle production in ultrastrong-coupling waveguide {QED}},\ }\href
  {https://doi.org/10.1103/PhysRevA.98.043816} {\bibfield  {journal} {\bibinfo
  {journal} {Phys. Rev. A}\ }\textbf {\bibinfo {volume} {98}},\ \bibinfo
  {pages} {043816} (\bibinfo {year} {2018})}\BibitemShut {NoStop}%
\bibitem [{\citenamefont {Sanchez-Burillo}\ \emph {et~al.}(2014)\citenamefont
  {Sanchez-Burillo}, \citenamefont {Zueco}, \citenamefont {Garcia-Ripoll},\
  and\ \citenamefont {Martin-Moreno}}]{PhysRevLett.113.263604}%
  \BibitemOpen
  \bibfield  {author} {\bibinfo {author} {\bibfnamefont {E.}~\bibnamefont
  {Sanchez-Burillo}}, \bibinfo {author} {\bibfnamefont {D.}~\bibnamefont
  {Zueco}}, \bibinfo {author} {\bibfnamefont {J.~J.}\ \bibnamefont
  {Garcia-Ripoll}},\ and\ \bibinfo {author} {\bibfnamefont {L.}~\bibnamefont
  {Martin-Moreno}},\ }\bibfield  {title} {\bibinfo {title} {Scattering in the
  ultrastrong regime: Nonlinear optics with one photon},\ }\href
  {https://doi.org/10.1103/PhysRevLett.113.263604} {\bibfield  {journal}
  {\bibinfo  {journal} {Phys. Rev. Lett.}\ }\textbf {\bibinfo {volume} {113}},\
  \bibinfo {pages} {263604} (\bibinfo {year} {2014})}\BibitemShut {NoStop}%
\bibitem [{\citenamefont {Gonz\'alez-Guti\'errez}\ \emph
  {et~al.}(2021)\citenamefont {Gonz\'alez-Guti\'errez}, \citenamefont
  {Rom\'an-Roche},\ and\ \citenamefont {Zueco}}]{PhysRevA.104.053701}%
  \BibitemOpen
  \bibfield  {author} {\bibinfo {author} {\bibfnamefont {C.~A.}\ \bibnamefont
  {Gonz\'alez-Guti\'errez}}, \bibinfo {author} {\bibfnamefont {J.}~\bibnamefont
  {Rom\'an-Roche}},\ and\ \bibinfo {author} {\bibfnamefont {D.}~\bibnamefont
  {Zueco}},\ }\bibfield  {title} {\bibinfo {title} {Distant emitters in
  ultrastrong waveguide {QED}: Ground-state properties and {non-Markovian}
  dynamics},\ }\href {https://doi.org/10.1103/PhysRevA.104.053701} {\bibfield
  {journal} {\bibinfo  {journal} {Phys. Rev. A}\ }\textbf {\bibinfo {volume}
  {104}},\ \bibinfo {pages} {053701} (\bibinfo {year} {2021})}\BibitemShut
  {NoStop}%
\bibitem [{\citenamefont {Chang}\ \emph {et~al.}(2012)\citenamefont {Chang},
  \citenamefont {Jiang}, \citenamefont {Gorshkov},\ and\ \citenamefont
  {Kimble}}]{Chang_2012}%
  \BibitemOpen
  \bibfield  {author} {\bibinfo {author} {\bibfnamefont {D.~E.}\ \bibnamefont
  {Chang}}, \bibinfo {author} {\bibfnamefont {L.}~\bibnamefont {Jiang}},
  \bibinfo {author} {\bibfnamefont {A.~V.}\ \bibnamefont {Gorshkov}},\ and\
  \bibinfo {author} {\bibfnamefont {H.~J.}\ \bibnamefont {Kimble}},\ }\bibfield
   {title} {\bibinfo {title} {Cavity {QED} with atomic mirrors},\ }\href
  {https://doi.org/10.1088/1367-2630/14/6/063003} {\bibfield  {journal}
  {\bibinfo  {journal} {New Journal of Physics}\ }\textbf {\bibinfo {volume}
  {14}},\ \bibinfo {pages} {063003} (\bibinfo {year} {2012})}\BibitemShut
  {NoStop}%
\bibitem [{\citenamefont {Fan}\ \emph {et~al.}(2010)\citenamefont {Fan},
  \citenamefont {Kocaba\mbox{\c{s}}},\ and\ \citenamefont
  {Shen}}]{PhysRevA.82.063821}%
  \BibitemOpen
  \bibfield  {author} {\bibinfo {author} {\bibfnamefont {S.}~\bibnamefont
  {Fan}}, \bibinfo {author} {\bibfnamefont {S.~E.}\ \bibnamefont
  {Kocaba\mbox{\c{s}}}},\ and\ \bibinfo {author} {\bibfnamefont {J.-T.}\
  \bibnamefont {Shen}},\ }\bibfield  {title} {\bibinfo {title} {Input-output
  formalism for few-photon transport in one-dimensional nanophotonic waveguides
  coupled to a qubit},\ }\href {https://doi.org/10.1103/PhysRevA.82.063821}
  {\bibfield  {journal} {\bibinfo  {journal} {Phys. Rev. A}\ }\textbf {\bibinfo
  {volume} {82}},\ \bibinfo {pages} {063821} (\bibinfo {year}
  {2010})}\BibitemShut {NoStop}%
\bibitem [{\citenamefont {Shen}\ and\ \citenamefont
  {Fan}(2007{\natexlab{a}})}]{PhysRevLett.98.153003}%
  \BibitemOpen
  \bibfield  {author} {\bibinfo {author} {\bibfnamefont {J.-T.}\ \bibnamefont
  {Shen}}\ and\ \bibinfo {author} {\bibfnamefont {S.}~\bibnamefont {Fan}},\
  }\bibfield  {title} {\bibinfo {title} {Strongly correlated two-photon
  transport in a one-dimensional waveguide coupled to a two-level system},\
  }\href {https://doi.org/10.1103/PhysRevLett.98.153003} {\bibfield  {journal}
  {\bibinfo  {journal} {Phys. Rev. Lett.}\ }\textbf {\bibinfo {volume} {98}},\
  \bibinfo {pages} {153003} (\bibinfo {year} {2007}{\natexlab{a}})}\BibitemShut
  {NoStop}%
\bibitem [{\citenamefont {Shen}\ and\ \citenamefont
  {Fan}(2007{\natexlab{b}})}]{PhysRevA.76.062709}%
  \BibitemOpen
  \bibfield  {author} {\bibinfo {author} {\bibfnamefont {J.-T.}\ \bibnamefont
  {Shen}}\ and\ \bibinfo {author} {\bibfnamefont {S.}~\bibnamefont {Fan}},\
  }\bibfield  {title} {\bibinfo {title} {Strongly correlated multiparticle
  transport in one dimension through a quantum impurity},\ }\href
  {https://doi.org/10.1103/PhysRevA.76.062709} {\bibfield  {journal} {\bibinfo
  {journal} {Phys. Rev. A}\ }\textbf {\bibinfo {volume} {76}},\ \bibinfo
  {pages} {062709} (\bibinfo {year} {2007}{\natexlab{b}})}\BibitemShut
  {NoStop}%
\bibitem [{\citenamefont {Rephaeli}\ and\ \citenamefont
  {Fan}(2012)}]{Rephaeli2012FewPhotonSC}%
  \BibitemOpen
  \bibfield  {author} {\bibinfo {author} {\bibfnamefont {E.}~\bibnamefont
  {Rephaeli}}\ and\ \bibinfo {author} {\bibfnamefont {S.}~\bibnamefont {Fan}},\
  }\bibfield  {title} {\bibinfo {title} {Few-photon single-atom cavity {QED}
  with input-output formalism in fock space},\ }\href
  {https://api.semanticscholar.org/CorpusID:40913017} {\bibfield  {journal}
  {\bibinfo  {journal} {IEEE Journal of Selected Topics in Quantum
  Electronics}\ }\textbf {\bibinfo {volume} {18}},\ \bibinfo {pages} {1754}
  (\bibinfo {year} {2012})}\BibitemShut {NoStop}%
\bibitem [{\citenamefont {Rephaeli}\ \emph {et~al.}(2010)\citenamefont
  {Rephaeli}, \citenamefont {Shen},\ and\ \citenamefont
  {Fan}}]{PhysRevA.82.033804}%
  \BibitemOpen
  \bibfield  {author} {\bibinfo {author} {\bibfnamefont {E.}~\bibnamefont
  {Rephaeli}}, \bibinfo {author} {\bibfnamefont {J.-T.}\ \bibnamefont {Shen}},\
  and\ \bibinfo {author} {\bibfnamefont {S.}~\bibnamefont {Fan}},\ }\bibfield
  {title} {\bibinfo {title} {Full inversion of a two-level atom with a
  single-photon pulse in one-dimensional geometries},\ }\href
  {https://doi.org/10.1103/PhysRevA.82.033804} {\bibfield  {journal} {\bibinfo
  {journal} {Phys. Rev. A}\ }\textbf {\bibinfo {volume} {82}},\ \bibinfo
  {pages} {033804} (\bibinfo {year} {2010})}\BibitemShut {NoStop}%
\bibitem [{\citenamefont {Nysteen}\ \emph {et~al.}(2015)\citenamefont
  {Nysteen}, \citenamefont {Kristensen}, \citenamefont {McCutcheon},
  \citenamefont {Kaer},\ and\ \citenamefont {Mørk}}]{Nysteen2015}%
  \BibitemOpen
  \bibfield  {author} {\bibinfo {author} {\bibfnamefont {A.}~\bibnamefont
  {Nysteen}}, \bibinfo {author} {\bibfnamefont {P.~T.}\ \bibnamefont
  {Kristensen}}, \bibinfo {author} {\bibfnamefont {D.~P.~S.}\ \bibnamefont
  {McCutcheon}}, \bibinfo {author} {\bibfnamefont {P.}~\bibnamefont {Kaer}},\
  and\ \bibinfo {author} {\bibfnamefont {J.}~\bibnamefont {Mørk}},\ }\bibfield
   {title} {\bibinfo {title} {Scattering of two photons on a quantum emitter in
  a one-dimensional waveguide: exact dynamics and induced correlations},\
  }\href {https://doi.org/10.1088/1367-2630/17/2/023030} {\bibfield  {journal}
  {\bibinfo  {journal} {New Journal of Physics}\ }\textbf {\bibinfo {volume}
  {17}},\ \bibinfo {pages} {023030} (\bibinfo {year} {2015})}\BibitemShut
  {NoStop}%
\bibitem [{\citenamefont {Chen}\ \emph {et~al.}(2011)\citenamefont {Chen},
  \citenamefont {Wubs}, \citenamefont {Mørk},\ and\ \citenamefont
  {Koenderink}}]{Chen_2011}%
  \BibitemOpen
  \bibfield  {author} {\bibinfo {author} {\bibfnamefont {Y.}~\bibnamefont
  {Chen}}, \bibinfo {author} {\bibfnamefont {M.}~\bibnamefont {Wubs}}, \bibinfo
  {author} {\bibfnamefont {J.}~\bibnamefont {Mørk}},\ and\ \bibinfo {author}
  {\bibfnamefont {A.~F.}\ \bibnamefont {Koenderink}},\ }\bibfield  {title}
  {\bibinfo {title} {Coherent single-photon absorption by single emitters
  coupled to one-dimensional nanophotonic waveguides},\ }\href
  {https://doi.org/10.1088/1367-2630/13/10/103010} {\bibfield  {journal}
  {\bibinfo  {journal} {New Journal of Physics}\ }\textbf {\bibinfo {volume}
  {13}},\ \bibinfo {pages} {103010} (\bibinfo {year} {2011})}\BibitemShut
  {NoStop}%
\bibitem [{\citenamefont {Gardiner}\ and\ \citenamefont
  {Zoller}(2010)}]{gardiner_zoller_2010}%
  \BibitemOpen
  \bibfield  {author} {\bibinfo {author} {\bibfnamefont {C.~W.}\ \bibnamefont
  {Gardiner}}\ and\ \bibinfo {author} {\bibfnamefont {P.}~\bibnamefont
  {Zoller}},\ }\href@noop {} {\emph {\bibinfo {title} {Quantum noise: a
  handbook of {Markovian} and {non-Markovian} quantum stochastic methods with
  applications to quantum optics}}}\ (\bibinfo  {publisher} {Springer},\
  \bibinfo {address} {Berlin},\ \bibinfo {year} {2010})\BibitemShut {NoStop}%
\bibitem [{\citenamefont {Barkemeyer}\ \emph {et~al.}(2022)\citenamefont
  {Barkemeyer}, \citenamefont {Knorr},\ and\ \citenamefont
  {Carmele}}]{PhysRevA.106.023708}%
  \BibitemOpen
  \bibfield  {author} {\bibinfo {author} {\bibfnamefont {K.}~\bibnamefont
  {Barkemeyer}}, \bibinfo {author} {\bibfnamefont {A.}~\bibnamefont {Knorr}},\
  and\ \bibinfo {author} {\bibfnamefont {A.}~\bibnamefont {Carmele}},\
  }\bibfield  {title} {\bibinfo {title} {Heisenberg treatment of multiphoton
  pulses in waveguide {QED} with time-delayed feedback},\ }\href
  {https://doi.org/10.1103/PhysRevA.106.023708} {\bibfield  {journal} {\bibinfo
   {journal} {Phys. Rev. A}\ }\textbf {\bibinfo {volume} {106}},\ \bibinfo
  {pages} {023708} (\bibinfo {year} {2022})}\BibitemShut {NoStop}%
\bibitem [{\citenamefont {Iblisdir}\ \emph {et~al.}(2007)\citenamefont
  {Iblisdir}, \citenamefont {Orús},\ and\ \citenamefont
  {Latorre}}]{iblisdir_matrix_2007}%
  \BibitemOpen
  \bibfield  {author} {\bibinfo {author} {\bibfnamefont {S.}~\bibnamefont
  {Iblisdir}}, \bibinfo {author} {\bibfnamefont {R.}~\bibnamefont {Orús}},\
  and\ \bibinfo {author} {\bibfnamefont {J.~I.}\ \bibnamefont {Latorre}},\
  }\bibfield  {title} {\bibinfo {title} {Matrix product states algorithms and
  continuous systems},\ }\href {https://doi.org/10.1103/PhysRevB.75.104305}
  {\bibfield  {journal} {\bibinfo  {journal} {Physical Review B}\ }\textbf
  {\bibinfo {volume} {75}},\ \bibinfo {pages} {104305} (\bibinfo {year}
  {2007})}\BibitemShut {NoStop}%
\bibitem [{\citenamefont {Guimond}\ \emph {et~al.}(2017)\citenamefont
  {Guimond}, \citenamefont {Pletyukhov}, \citenamefont {Pichler},\ and\
  \citenamefont {Zoller}}]{Guimond_2017}%
  \BibitemOpen
  \bibfield  {author} {\bibinfo {author} {\bibfnamefont {P.-O.}\ \bibnamefont
  {Guimond}}, \bibinfo {author} {\bibfnamefont {M.}~\bibnamefont {Pletyukhov}},
  \bibinfo {author} {\bibfnamefont {H.}~\bibnamefont {Pichler}},\ and\ \bibinfo
  {author} {\bibfnamefont {P.}~\bibnamefont {Zoller}},\ }\bibfield  {title}
  {\bibinfo {title} {Delayed coherent quantum feedback from a scattering theory
  and a matrix product state perspective},\ }\href
  {https://doi.org/10.1088/2058-9565/aa7f03} {\bibfield  {journal} {\bibinfo
  {journal} {Quantum Science and Technology}\ }\textbf {\bibinfo {volume}
  {2}},\ \bibinfo {pages} {044012} (\bibinfo {year} {2017})}\BibitemShut
  {NoStop}%
\bibitem [{\citenamefont {Barkemeyer}\ \emph {et~al.}(2021)\citenamefont
  {Barkemeyer}, \citenamefont {Knorr},\ and\ \citenamefont
  {Carmele}}]{PhysRevA.103.033704}%
  \BibitemOpen
  \bibfield  {author} {\bibinfo {author} {\bibfnamefont {K.}~\bibnamefont
  {Barkemeyer}}, \bibinfo {author} {\bibfnamefont {A.}~\bibnamefont {Knorr}},\
  and\ \bibinfo {author} {\bibfnamefont {A.}~\bibnamefont {Carmele}},\
  }\bibfield  {title} {\bibinfo {title} {Strongly entangled system-reservoir
  dynamics with multiphoton pulses beyond the two-excitation limit: Exciting
  the atom-photon bound state},\ }\href
  {https://doi.org/10.1103/PhysRevA.103.033704} {\bibfield  {journal} {\bibinfo
   {journal} {Phys. Rev. A}\ }\textbf {\bibinfo {volume} {103}},\ \bibinfo
  {pages} {033704} (\bibinfo {year} {2021})}\BibitemShut {NoStop}%
\bibitem [{\citenamefont {Vanderstraeten}\ \emph {et~al.}(2019)\citenamefont
  {Vanderstraeten}, \citenamefont {Haegeman},\ and\ \citenamefont
  {Verstraete}}]{vanderstraeten_tangent-space_2019}%
  \BibitemOpen
  \bibfield  {author} {\bibinfo {author} {\bibfnamefont {L.}~\bibnamefont
  {Vanderstraeten}}, \bibinfo {author} {\bibfnamefont {J.}~\bibnamefont
  {Haegeman}},\ and\ \bibinfo {author} {\bibfnamefont {F.}~\bibnamefont
  {Verstraete}},\ }\bibfield  {title} {\bibinfo {title} {Tangent-space methods
  for uniform matrix product states},\ }\href
  {https://doi.org/10.21468/SciPostPhysLectNotes.7} {\bibfield  {journal}
  {\bibinfo  {journal} {SciPost Physics Lecture Notes}\ ,\ \bibinfo {pages}
  {7}} (\bibinfo {year} {2019})}\BibitemShut {NoStop}%
\bibitem [{\citenamefont {Orús}(2014)}]{orus_practical_2014}%
  \BibitemOpen
  \bibfield  {author} {\bibinfo {author} {\bibfnamefont {R.}~\bibnamefont
  {Orús}},\ }\bibfield  {title} {\bibinfo {title} {A practical introduction to
  tensor networks: Matrix product states and projected entangled pair states},\
  }\href {https://doi.org/10.1016/j.aop.2014.06.013} {\bibfield  {journal}
  {\bibinfo  {journal} {Annals of Physics}\ }\textbf {\bibinfo {volume}
  {349}},\ \bibinfo {pages} {117} (\bibinfo {year} {2014})}\BibitemShut
  {NoStop}%
\bibitem [{\citenamefont {Yang}\ \emph {et~al.}(2018)\citenamefont {Yang},
  \citenamefont {Binder}, \citenamefont {Narasimhachar},\ and\ \citenamefont
  {Gu}}]{yang_matrix_2018}%
  \BibitemOpen
  \bibfield  {author} {\bibinfo {author} {\bibfnamefont {C.}~\bibnamefont
  {Yang}}, \bibinfo {author} {\bibfnamefont {F.~C.}\ \bibnamefont {Binder}},
  \bibinfo {author} {\bibfnamefont {V.}~\bibnamefont {Narasimhachar}},\ and\
  \bibinfo {author} {\bibfnamefont {M.}~\bibnamefont {Gu}},\ }\bibfield
  {title} {\bibinfo {title} {Matrix product states for quantum stochastic
  modeling},\ }\href {https://doi.org/10.1103/PhysRevLett.121.260602}
  {\bibfield  {journal} {\bibinfo  {journal} {Physical Review Letters}\
  }\textbf {\bibinfo {volume} {121}},\ \bibinfo {pages} {260602} (\bibinfo
  {year} {2018})}\BibitemShut {NoStop}%
\bibitem [{\citenamefont {Arranz~Regidor}\ \emph {et~al.}(2024)\citenamefont
  {Arranz~Regidor}, \citenamefont {Knorr},\ and\ \citenamefont
  {Hughes}}]{sofia2024letter}%
  \BibitemOpen
  \bibfield  {author} {\bibinfo {author} {\bibfnamefont {S.}~\bibnamefont
  {Arranz~Regidor}}, \bibinfo {author} {\bibfnamefont {A.}~\bibnamefont
  {Knorr}},\ and\ \bibinfo {author} {\bibfnamefont {S.}~\bibnamefont
  {Hughes}},\ }\bibfield  {title} {\bibinfo {title} {Dynamical spectra from one
  and two-photon {Fock} state pulses exciting a single chiral qubit in a
  waveguide}} (\bibinfo {year} {2024}),\ \bibinfo {note} {manuscript in
  preparation}\BibitemShut {NoStop}%
\bibitem [{\citenamefont {Liu}\ \emph {et~al.}(2024)\citenamefont {Liu},
  \citenamefont {Gustin}, \citenamefont {Liu}, \citenamefont {Li},
  \citenamefont {Yu}, \citenamefont {Ni}, \citenamefont {Niu}, \citenamefont
  {Hughes}, \citenamefont {Wang},\ and\ \citenamefont {Liu}}]{Liu2024}%
  \BibitemOpen
  \bibfield  {author} {\bibinfo {author} {\bibfnamefont {S.}~\bibnamefont
  {Liu}}, \bibinfo {author} {\bibfnamefont {C.}~\bibnamefont {Gustin}},
  \bibinfo {author} {\bibfnamefont {H.}~\bibnamefont {Liu}}, \bibinfo {author}
  {\bibfnamefont {X.}~\bibnamefont {Li}}, \bibinfo {author} {\bibfnamefont
  {Y.}~\bibnamefont {Yu}}, \bibinfo {author} {\bibfnamefont {H.}~\bibnamefont
  {Ni}}, \bibinfo {author} {\bibfnamefont {Z.}~\bibnamefont {Niu}}, \bibinfo
  {author} {\bibfnamefont {S.}~\bibnamefont {Hughes}}, \bibinfo {author}
  {\bibfnamefont {X.}~\bibnamefont {Wang}},\ and\ \bibinfo {author}
  {\bibfnamefont {J.}~\bibnamefont {Liu}},\ }\bibfield  {title} {\bibinfo
  {title} {Dynamic resonance fluorescence in solid-state cavity quantum
  electrodynamics},\ }\bibfield  {journal} {\bibinfo  {journal} {Nature
  Photonics}\ }\href {https://doi.org/10.1038/s41566-023-01359-x}
  {10.1038/s41566-023-01359-x} (\bibinfo {year} {2024})\BibitemShut {NoStop}%
\bibitem [{\citenamefont {Boos}\ \emph {et~al.}(2024)\citenamefont {Boos},
  \citenamefont {Kim}, \citenamefont {Bracht}, \citenamefont {Sbresny},
  \citenamefont {Kaspari}, \citenamefont {Cygorek}, \citenamefont {Riedl},
  \citenamefont {Bopp}, \citenamefont {Rauhaus}, \citenamefont {Calcagno},
  \citenamefont {Finley}, \citenamefont {Reiter},\ and\ \citenamefont
  {M\"uller}}]{PhysRevLett.132.053602}%
  \BibitemOpen
  \bibfield  {author} {\bibinfo {author} {\bibfnamefont {K.}~\bibnamefont
  {Boos}}, \bibinfo {author} {\bibfnamefont {S.~K.}\ \bibnamefont {Kim}},
  \bibinfo {author} {\bibfnamefont {T.}~\bibnamefont {Bracht}}, \bibinfo
  {author} {\bibfnamefont {F.}~\bibnamefont {Sbresny}}, \bibinfo {author}
  {\bibfnamefont {J.~M.}\ \bibnamefont {Kaspari}}, \bibinfo {author}
  {\bibfnamefont {M.}~\bibnamefont {Cygorek}}, \bibinfo {author} {\bibfnamefont
  {H.}~\bibnamefont {Riedl}}, \bibinfo {author} {\bibfnamefont {F.~W.}\
  \bibnamefont {Bopp}}, \bibinfo {author} {\bibfnamefont {W.}~\bibnamefont
  {Rauhaus}}, \bibinfo {author} {\bibfnamefont {C.}~\bibnamefont {Calcagno}},
  \bibinfo {author} {\bibfnamefont {J.~J.}\ \bibnamefont {Finley}}, \bibinfo
  {author} {\bibfnamefont {D.~E.}\ \bibnamefont {Reiter}},\ and\ \bibinfo
  {author} {\bibfnamefont {K.}~\bibnamefont {M\"uller}},\ }\bibfield  {title}
  {\bibinfo {title} {Signatures of dynamically dressed states},\ }\href
  {https://doi.org/10.1103/PhysRevLett.132.053602} {\bibfield  {journal}
  {\bibinfo  {journal} {Phys. Rev. Lett.}\ }\textbf {\bibinfo {volume} {132}},\
  \bibinfo {pages} {053602} (\bibinfo {year} {2024})}\BibitemShut {NoStop}%
\bibitem [{\citenamefont {Mehrabad}\ \emph {et~al.}(2020)\citenamefont
  {Mehrabad}, \citenamefont {Foster}, \citenamefont {Dost}, \citenamefont
  {Clarke}, \citenamefont {Patil}, \citenamefont {Fox}, \citenamefont
  {Skolnick},\ and\ \citenamefont {Wilson}}]{JalaliMehrabad:20}%
  \BibitemOpen
  \bibfield  {author} {\bibinfo {author} {\bibfnamefont {M.~J.}\ \bibnamefont
  {Mehrabad}}, \bibinfo {author} {\bibfnamefont {A.~P.}\ \bibnamefont
  {Foster}}, \bibinfo {author} {\bibfnamefont {R.}~\bibnamefont {Dost}},
  \bibinfo {author} {\bibfnamefont {E.}~\bibnamefont {Clarke}}, \bibinfo
  {author} {\bibfnamefont {P.~K.}\ \bibnamefont {Patil}}, \bibinfo {author}
  {\bibfnamefont {A.~M.}\ \bibnamefont {Fox}}, \bibinfo {author} {\bibfnamefont
  {M.~S.}\ \bibnamefont {Skolnick}},\ and\ \bibinfo {author} {\bibfnamefont
  {L.~R.}\ \bibnamefont {Wilson}},\ }\bibfield  {title} {\bibinfo {title}
  {Chiral topological photonics with an embedded quantum emitter},\ }\href
  {https://doi.org/10.1364/OPTICA.393035} {\bibfield  {journal} {\bibinfo
  {journal} {Optica}\ }\textbf {\bibinfo {volume} {7}},\ \bibinfo {pages}
  {1690} (\bibinfo {year} {2020})}\BibitemShut {NoStop}%
\bibitem [{\citenamefont {Lodahl}\ \emph {et~al.}(2015)\citenamefont {Lodahl},
  \citenamefont {Mahmoodian},\ and\ \citenamefont
  {Stobbe}}]{RevModPhys.87.347}%
  \BibitemOpen
  \bibfield  {author} {\bibinfo {author} {\bibfnamefont {P.}~\bibnamefont
  {Lodahl}}, \bibinfo {author} {\bibfnamefont {S.}~\bibnamefont {Mahmoodian}},\
  and\ \bibinfo {author} {\bibfnamefont {S.}~\bibnamefont {Stobbe}},\
  }\bibfield  {title} {\bibinfo {title} {Interfacing single photons and single
  quantum dots with photonic nanostructures},\ }\href
  {https://doi.org/10.1103/RevModPhys.87.347} {\bibfield  {journal} {\bibinfo
  {journal} {Rev. Mod. Phys.}\ }\textbf {\bibinfo {volume} {87}},\ \bibinfo
  {pages} {347} (\bibinfo {year} {2015})}\BibitemShut {NoStop}%
\bibitem [{\citenamefont {Mnaymneh}\ \emph {et~al.}(2019)\citenamefont
  {Mnaymneh}, \citenamefont {Dalacu}, \citenamefont {McKee}, \citenamefont
  {Lapointe}, \citenamefont {Haffouz}, \citenamefont {Weber}, \citenamefont
  {Northeast}, \citenamefont {Poole}, \citenamefont {Aers},\ and\ \citenamefont
  {Williams}}]{Mnaymneh2019}%
  \BibitemOpen
  \bibfield  {author} {\bibinfo {author} {\bibfnamefont {K.}~\bibnamefont
  {Mnaymneh}}, \bibinfo {author} {\bibfnamefont {D.}~\bibnamefont {Dalacu}},
  \bibinfo {author} {\bibfnamefont {J.}~\bibnamefont {McKee}}, \bibinfo
  {author} {\bibfnamefont {J.}~\bibnamefont {Lapointe}}, \bibinfo {author}
  {\bibfnamefont {S.}~\bibnamefont {Haffouz}}, \bibinfo {author} {\bibfnamefont
  {J.~F.}\ \bibnamefont {Weber}}, \bibinfo {author} {\bibfnamefont {D.~B.}\
  \bibnamefont {Northeast}}, \bibinfo {author} {\bibfnamefont {P.~J.}\
  \bibnamefont {Poole}}, \bibinfo {author} {\bibfnamefont {G.~C.}\ \bibnamefont
  {Aers}},\ and\ \bibinfo {author} {\bibfnamefont {R.~L.}\ \bibnamefont
  {Williams}},\ }\bibfield  {title} {\bibinfo {title} {On-chip integration of
  single photon sources via evanescent coupling of tapered nanowires to {SiN}
  waveguides},\ }\href {https://doi.org/10.1002/qute.201900021} {\bibfield
  {journal} {\bibinfo  {journal} {Advanced Quantum Technologies}\ }\textbf
  {\bibinfo {volume} {3}},\ \bibinfo {pages} {1900021} (\bibinfo {year}
  {2019})}\BibitemShut {NoStop}%
\bibitem [{\citenamefont {Yao}\ \emph {et~al.}(2010)\citenamefont {Yao},
  \citenamefont {Manga~Rao},\ and\ \citenamefont
  {Hughes}}]{https://doi.org/10.1002/lpor.200810081}%
  \BibitemOpen
  \bibfield  {author} {\bibinfo {author} {\bibfnamefont {P.}~\bibnamefont
  {Yao}}, \bibinfo {author} {\bibfnamefont {V.~S.~C.}\ \bibnamefont
  {Manga~Rao}},\ and\ \bibinfo {author} {\bibfnamefont {S.}~\bibnamefont
  {Hughes}},\ }\bibfield  {title} {\bibinfo {title} {On-chip single photon
  sources using planar photonic crystals and single quantum dots},\ }\href
  {https://doi.org/https://doi.org/10.1002/lpor.200810081} {\bibfield
  {journal} {\bibinfo  {journal} {Laser \& Photonics Reviews}\ }\textbf
  {\bibinfo {volume} {4}},\ \bibinfo {pages} {499} (\bibinfo {year}
  {2010})}\BibitemShut {NoStop}%
\bibitem [{\citenamefont {Laferrière}\ \emph {et~al.}(2020)\citenamefont
  {Laferrière}, \citenamefont {Yeung}, \citenamefont {Giner}, \citenamefont
  {Haffouz}, \citenamefont {Lapointe}, \citenamefont {Aers}, \citenamefont
  {Poole}, \citenamefont {Williams},\ and\ \citenamefont
  {Dalacu}}]{doi:10.1021/acs.nanolett.0c00607}%
  \BibitemOpen
  \bibfield  {author} {\bibinfo {author} {\bibfnamefont {P.}~\bibnamefont
  {Laferrière}}, \bibinfo {author} {\bibfnamefont {E.}~\bibnamefont {Yeung}},
  \bibinfo {author} {\bibfnamefont {L.}~\bibnamefont {Giner}}, \bibinfo
  {author} {\bibfnamefont {S.}~\bibnamefont {Haffouz}}, \bibinfo {author}
  {\bibfnamefont {J.}~\bibnamefont {Lapointe}}, \bibinfo {author}
  {\bibfnamefont {G.~C.}\ \bibnamefont {Aers}}, \bibinfo {author}
  {\bibfnamefont {P.~J.}\ \bibnamefont {Poole}}, \bibinfo {author}
  {\bibfnamefont {R.~L.}\ \bibnamefont {Williams}},\ and\ \bibinfo {author}
  {\bibfnamefont {D.}~\bibnamefont {Dalacu}},\ }\bibfield  {title} {\bibinfo
  {title} {Multiplexed single-photon source based on multiple quantum dots
  embedded within a single nanowire},\ }\href
  {https://doi.org/10.1021/acs.nanolett.0c00607} {\bibfield  {journal}
  {\bibinfo  {journal} {Nano Letters}\ }\textbf {\bibinfo {volume} {20}},\
  \bibinfo {pages} {3688} (\bibinfo {year} {2020})}\BibitemShut {NoStop}%
\bibitem [{\citenamefont {Arranz~Regidor}\ \emph {et~al.}(2021)\citenamefont
  {Arranz~Regidor}, \citenamefont {Crowder}, \citenamefont {Carmichael},\ and\
  \citenamefont {Hughes}}]{PhysRevResearch.3.023030}%
  \BibitemOpen
  \bibfield  {author} {\bibinfo {author} {\bibfnamefont {S.}~\bibnamefont
  {Arranz~Regidor}}, \bibinfo {author} {\bibfnamefont {G.}~\bibnamefont
  {Crowder}}, \bibinfo {author} {\bibfnamefont {H.}~\bibnamefont
  {Carmichael}},\ and\ \bibinfo {author} {\bibfnamefont {S.}~\bibnamefont
  {Hughes}},\ }\bibfield  {title} {\bibinfo {title} {Modeling quantum
  light-matter interactions in waveguide {QED} with retardation, nonlinear
  interactions, and a time-delayed feedback: Matrix product states versus a
  space-discretized waveguide model},\ }\href
  {https://doi.org/10.1103/PhysRevResearch.3.023030} {\bibfield  {journal}
  {\bibinfo  {journal} {Phys. Rev. Res.}\ }\textbf {\bibinfo {volume} {3}},\
  \bibinfo {pages} {023030} (\bibinfo {year} {2021})}\BibitemShut {NoStop}%
\bibitem [{\citenamefont {Pichler}\ and\ \citenamefont
  {Zoller}(2016)}]{PhysRevLett.116.093601}%
  \BibitemOpen
  \bibfield  {author} {\bibinfo {author} {\bibfnamefont {H.}~\bibnamefont
  {Pichler}}\ and\ \bibinfo {author} {\bibfnamefont {P.}~\bibnamefont
  {Zoller}},\ }\bibfield  {title} {\bibinfo {title} {Photonic circuits with
  time delays and quantum feedback},\ }\href
  {https://doi.org/10.1103/PhysRevLett.116.093601} {\bibfield  {journal}
  {\bibinfo  {journal} {Phys. Rev. Lett.}\ }\textbf {\bibinfo {volume} {116}},\
  \bibinfo {pages} {093601} (\bibinfo {year} {2016})}\BibitemShut {NoStop}%
\end{thebibliography}%

\end{document}